\RequirePackage[l2tabu, orthodox]{nag} 

\documentclass[twoside]{um} 
\PassOptionsToPackage{xcdraw}{xcolor}
\usepackage[table]{xcolor} 
\usepackage{blindtext} 
\usepackage{coffee4}    
\usepackage{float}
\usepackage{listings}
\usepackage{pgffor}
\usepackage{tabularx}
\usepackage{hyperref}
\usepackage{xurl}
\newcounter{SortListTotal}
\usepackage{multirow}
\usepackage{tikz}
\usepackage{pgfplots}
\usepackage{soul}
\pgfplotsset{compat=1.17}

\usetikzlibrary{shapes.geometric, arrows}
\usepackage[T1]{fontenc}
\usepackage[numbers]{natbib}  

\newcommand{\mysection}[2]
  {{\renewcommand{\sectionmark}[1]{}
    \section{#1}}\sectionmark{#2}}

\tikzstyle{startstop} = [rectangle, rounded corners, minimum width=3cm, minimum height=1cm,text centered, draw=black, fill=red!30]
\tikzstyle{process} = [rectangle, minimum width=3cm, minimum height=1cm, text centered, draw=black, fill=blue!30]
\tikzstyle{decision} = [diamond, minimum width=3cm, minimum height=1cm, text centered, draw=black, fill=green!30]
\tikzstyle{arrow} = [thick,->,>=stealth]

\DeclareFontShape{T1}{lato-TLF}{b}{sc}{ <-> ssub * lato-TLF/b/n }{}

\newcommand{\sortitem}[2]{\expandafter\def\csname SortListItem#1\endcsname{#2}\stepcounter{SortListTotal}}

\newcommand{\printsortlist}{\foreach\currentlistitem in{1,2,...,\value{SortListTotal}}{\item[\currentlistitem]\csname SortListItem\currentlistitem\endcsname}\setcounter{SortListTotal}{0}}

\colorlet{punct}{red!60!black}
\definecolor{background}{HTML}{EEEEEE}
\definecolor{delim}{RGB}{20,105,176}
\colorlet{numb}{magenta!60!black}
\usepackage{verbatimbox,caption,float,lipsum}

\lstdefinelanguage{json}{
    basicstyle=\small\ttfamily,
    numbers=left,
    numberstyle=\scriptsize,
    stepnumber=1,
    numbersep=8pt,
    showstringspaces=false,
    breaklines=true,
    frame=lines,
    backgroundcolor=\color{background},
    literate=
     *{0}{{{\color{numb}0}}}{1}
      {1}{{{\color{numb}1}}}{1}
      {2}{{{\color{numb}2}}}{1}
      {3}{{{\color{numb}3}}}{1}
      {4}{{{\color{numb}4}}}{1}
      {5}{{{\color{numb}5}}}{1}
      {6}{{{\color{numb}6}}}{1}
      {7}{{{\color{numb}7}}}{1}
      {8}{{{\color{numb}8}}}{1}
      {9}{{{\color{numb}9}}}{1}
      {:}{{{\color{punct}{:}}}}{1}
      {,}{{{\color{punct}{,}}}}{1}
      {\{}{{{\color{delim}{\{}}}}{1}
      {\}}{{{\color{delim}{\}}}}}{1}
      {[}{{{\color{delim}{[}}}}{1}
      {]}{{{\color{delim}{]}}}}{1},
}

\title{Decentralized Infrastructure\\ for Digital Notarizing, \\Signing and Sharing Files \\using Blockchain} 

\author{Cosmin-Iulian Irimia}                          
\supervisor{Prof. Dr. Adrian Iftene}      
\department{Faculty of Computer Science}    
\faculty{Faculty of Computer Science}    
\degree{Doctor of Philosophy}
\degreeFirstLine{A thesis submitted for the degree of} 
\degreeSecondLine{Doctor of Philosophy}     
\doctype{dissertation}       
\tagline{}
\degreedate{November, 2025}                        

\graphicspath{{./images/}{./chap1-introduction/images/}}   

\makeindex

\begin{document}

\frontmatter 
    \maketitle
    \begin{copyrightenv}
\end{copyrightenv}
       
    \begin{dedication}\\

{\large{To my family}}\\[5mm]
\hspace{10em} To my fiancée and my parents, for their love, patience\\
\hspace{10em} and belief in me through every step of this journey.\\[25mm]

{\large{To my teachers}}\\[5mm]
\hspace{10em} To my educators and teachers whose wisdom and passion\\
\hspace{10em} for knowledge have inspired my journey and shaped my\\ \hspace{10em} understanding of the world.\\[25mm]

{\large{To my students}}\\[5mm]
\hspace{10em} To my students, past and present, who have taught me as\\
\hspace{10em} much as I have taught them, whose successes have been my\\ \hspace{10em} greatest rewards, and whose struggles have reminded me \\
\hspace{10em} of the importance of patience, empathy, and perseverance.\\[15mm]

\end{dedication}

    
\begin{adjustwidth}{-1.2cm}{-1.2cm}
\begin{acknowledgements}
\vspace{0.5cm}
First of all, I would like to thank my supervisor, \textbf{Prof. Dr. Adrian Iftene}, for all the help, advice, and motivation that he has given me not just now, but throughout all these years. In addition to being my supervisor throughout all my academic degrees, he is also my mentor, idol and a true friend. Without his help, not only would I not have completed this work, but I don't think I would have even started it. Thank you from the bottom of my heart!

Secondly, I would like to express my gratitude to my guidance committee: \textbf{Prof. Dr. Lenuța Alboaie}, \textbf{Conf. Dr. Andrei Arusoaie}, and \textbf{Conf. Dr. Emanuel Onica}. Their insightful feedback, constant support, and deep expertise in distributed systems, blockchain, and cloud computing have been instrumental to my progress. I have always considered them an outstanding team of mentors, and I am truly grateful for the opportunity to learn from them.

Most importantly, I wish to express my deepest gratitude to the three most important people in my life: my fiancée, \textbf{Diana Paveliuc}, and my parents, \textbf{Constantin} and \textbf{Dumitrița}. Their unwavering belief in me, constant care, boundless empathy, and steadfast support, even in the most challenging circumstances, have been my greatest source of strength and inspiration.

I would like to extend my appreciation to \textbf{Lect. Dr. Cristian Frasinaru} for granting me the life-changing opportunity to teach Advanced Programming. His continuous support, confidence in my abilities, and trust in my perspectives have been invaluable throughout this journey.

Equally important, I want to thank \textbf{Mr. Florin Olariu}, a former teacher who has since become a valued friend. His exemplary approach to teaching profoundly influenced me and inspired me to pursue this career path, motivating me to pass on the same dedication and passion to future generations of students.

Lastly, I want to thank my dear friend \textbf{Mădălin Matei}, my closest friend throughout these years. Having someone I deeply respect and connect with on both personal and intellectual levels has been invaluable to my growth. I also want to thank \textbf{Cristian Simionescu}, a trusted colleague and friend, especially during our time together in the Doctoral School Council. His support and steady presence have meant a great deal, and I know they always will.

Finally, I would like to mention that the data processing and analysis in this thesis were supported by the Research Center with Integrated Techniques for the Investigation of Atmospheric Aerosols in Romania, under project SMIS 127324 - RECENT AIR (RA). I would like to thank \textbf{Prof. Dr. Cecilia Arsene} for her support, guidance, and patience during the last three years.

\end{acknowledgements}
\end{adjustwidth}   

\begin{abstract}

Traditional paper-based document management has long posed challenges related to security, authenticity, and efficiency. Despite advances in digitalization, official documents remain vulnerable to forgery, loss, and unauthorized access. This thesis proposes a decentralized infrastructure for  digital notarization, signing, and sharing of documents using blockchain technology. The research addresses key issues of transparency, immutability, and feasibility by defining system requirements, evaluating existing solutions, and proposing a novel architecture based on distributed systems.

The Introduction chapter outlines the motivation behind this research, highlighting the inefficiencies of current bureaucratic processes and the need for a secure, digital alternative. The Background \& Literature Overview presents an in-depth review of existing technologies, such as optical character recognition, digital signatures, and blockchain-based solutions, assessing their advantages and limitations. The System Requirements chapter defines the functional and non-functional requirements necessary for the proposed infrastructure, detailing key use cases and constraints.

The Proposed Architecture chapter introduces a distributed system design based on microservices, blockchain, and decentralized storage, ensuring scalability and flexibility. The Prototype Implementation discusses the practical development of core system components, including data extraction, document obfuscation, and cryptographic notarization. The Evaluation and Comparison chapter assesses the system’s performance, security, and usability, comparing it against existing solutions to demonstrate its advantages and areas for improvement. Finally, the Conclusions chapter reflects on the research contributions, critiques the system’s limitations, and proposes future directions for enhancing digital notarization and document management.

By combining cryptographic techniques with decentralized storage, this research contributes to the development of a more secure and efficient framework for managing official documents. The findings highlight the potential of blockchain-based digital notarization to streamline bureaucratic processes, mitigate security risks, and enhance user trust in digital document management.

\end{abstract}\if@openright\cleardoublepage\else\clearpage\fi
    \if@openright\cleardoublepage\else\clearpage\fi
    \thispagestyle{empty}
    \mbox{}
    \if@openright\cleardoublepage\else\clearpage\fi

    \tableofcontents*\if@openright\cleardoublepage\else\clearpage\fi

\listoffigures*

\listoftables* 


\chapter*{List of Abbreviations}
\markboth{List of Abbreviations}{List of Abbreviations}
               
\begin{acronym}\itemsep-20pt\parsep-20pt 

\acro{ANN}{Artificial Neural Networks}
\acro{API}{Application programming interface}
\acro{BLSTM}{Bidirectional Long-Short Term Memory}
\acro{CCPA}{California Consumer Privacy Act}
\acro{CER}{Character Error Rate}
\acro{CNN}{WordErrorRate}
\acro{DSA}{Digital Signature Algorithm}
\acro{DWT}{Discrete Wavelet Transform}
\acro{EDM}{Euclidean Distance Metric}
\acro{eIDAS}{electronic IDentification, Authentication and trust Services}
\acro{GDPR}{General Data Protection Regulation}
\acro{HDR}{High Dynamic Range}
\acro{IoT}{Internet of Things}
\acro{IPFS}{InterPlanetary File System}
\acro{JPEG}{Joint Photographic Experts Group}
\acro{JWT}{JSON Web Token}
\acro{LLM}{Large Language Model}
\acro{OCR}{Optical Character Recognition}
\acro{P2P}{Peer-To-Peer}
\acro{PDF}{Portable Document Format}
\acro{PKI}{Public key infrastructure}
\acro{POW}{Proof Of Work}
\acro{PSOA}{Particle Swarm Optimization Approach}
\acro{QR}{Quick-Response code}
\acro{RNN}{Recurrent Neural Network}
\acro{RSA}{Rivest–Shamir–Adleman}
\acro{SD}{Secure Digital}
\acro{SQL}{Structured Query Language}
\acro{SVM}{Support Vector Machines}
\acro{TLS}{Transport Layer Security}
\acro{UI}{User interface}
\acro{URL}{Uniform Resource Locator}
\acro{UX}{User experience}
\acro{WER}{Word Error Rate}
\acro{XML}{Extensible Markup Language}
\acro{XOR}{Exclusive or}
\end{acronym}


\chapter*{List of Publications}
\markboth{List of Publications}{List of Publications}

\begin{itemize}

\subsection{Journals}

\item 
    \textbf{Irimia Cosmin-Iulian}, Iftene Adrian. 2024. \\
    \textit{Decentralized Infrastructure for Digital Notarizing, Signing, and Sharing Documents Securely Using Microservices and Blockchain} \\
    IEEE Access Journal - (Ranked Scopus quartile Q2) - 4 points\\
    \url{https://ieeexplore.ieee.org/document/10804153}\\



\subsection{Conferences}
\subsubsection{Ranked B}

\item 
    Ungureanu Delia-Elena, Aflori Irina, \textbf{Irimia Cosmin-Iulian}. 2024. \\
    \textit{Document Identity Provider: an Anti-Tampering Solution.} \\
    28th International Conference on Knowledge-Based and Intelligent Information \& Engineering Systems\\
    (KES 2024) 11-13 September, 2024, Seville, Spain - (Ranked B) - 4 points\\ 
    \url{www.sciencedirect.com/science/article/pii/S1877050924023974}\\

\item 
    \textbf{Irimia Cosmin-Iulian}, Bejan Luciana, Iftene Adrian. 2023. \\
    \textit{BlockchainPedia: A Comprehensive Framework for Blockchain Network Comparison.} \\
    In Proceedings of 27th International Conference on Knowledge-Based and Intelligent Information \& Engineering Systems\\
    (KES 2023) 6-8 September 2023, Athens, Greece - (Ranked B) - 4 points\\
    \url{www.sciencedirect.com/science/article/pii/S1877050923013790} \\

\item 
    \textbf{Irimia Cosmin-Iulian}, Harbuzariu Florin, Hazi Ionut, Iftene Adrian. 2022. \\
    \textit{Official Document Identification and Data Extraction using Templates and OCR.} \\
    In Proceedings of 26th International Conference on Knowledge-Based and Intelligent Information \& Engineering Systems\\
    (KES 2022) 7-9 September 2022, Verona, Italy - (Ranked B) - 2 points\\
    \url{www.sciencedirect.com/science/article/pii/S1877050922010973} \\

\item 
    \textbf{Irimia Cosmin-Iulian}, Irimia Roxana, Milea Robert, Ilas Silviu, Vasiliu Ana, Iftene Adrian. 2022. \\
    \textit{Obfuscation of Documents using Randomly Generated Steps.} \\
    In Proceedings of 26th International Conference on Knowledge-Based and Intelligent Information \& Engineering Systems\\
    (KES 2022) 7-9 September 2022, Verona, Italy - (Ranked B) - 1 point\\
    
\subsubsection{Ranked C}


\item 
    Harbuzariu Florin, \textbf{Irimia Cosmin-Iulian}, Iftene Adrian. 2023. \\
    \textit{Official Document Text Extraction using Templates and Optical Character Recognition.} \\
    In International Conference on INnovations in Intelligent SysTems and Applications\\
    (INISTA 2023) Hammamet, Tunisia 20-23 September - (Ranked C) - 2 points\\
    \url{https://ieeexplore.ieee.org/document/10310514} \\

\subsubsection{Ranked D}





 \item 
    \textbf{Irimia Cosmin-Iulian}. 2022. \\
    \textit{Decentralized infrastructure for digital notarizing, signing and sharing files securely using Blockchain} \\
    In Proceedings of Symposium on Logic and Artificial Intelligence (SLAI)\\ January 12-16, 2022, Louisiana, USA - (Ranked D) - 1 point\\



\subsubsection{Points}
Total (D rank capped at 2 points): 18 points

\end{itemize}

\chapter*{Domain Related Publications}
\markboth{Domain Related Publications}{Domain Related Publications}

\begin{itemize}

\item 
Jufa Emanuel, \textbf{Irimia Cosmin-Iulian}, Matei Madalin. 2025. \\
\textit{DApp for High-Value Product Authentication} \\
In Proceedings of the 2025 IEEE International Symposium on INnovations in Intelligent SysTems and Applications (INISTA-25) (Ranked C) - 2 points \\
Research used in the Decentralized Product Validation System \\

\item 
Cărăușu Ana-Mădălina, \textbf{Irimia Cosmin-Iulian}, Iftene Adrian. 2025. \\
\textit{Towards Digital Sovereignty: Inter-cloud Migration To European Infrastructure} \\
In Proceedings of the 2025 IEEE International Symposium on INnovations in Intelligent SysTems and Applications (INISTA-25)
(Ranked C) - 2 points\\
Research used in the Cloud Infrastructure Design for DocChain \\

\item 
Petrea Daniela, \textbf{Irimia Cosmin-Iulian}, Iftene Adrian. 2025. \\
\textit{Seeing Beyond The Data: Enhancing Data Visualization With Natural Language Understanding} \\
In Proceedings of the 2025 IEEE International Symposium on INnovations in Intelligent SysTems and Applications (INISTA-25) 
(Ranked C) - 2 points \\
Research used in the Log Analysis \\

\newpage

\item 
\textbf{Irimia Cosmin-Iulian}, Matei Andrei Madalin. 2025. \\
\textit{Unsupervised Deep Learning for Web Server Log Anomaly Detection} \\
In Proceedings of the 2025 International Workshop on Information and Intelligent Systems (WIIS-25) 
(Ranked D) - 1 point \\
Research used in the Security Module for Anomaly Detection in Log Streams \\

\item 
    Ivanușcă Teodor, \textbf{Irimia Cosmin-Iulian}. 2024. \\
    \textit{The Impact of Prompting Techniques on the Security of the LLMs and the Systems to Which They Belong} \\
    Applied Sciences Journal, MDPI - (Ranked Scopus quartile Q1) - 8 point \\
    Research used in the Text-Extraction from Official Documents \\

\item 
    Cărăușu Madalina, Haiura Isabela, Petrea Daniela, \textbf{Irimia Cosmin-Iulian}, Iftene Adrian. 2024. \\
    \textit{ShareFactory - Efficient and Accessible Content-Sharing Platform.} \\
    In Proceedings of the 10th Balkan Conference in Informatics \\
    (BCI 2024), 4-6 September 2024, Craiova, Romania - (Ranked D) - 1 point \\
    Research used in the Time-based sharing for the Deobfucated documents \\

 \item 
    \textbf{Irimia Cosmin-Iulian}, Iftene Adrian, Gîfu Daniela. 2021. \\
    \textit{A Large-Scale E-voting System Based on Blockchain.} \\
    In 29th International Conference on Information Systems Development (ISD2021)\\ 
    September 8-10, Valencia, Spain - (Ranked D) - 1 point \\
    Research used in the Ethereum Smart Contracts Deployment \\
    
 \end{itemize}

\subsubsection{Points}
Total (D rank capped at 2 points): 15 points

\if@openright\cleardoublepage\else\clearpage\fi

\pagestyle{umpage}
\mainmatter 
\if@openright\cleardoublepage\else\clearpage\fi
\thispagestyle{empty}
\mbox{}

\chapter{Introduction}

\section{Motivation} 

The big advantage of electronic documents is their ease of access compared to physical documents. They can exist as physical memories or be stored in the cloud, allowing us to access them from multiple devices and share them instantly with anyone we want. Aside from the possibility of quick access, even searching for a lost electronic document is easier than searching for a physical one. Modern software allows searching by name, content, and extension, and a deleted document can be recovered as long as the memory area has not already been overwritten - something that cannot be said about a burned or a damaged document. Another advantage of electronic documents \cite{bibitem_18} is the ability to store large amounts of data in very little physical space. In just a few decades, digital storage has proven its superiority, evolving from hard drives that occupied several meters of space in a room to \ac{SD} cards measuring only a few millimeters and capable of storing up to a terabyte of data. For instance, a memory the size of an A4 page can store around 75 million digital pages \cite{bibitem_3}, saving approximately 50,000 trees and reducing the weight of processed paper by around 250 tons. 

The pandemic which profoundly impacted the world for more than two years has transformed many aspects of daily life. Among the most notable changes are those in communication practices and how official entities handle documentation. Rapid adaptation to changing circumstances has driven an acceleration toward electronic formats, enabling remote process management by individuals and organizations. This shift has not only removed the need for a physical presence but also redefined the standards of efficiency, accessibility, and convenience in professional and administrative interactions. The influence of this period still shapes how we communicate and document in an increasingly digital world.

One of the most common yet sensitive scenarios in digital identity is age verification. For example, a person may need to prove they are over 18 to access a restricted service—such as buying alcohol, registering for a platform, or entering a venue—without disclosing unnecessary information like their full address, ID number, or document series. Traditional identity verification often exposes the entire identity document, creating privacy risks. A better approach is to enable selective disclosure, where only the relevant facts (e.g., “user is over 18”) are revealed in a verifiable, privacy-preserving way. This is the type of use case our system is designed to support.

While simple solutions exist, this research aims to fundamentally transform digital file sharing by addressing privacy, security, and data minimization. One potential improvement would be the ability to share documents securely while revealing only the necessary information. For example, when visiting a bank to exchange currency, an individual could present a digitally signed version of their ID card that includes only their name, photograph and date of birth, omitting details like their address. 

Addressing these issues requires tackling them incrementally, starting with the challenge of sharing only parts of a file's content. For instance, in the case of photographs, one could apply layers of random visual transformations, such as adjusting contrast, adding effects like blur \cite{bibitem_9} or pixelation, or altering other properties. These transformations would be reversible, allowing the original content to be retrieved when needed. Additionally, mechanisms for verifying document integrity \cite{Mthethwa2018}, such as signatures of the original content, could be incorporated. The proposed solution involves obfuscating different parts of an image using multiple algorithms, each with separate encryption keys. This approach allows for the partial deobfuscation of content, enabling documents to be shared either fully or partially obfuscated, with the assurance that they can be securely restored to their original state. Thus, encrypted documents could be safely transmitted, even as email attachments, without risking exposure of sensitive information to attackers.

For document signing, existing algorithms or custom-built solutions could be employed, though a detailed comparison of available options is necessary. Another critical component is the development of a distributed system for document storage. Such a system would use a ledger-like database to ensure transparency and establish proof of ownership.

\section{Aims and Objectives} 

This research aims at designing and developing a hybrid architecture \cite{Rachmawati2023} that leverages decentralized infrastructure components—such as blockchain for notarization and IPFS for document storage\cite{Ref006}—alongside distributed microservice-based components for document processing, obfuscation, and orchestration. The goal is to combine the security, transparency, and immutability of decentralization with the scalability and modularity of distributed computing.

Main objectives for this work are:

\begin{itemize}
    \item \textbf{Design a distributed microservice-based system:} Architect independent services responsible for core functionalities such as data extraction, obfuscation, notarization coordination, and metadata management, ensuring modularity and scalability.
    \item \textbf{Integrate decentralized technologies for integrity and transparency:} Use blockchain to store document fingerprints and notarized facts, and leverage decentralized storage (e.g., IPFS) for immutable document access, ensuring auditability without centralized control.
    \item \textbf{Implement privacy-preserving mechanisms:} Create selective obfuscation methods and controlled access strategies that prevent disclosure of full documents while enabling secure partial verification.
    \item \textbf{Address scalability and usability challenges:} Ensure the system supports high transaction throughput and provides intuitive interfaces for both technical and non-technical users, including notaries and verifiers.
    \item \textbf{Enhance security and trust:} Apply strong encryption, secure identity verification (including video-based workflows), and integrate a secure key management solution.
    \item \textbf{Ensure compliance with data protection regulations:} Follow principles of \ac{GDPR} and other data privacy standards by implementing mechanisms such as on-premise processing, ephemeral storage, and optional zero-knowledge proof schemes.
\end{itemize}

Given the need to automatically extract structured information from official documents, Optical Character Recognition (OCR) techniques are a critical component of the proposed system. These techniques enable the interpretation of text from scanned or photographed documents, serving as the first step in the document processing pipeline. A detailed overview of OCR and its evolution is provided in Chapter 2, along with the review of supporting technologies such as digital signatures, obfuscation, and blockchain.

\section{Document Structure}
This thesis is composed of several interconnected chapters that contribute to the overall goal of designing a secure, decentralized \cite{Shah2020} system for notarizing, signing, and sharing documents. 

The second chapter presents a review of the literature. This section explores key technologies and concepts supporting the proposed solution, such as OCR \cite{bibitem_6, bibitem_7}, document identification and obfuscation techniques, digital signatures \cite{Subramanya2006}, and blockchain technology. It evaluates existing solutions and methodologies, highlighting their strengths and limitations while laying the foundation for the proposed system. The chapter also introduces the Blockchain Data Management Stack as an important component for future discussions.

The third chapter outlines the system's initial vision and conceptual framework, as well as practical use cases. Scenarios such as automating criminal records checks and streamlining college registrations are discussed to demonstrate the real-world applicability of the proposed solution. A preliminary system design is presented, along with a discussion of constraints and challenges that may arise during implementation. This chapter serves as a bridge between theoretical foundations and practical considerations.

The fourth chapter provides a detailed explanation of the proposed architecture of the system. It examines architectural models, comparing centralized and distributed approaches, before presenting the chosen system design. The chapter describes various components of the architecture, including data extraction mechanisms, obfuscation \cite{Fan2019b} algorithms, blockchain integration and user interaction modules. Privacy and security considerations are emphasized throughout to address current data protection \cite{Xu2021} concerns.

Building on this architectural blueprint, the fifth chapter discusses the potential implementation of the system. It explores practical aspects such as data extraction using template-based and \ac{LLM}-based approaches, notarization \cite{bibitem_16} processes and advanced obfuscation algorithms. Additionally, secure data persistence strategies, including the use of blockchain technology, are discussed. This chapter focuses on the feasibility of the solution, offering insights into how the theoretical design can be transformed into a working system.

The sixth chapter focuses on evaluation and comparison, analyzing the proposed system's performance and benchmarking it against existing solutions. Metrics such as accuracy, scalability and usability are used to evaluate system components, including data extractors \cite{bibitem_25} and obfuscation mechanisms. This analysis highlights the strengths and identifies areas for improvement in the proposed solution.

The thesis concludes with a reflective chapter that revisits the aims and objectives of the research, summarizing the key contributions, achievements, acknowledging its limitations and proposes directions for future work. The final remarks highlight the broader implications of the study and its potential to transform digital document management.

In summary, the structure of this thesis is designed to guide the reader through a logical and coherent exploration of the problem, the proposed solution, and its evaluation. Each chapter builds upon the previous one, creating a cohesive narrative that demonstrates the significance and impact of the research.
\if@openright\cleardoublepage\else\clearpage\fi
\thispagestyle{empty}
\mbox{}

    \chapter{Background \& Literature Overview}

\section{Optical Character Recognition}
\index{OCR|(}

In the 1920s, Austrian engineer Gustav Tauschek developed the first OCR device and he managed to patent his "Reading Machine" in Germany in 1929. While early OCR techniques required training with images of each character for every font, today's systems can recognize most fonts with a high degree of accuracy. Many OCR solutions pre-process images to improve the likelihood of successful recognition using techniques such as deskewing (aligning lines of text to be perfectly horizontal or vertical), binarization (converting an image from color or grayscale to black-and-white to effectively separate text from the background), line removal (eliminating non-glyph boxes and lines), and zoning (identifying columns, paragraphs, etc.). The core OCR algorithms can be categorized into two main types:

\begin{figure}[H]
\centering
\includegraphics[width=0.5\linewidth]{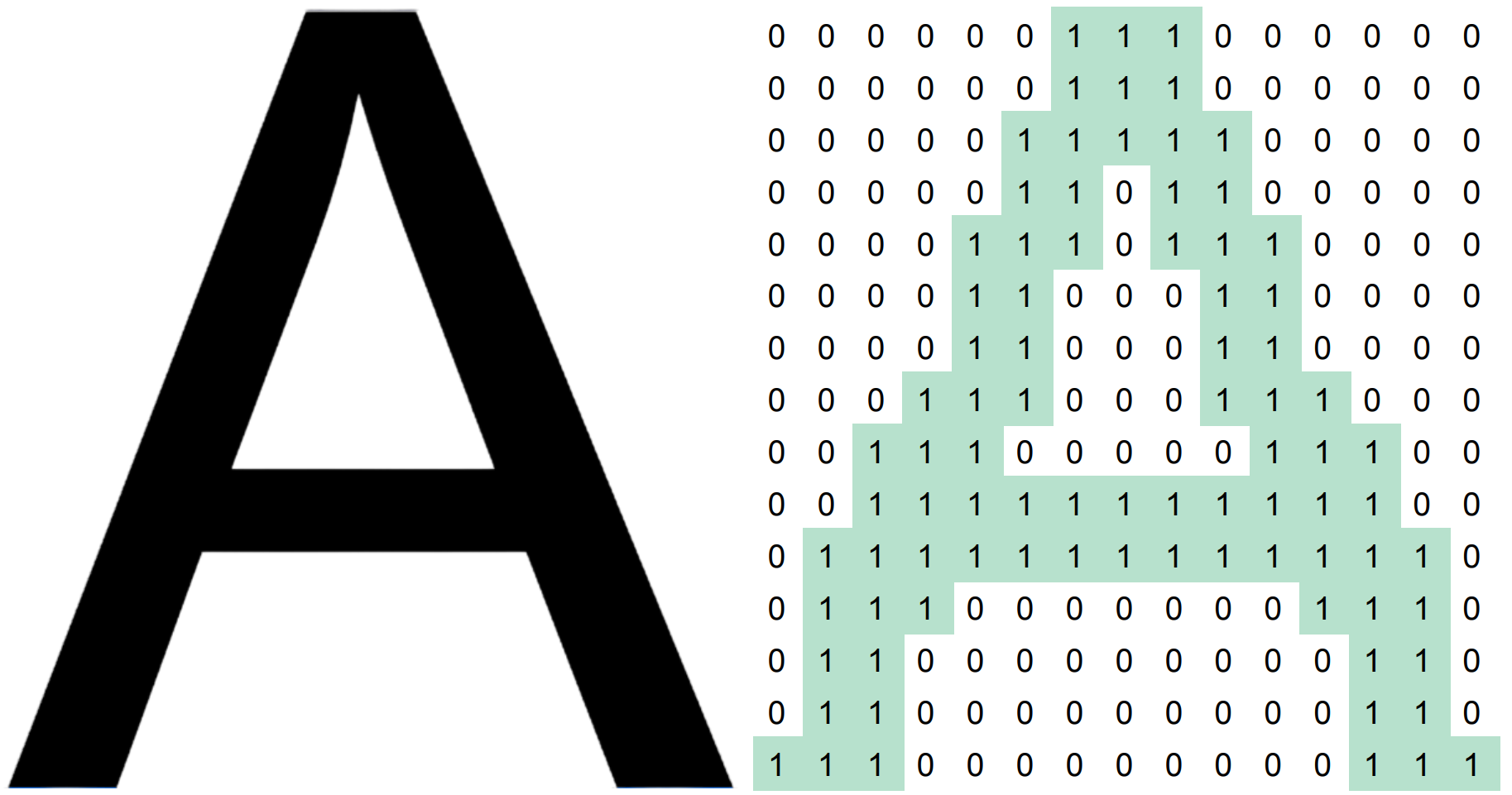}
\caption{OCR Matrix Matching}
\end{figure}

\begin{itemize}
 \item \textit{Matrix matching} - the image is compared to a stored glyph on a pixel-by-pixel basis. This type works best with typewritten text and not very well when encountering new fonts, it was especially used in older OCR solutions.

 \item \textit{Feature extraction} \cite{Vamvakas2007} - detected glyphs are decomposed into “features” (lines, closed loops, line directions, line intersections, etc.), reducing the dimensionality of the representation and making the recognition computationally efficient. This type \cite{shah2016} is also suitable for handwriting recognition and is found in most modern OCR software.

\begin{figure}[H]
\centering
\includegraphics[width=0.9\linewidth]{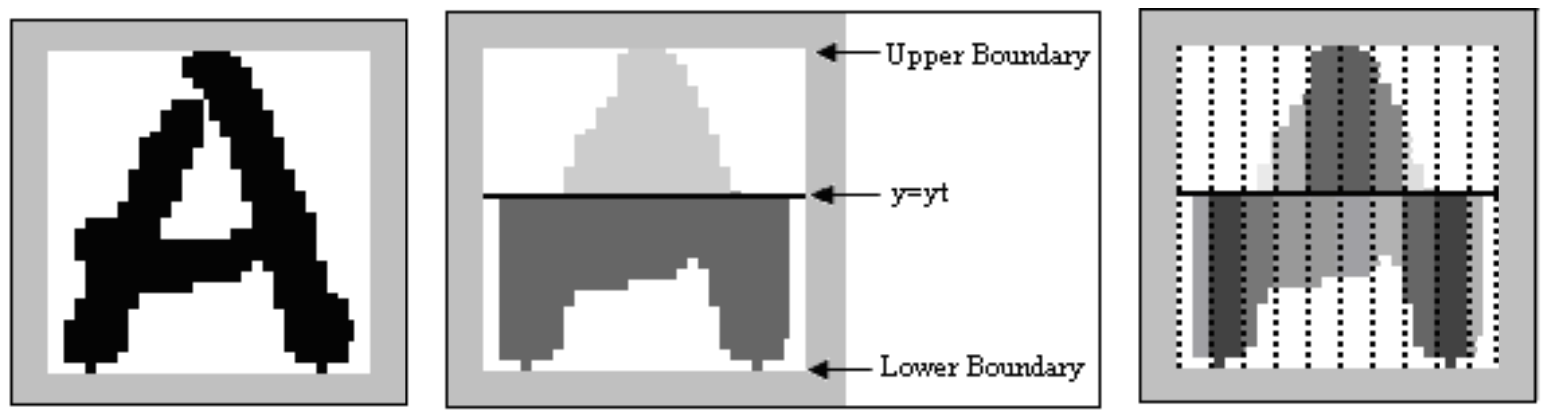}
\caption{Feature Extraction of a Character Based on Upper and Lower Character Profile Projections. \protect\footnotemark}
\end{figure}

\begin{figure}[H]
\centering
\includegraphics[width=0.8\linewidth]{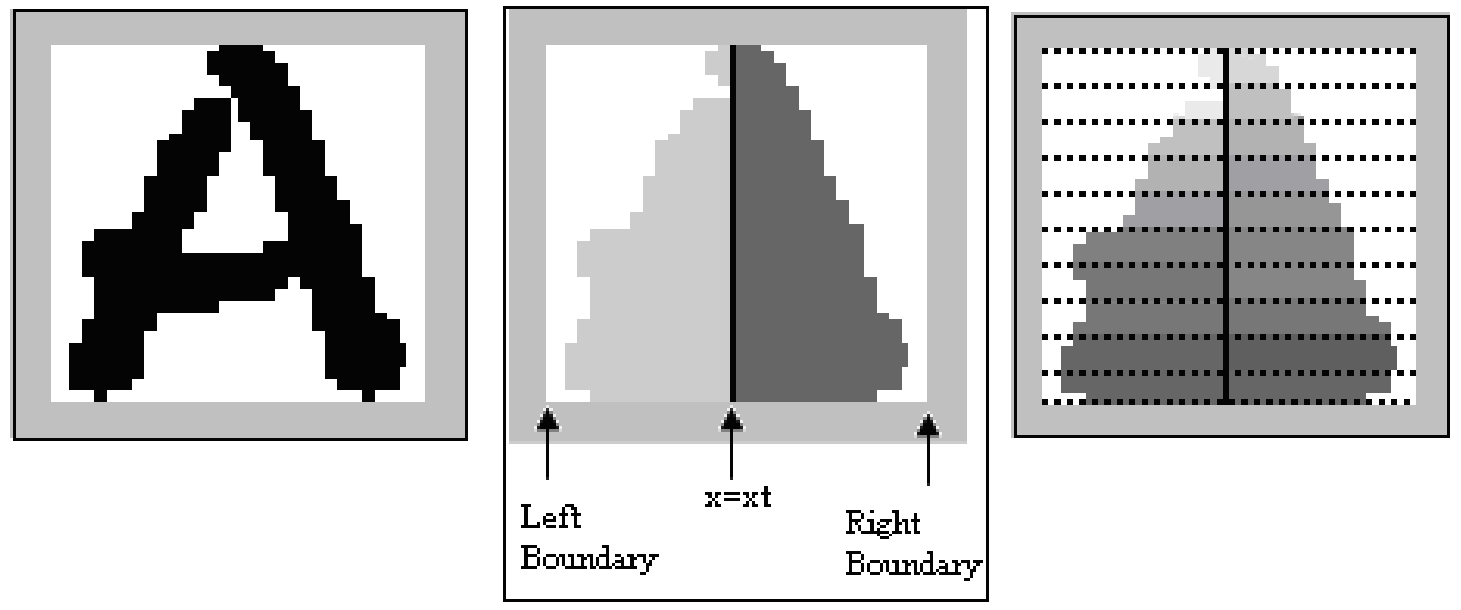}
\caption{Feature Extraction of a Character Based on Left and Right Character Profile Projections.\protect\footnotemark}
\end{figure}

\footnotetext{Source: \href{https://ieeexplore.ieee.org/document/4377080}{An Efficient Feature Extraction and Dimensionality Reduction Scheme for Isolated Greek Handwritten Character Recognition (https://ieeexplore.ieee.org/document/4377080)}}

\end{itemize}

Modern OCRs \cite{Natarajan1999} also rely heavily on neural networks \cite{Dinghang2024} (e.g., OCRopus, Tesseract), which are trained to recognize whole lines of text rather than single characters. There are various ways to evaluate the performance of OCR solutions. A simple method to assess the prediction output is using the accuracy metric (which indicates a match or a no-match), but this alone is insufficient to evaluate OCR performance effectively. Instead, error rates are commonly used to measure the differences between the OCR output text and the reference text. Two key metrics are: \ac{CER}, which is based on the Levenshtein distance and calculates the minimum number of character-level operations required to transform the reference text into the OCR output. This metric is better suited for use cases involving the transcription of specific sequences (IDs, phone numbers, etc.). The other metric is \ac{WER}, which uses the same formula as CER but operates at the word level instead. This is more applicable for transcribing paragraphs and sentences with meaningful content (pages of books, newspapers, etc.).

Later, in their work, Pansare et al. discussed the approaches used in the design of modern OCR \cite{Chumwatana2021} systems and classified them into five categories:

\begin{itemize}
 \item \textit{Matrix Matching}: Matrix Matching transforms each character into a pattern inside a matrix, which is then compared to an index of known characters. It is most noticeable on monotype and homogenous single column pages.

 \item \textit{Fuzzy Logic}: Fuzzy logic is a multi-valued logic that allows for the definition of intermediate values between standard evaluations such as yes/no, true/false, black/white, and so on. 

 \item \textit{Feature Extraction}: Each character is defined by the presence or absence of important elements such as height, width, density, loops, lines, stems, and other character attributes. Feature extraction is an excellent method for OCR of publications, laser print, and high-resolution photos.
 
 \item \textit{Structural Analysis}: Characters are identified via structural analysis by studying their subfeatures, picture form, sub-vertical and horizontal histograms. Its character restoration \cite{Huang2014} function is particularly useful for low-quality text and newspapers.\\
 
 \item \textit{Neural Networks}: This strategy mimics the operation of the human brain system. It takes a sample of each image's pixels and compares them to a known index of character pixel patterns. Character recognition via abstraction is useful for faxed documents and damaged text. 

\end{itemize}

Chirag I Patel and his team \cite{Patel2011} studied the effects of modifying artificial neural network models for character recognition in documents. They examined parameters such as the number of hidden layers, the size of these hidden layers, and the number of epochs. Backpropagation is applied using a multilayer feed-forward network. During preprocessing, they used basic methods for character segmentation, normalization, and de-skewing. Various neural network models were employed, and each model was tested for correctness and accuracy.

As noted by Gur et al. \cite{Gur2012}, automatic \ac{OCR} methods often fall short of providing a complete solution, and in most cases, human intervention is necessary. The authors propose a novel text recognition method based on fuzzy logic principles and statistical data from the studied font. This method combines letter statistics and correlation coefficients into fuzzy-based rules to detect deformed letters that would otherwise be unrecoverable. The authors focused on Rashi typefaces used in Bible commentaries, which are essentially handwritten calligraphy.

Dileep Kumar Patel et al. \cite{Kumar2012} addressed the challenge of handwritten character recognition \cite{Junker1997} using multiresolution techniques such as \ac{DWT} and \ac{EDM}. The new method was evaluated and found to be more accurate and faster than existing approaches. Characters are categorized into 26 pattern classes based on their attributes. Features are extracted from handwritten character images using DWT with an appropriate multi-resolution approach, and each pattern class is described by a mean vector. As demonstrated in this study, distances from input pattern vectors to all other mean vectors are computed using \ac{EDM}. The class membership of the input pattern vector is determined by the minimum distance. This solution achieves a 90\% accuracy rate for handwritten character recognition \cite{Wu2015}.

Jawahar et al. \cite{Sankaran2012} proposed a system for recognizing the Indian script Devanagari. The Devanagari script does not yet achieve the same level of character recognition accuracy as its Roman counterparts due to its complexity and unique writing style. The proposed approach uses a \ac{RNN} called \ac{BLSTM}. This method eliminates the need for word-to-character segmentation, which is a common cause of high word error rates \cite{Lee2019}. Compared to the best available OCR \cite{Manna1999} system, their results showed a reduction of more than 20\% in word error rate and over 9\% in \ac{CER}.

Majida Ali Abed et al. \cite{Ali2013} proposed \ac{PSOA} to simplify handwritten character recognition by simulating the behavior of schools of fish and flocks of birds. This thesis outlines the proposed methodologies, which were developed and evaluated through tests on various handwritten characters. The experimental results demonstrate the superior performance of the proposed methodologies. \ac{PSOA} creates an optimal comparison between input samples and database samples, improving the overall recognition rate. The experimental results also show that \ac{PSOA} is more convergent and accurate, reducing the error recognition rate.

According to Laxmi Sahu et al. \cite{Sahu2013}, classification approaches based on learning from examples have been widely used for character recognition since the 1990s, leading to significant improvements in recognition accuracy. These approaches include statistical methods, \ac{ANN}, \ac{SVM}, and various classifier combinations. This work reviews the characteristics of classification approaches that have been successfully applied to character recognition, as well as the remaining challenges that could potentially be addressed by learning methods.

Shalin A. Chopra et al. \cite{Chopra2014} describe a simple, effective, and low-cost method for developing OCR to read text with a fixed font size and handwriting style. In this study, \ac{OCR} uses a database to detect English characters, making it relatively simple to operate and enabling higher efficiency with lower computational costs. The most critical phase in \ac{OCR} is feature extraction. This method can be integrated with other existing OCR algorithms to recognize English text. 

\section{Document Identification \& Extraction}
\index{DIDE|(}

Identification of documents and data extraction are not new things but out there exist many methods  for doing this and of the most important one is found in [“Application of document type identification in medical handwritten texts recognition” \cite{appDTI}]. This thesis presents a frame-based algorithm for document identification by using characteristic graphical elements, position of the text in the page and the frame or grid printed in the document form. This thesis measures the similarity based on relation between fixed graphical elements extracted from the image and fixed elements of the document template.\\

The document type recognition procedure presented in this thesis is as follows: 
\begin{itemize}
 \item identify dominant lines on the document image using classical Hough transform for lines detection \cite{Yao2014}.

 \item correct detected lines position to obtain two sets of parallel lines – one set contains lines corresponding to vertical lines on the template, the second one contains lines which are horizontal.

 \item rotate the grid of recognized lines so as to obtain perfectly vertical and horizontal lines.
 
 \item for each template $k$ find such translation $(tx, ty)$, which minimizes certain dissimilarity measure between $k$-th template and processed document image.
 
 \item finally recognize the template $k*$ having minimal dissimilarity measure.

\end{itemize}

The dissimilarity measure is in fact dissimilarity measure between original frame in the template and set of extracted lines on appropriately rotated document image. This type of document identification algorithm will be further use in development of our system, of course with adjustments that better suit our needs.
A review of community practices and recent implementations reveals practical variations of this identification method.:

Fan, Kuo-Chin et al. \cite{Fan2001} proposed a unique form of  document recognition approach based on an analysis of the line structure encoded in an input form document. The first step is to remove all vertical and horizontal lines included in the form picture. A line crossing relationship matrix may be constructed by evaluating the crossing relationships between horizontal and vertical lines, with each row referring to one horizontal line and each column corresponding to one vertical line. Furthermore, by examining the distance correlations between horizontal and vertical lines, two line distance relationship matrices, horizontal and vertical line distance relationship matrices, are constructed. Finally, the recognition challenge is accomplished by matching these three matrices. The experimental findings show that our suggested approach for identifying form documents is feasible and efficient.

Nidhish, Shriyans et al. \cite{Nidhish2020} developed a method that converts printed text into an editable digital version, making identification easier and more efficient. The system will depend on well-known technologies including image processing, pattern recognition, and artificial intelligence, as well as a novel methodology called natural image processing. The employment of the CNN algorithm successfully would decrease image blur while efficiently identifying text. The system will consist of three steps: producing a digital document from a physical copy, scanning the digital document, and detecting templates in the document for accurate, error-free outcomes. 

S. Yacoub et al. \cite{Yacoub2005} described a general method for identifying table of contents (TOC) pages in scanned texts. Using different information sources, they evaluated the TOC pages and the location of articles. Three approaches were used to generate these sources: title matching, section keyword matching, and numeric content. Finally, a combination component was employed to evaluate prospective TOC pages and select the best options. The technology is used to identify the table of contents, locate the start of articles, detect advertisements (where present), and more broadly, determine the structure of scanned documents to facilitate article extraction and online deployment of digital content.

Ahmad Montaser \cite{Awal2017} and his team attempted to classify documents containing limited text and complex background information, such as identity documents. Unlike most existing systems, the proposed method not only locates the document but also identifies its class. The class is determined by the document type (passport, ID, etc.), issuing nation, version, and visible side (main or back). Given unrestricted capturing conditions, limited textual information, and variable elements unrelated to categorization, such as photos, names, addresses, etc., this task is particularly challenging. First, a set of document models is created using reference photos. They demonstrated that no training images are required, and a document model can be built from just one reference image. The query image is then compared against all models in the database. Unknown documents are rejected based on a quality estimation derived from the extracted document. The matching process was optimized to ensure that the execution time is independent of the number of document models. Once the document model is located, a more precise matching process is conducted to facilitate information extraction.

A technique proposed by Ahmad Montaser et al. \cite{Asim2019} employs linguistic and visual modalities to classify document images into 10 categories, such as letters, memos, news stories, etc. They used a filter-based feature-ranking technique to reduce the textual stream's reliance on the performance of the underlying OCR \cite{Kuhl2008} (as is typical in content-based document image classifiers). This method scores each class's features based on their ability to differentiate document images and selects a set of top 'K' features for further processing. In parallel, the visual stream extracts structural characteristics from document images using deep CNN models. Finally, the textual and visual streams are combined using an average ensemble approach. Experimental results using the publicly available in the Tobacco-3482 dataset show that the proposed technique outperforms the state-of-the-art system by a significant margin of 4.5\%.

Shweta Joshi \cite{Joshi2011} and her team investigated the challenge of multi-class document classification and explored how to achieve high classification accuracy in text documents. The Naive Bayes approach is used to address the document categorization problem with a deceptively simple model: it assumes all features are independent of one another and assigns a document's class based on the maximum probability. The Naive Bayes technique is applied to improve the performance of the classification model in both flat (linear) and hierarchical approaches. It was found that the hierarchical classification approach is more effective than the flat classification approach. It also performs better in multi-label document categorization. The evaluation dataset is derived from the UCI library dataset, with minor modifications made by the authors.

\section{Document Obfuscation}
\index{DO|(}

There are multiple research papers on obfuscation/deobfuscation topics that differ in the real-life problem they are trying to resolve, but ultimately they have the same fundamentals, securely storing sensitive data and being able to recover the original image when needed. For example, [“Understanding Obfuscation.”, Brunton et al. \cite{Brunton2015}]  proposes a way of solving the previously described problem but it takes a heavy focus on the social media part. It uses several techniques for obfuscating the data, and categorize them into two separate groups:	

\begin{itemize}
 \item \textit{Pixel replacement}: this technique replaces pixels of an original image with other masks, distortions or patterns. The original pixels are encrypted and embedded together with information about the position and shape of the area.

 \item \textit{Data manipulation}: this technique does not replace the original pixels but changes them in a specific way. Typical examples include image encryption and scrambling. Only information about the shape of the protected regions needs to be embedded together.
\end{itemize}

Three of the most common techniques used to obfuscate data are encryption, tokenization, and data masking. Surprisingly, even if data privacy is a highly discussed topic nowadays, there are not many tools that can solve the problem of document obfuscation \cite{bibitem_24} and almost none that take into account reversible document obfuscation \cite{bibitem_27}. The existing tools focus on censoring and encrypting the images using simple algorithms. The algorithms cannot be combined and only one region or the whole image can be censored or encrypted. 
Lecturing about this topic, I found some interesting takes on obfuscation:

Subramaniam Sarith et al. \cite{Subramaniam2019} evaluated glare using obfuscated \ac{HDR} photos. The authors conducted a pilot study in which visual conditions within an office space were recreated and captured as HDR photos using a verified physically-based renderer. These photos were then obscured to varying degrees using blur filters. When glare measurements from the obfuscated photos were compared to metrics derived from the original HDR images, a relative inaccuracy of 2–12\% was observed. This proof-of-concept study lays the groundwork for field-testing an HDR-based lighting management system in real office environments.

Tekli Jimmy et al. \cite{Tekli2019} proposed a methodology for evaluating and recommending the most resilient obfuscation \cite{bibitem_28} strategies in a given context. Their system employs deep learning to reconstruct obfuscated faces and assesses the reconstructions using structural or identity-based criteria. They tested their technique using a publicly available celebrity face dataset. The obfuscation techniques examined include pixelation, blurring \cite{Bahrami2015}, and masking. The reconstructions were tested against five deep learning-assisted privacy attacks. Based on structural and identity-based measures, the most robust obfuscation strategy was recommended.

Richard McPherson et al. \cite{McPherson2016} demonstrated that modern image identification systems based on artificial neural networks (ANNs) can extract hidden information from pictures obfuscated through various methods. The obfuscation \cite{Thavalengal2014} techniques explored in their work include mosaicing (pixelation), blurring (as used by YouTube), and P3—a recently proposed method for privacy-preserving photo sharing that encrypts critical \ac{JPEG} coefficients to make photos unrecognizable to humans. They also showed that ANNs can be trained to accurately identify individuals, recognize objects, and detect handwritten numbers, even when the images are protected using the aforementioned obfuscation techniques.

Liyue Fan et al. \cite{Fan2019_Practical} introduced metric privacy, a rigorous privacy framework generalized from differential privacy, as a novel approach to image obfuscation \cite{Fan2019}. Their solution offers a key advantage over traditional differential privacy methods by providing indistinguishability based on visual similarity, thereby improving utility. Empirical tests on real-world datasets demonstrate that their approach is highly effective while delivering verifiable privacy guarantees.

Croft, W.L. et al. \cite{Croft2021} developed a method that provides a robust privacy guarantee, avoiding the limitations of $k$-same obfuscation while producing photo-realistic obfuscated outputs. Through experimental comparisons, they demonstrated that their technique retains the utility of $k$-same obfuscation in preserving relevant image features. Additionally, they proposed a method to achieve differential privacy for any image (not limited to facial photos) by directly modifying pixel intensities. Although adding noise to pixel intensities did not achieve the high visual quality of generative machine learning models, it offered greater flexibility by eliminating the need for a trained model.

\section{Blockchain}
\index{BL|(}

Blockchain is an emerging technology in the domain of distributed systems, offering a form of decentralization unprecedented in prior architectures \cite{Ref003}. The term “blockchain” is often mistakenly conflated with Bitcoin; however, it is important to clarify that Bitcoin is merely one implementation of blockchain technology.

\begin{figure}[H]
\centering
\includegraphics[width=1\linewidth]{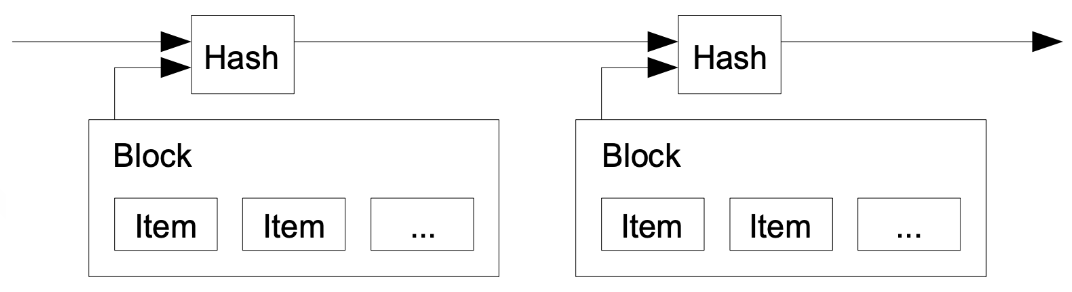}
\caption{Bitcoin Block Structure\protect\footnotemark}
\label{fig:no2}
\end{figure}

\footnotetext{Source: \href{https://bitcoin.org/bitcoin.pdf}{Bitcoin: A Peer-to-Peer Electronic Cash System (https://bitcoin.org/bitcoin.pdf)}}

Blockchain is an immutable, distributed ledger of transactions that operates without relying on a centralized authority to validate the authenticity and integrity of data \cite{Ref004, Ref005}. While transactions are often financial, the underlying system is capable of storing arbitrary data within its blocks. This ledger is composed of a chain of blocks, each connected to its predecessor through a cryptographic hash \cite{bibitem_12}. Each block contains a timestamp, a set of transactions, and the hash of the preceding block.

The chaining of blocks using cryptographic hashes ensures that once a block is incorporated into the chain, it cannot be modified without invalidating the entire chain. Altering any transaction would require recalculating the hashes of all subsequent blocks, a computationally infeasible task. This immutability \cite{bibitem_10} is one of blockchain’s defining characteristics. In the context of this thesis, this property is critical, as it guarantees that once a signed and obfuscated document is added to the blockchain, its integrity and authenticity are permanently preserved — ensuring the originality of the stored document, which is essential for the system’s intended use case.

Blockchain systems are essentially peer-to-peer (P2P) networks, consisting of interconnected nodes that do not require a centralized trusted authority. Each full node maintains a local copy of the blockchain ledger \cite{8635666}, including validated transactions and blocks relevant to its role in the network. Nodes propagate new transactions across the network, making tampering highly improbable, as an attacker would need to alter the data on a majority of nodes simultaneously. Even if an attacker compromises a single node, the broader network will reject invalid or inconsistent blocks.

The process of adding a new block requires the network to reach consensus on the chain’s validity. When a node proposes a new block, the network verifies it against the consensus protocol before inclusion. In cases where nodes maintain competing chains of varying lengths, the protocol typically resolves conflicts by selecting the longest valid chain and discarding alternatives.

In PoW, miners compete to discover a valid block by generating hash codes that meet predefined difficulty criteria. Due to the unpredictable nature of hash functions, miners must iteratively test numerous combinations until they satisfy the difficulty requirements. This process — known as mining — is computationally intensive and energy-consuming. The first miner to discover a valid block broadcasts it to the network, where it is verified by other nodes. Once accepted, miners are rewarded with cryptocurrency incentives, motivating participation despite the high resource costs.

However, PoW introduces notable drawbacks \cite{10707592}, including substantial energy consumption, environmental impact, and low transaction throughput. This performance gap underscores blockchain’s current limitations in supporting large-scale financial infrastructures \cite{Ref002}.



\section{Similar Applications}
\index{SAPP|(}

Among applications in secure document management \cite{Wei2021} and decentralized systems, some features and functionalities align closely with the principles of this research. We highlight five such systems and describe their relevance and unique contributions to secure document notarization, signing, and sharing.

\subsection{DocuSign}

DocuSign provides secure and efficient workflows for document signing, approval, and management. This platform ensures document integrity and confidentiality through advanced encryption standards and secure audit trails that record all interactions with a document. In addition to supporting legal compliance across jurisdictions, DocuSign integrates with numerous third-party applications and \ac{API}s, allowing organizations to customize workflows to fit their needs. While its centralized architecture contrasts with fully decentralized solutions, DocuSign's robust security mechanisms and commitment to protecting sensitive information make it a foundational example of secure digital document management. Its functionalities align with the current study's goal of establishing trustworthy systems for document verification and sharing, particularly in centralized and hybrid environments.

\subsection{BlockCerts}

BlockCerts is a decentralized, open-source platform for issuing and verifying digital certificates. Built on public blockchain \cite{Noh2020} infrastructures like Bitcoin and Ethereum, BlockCerts ensures that certificates are tamper-proof and publicly verifiable, enabling individuals and organizations to prove authenticity without relying on centralized authorities. This platform allows users to store their certificates locally while using blockchain as a decentralized ledger to validate the data, giving users full control over their credentials. BlockCerts is particularly valuable in educational and professional contexts, where trust in credentials is paramount. Its design principles—such as decentralized validation, user sovereignty, and immutability—resonate strongly with the goals of this study, particularly in ensuring privacy-preserving methods of verification without exposing private details.

\subsection{SignRequest}

SignRequest is a digital signing platform that prioritizes simplicity and security, while meeting international regulations such as the General Data Protection Regulation (GDPR) and the \ac{eIDAS} Regulation. It uses secure audit trails and role-based access controls to protect documents throughout their lifecycle. Additionally, it integrates with cloud storage and other enterprise tools, enabling seamless workflows for businesses of all sizes. SignRequest also supports tamper-proof metadata retention, ensuring that all changes to a document are tracked and verified. Its ability to embed directly into decentralized workflows via APIs makes it an adaptable solution for various use cases, including those explored in this study. By focusing on secure metadata handling and user-centric design, SignRequest demonstrates how privacy and verifiability can coexist in digital ecosystems.

\subsection{Jumio}

Jumio is an identity verification platform that leverages artificial intelligence and biometrics for secure, reliable, and scalable identity verification. The technology enables real-time document validation by cross-referencing user-submitted IDs, such as passports and driver's licenses, with government databases and facial recognition data. Jumio's strength lies in its advanced fraud detection capabilities, which use machine learning to identify document anomalies and suspicious patterns, significantly reducing the risk of identity theft. Moreover, it complies with global regulatory requirements, including GDPR, AML (Anti-Money Laundering), and KYC (Know Your Customer) standards, making it applicable in sectors such as finance, healthcare, and e-commerce. Jumio’s focus on real-time, automated verification aligns closely with the goals of this study, particularly in enabling the verification of facts without compromising data privacy or revealing sensitive details.

\subsection{Onfido}

Onfido is a trusted identity verification platform that uses artificial intelligence, document analysis \cite{Singh2018a, Kameshiro1999}, and biometric technologies to authenticate individuals. The system verifies identity documents against a global database and matches them with biometric data captured through facial recognition, ensuring the authenticity of the user. Onfido provides scalable solutions for industries requiring high levels of trust, including banking, insurance, and gig economy platforms. It integrates seamlessly with enterprise applications, offering APIs that allow organizations to create custom workflows for identity verification. A key feature of Onfido is its fraud detection system, which not only detects counterfeit documents but also analyzes behavioral data to flag suspicious activity. This focus on fraud prevention, combined with a commitment to user privacy and compliance with regulations such as GDPR and \ac{CCPA}, positions Onfido as a relevant example of how identity verification systems can maintain both security and user control—principles central to the objectives of this research.

\subsection{Trulioo}

Trulioo is a global identity verification platform that provides access to over 400 data sources in more than 195 countries, enabling organizations to perform reliable and compliant identity checks at scale. Its GlobalGateway product offers real-time verification of identity documents, business data, and watchlists, ensuring compliance with regulations such as AML and KYC while maintaining user privacy. Trulioo's key strength is its ability to handle cross-border identity verification, which is particularly challenging due to differing standards and regulations across regions. The platform’s modular design allows businesses to tailor verification processes to meet their specific needs, whether it’s verifying customer identities or conducting due diligence for business clients. Trulioo also emphasizes inclusivity, extending verification services to underserved populations through alternative data sources, such as mobile and utility records. This combination of scalability, privacy protection, and compliance makes Trulioo a valuable model for systems aiming to balance trust, inclusivity, and data security, echoing the goals of this research.

\subsection{Why Our System is Needed}

Despite the strengths and innovations presented by existing solutions like DocuSign, BlockCerts, SignRequest, Jumio, Onfido, and Trulioo, none fully satisfy the requirements of a decentralized, secure, and privacy-preserving document notarization, signing, and sharing system. The following limitations underscore the necessity for the solution proposed in this thesis:

\subsubsection{Centralization and Trust Dependencies}

While platforms like DocuSign and SignRequest offer robust security and seamless workflows, they rely on centralized architectures. This creates single points of failure and necessitates trust in the service providers to maintain integrity and confidentiality. Such trust dependencies reduce transparency and control for end users, and are not aligned with the principles of decentralization and user sovereignty that some systems prioritize in trustless or adversarial environments.

By leveraging a decentralized framework, our system eliminates the need for central authorities, distributing trust across multiple nodes. This ensures higher resilience against data breaches and enhances user sovereignty.

\subsubsection{Limited Decentralization and Privacy Concerns}

Blockcerts is an open standard for issuing and verifying blockchain-anchored credentials. While many implementations use public blockchains to ensure transparency and immutability, this design can raise privacy concerns, as metadata and credential hashes are publicly visible unless additional protection layers are applied. Jumio and Onfido, on the other hand, use centralized biometric verification systems that pose significant privacy risks, including irrevocable identity theft in case of data breaches.

Our system integrates privacy-preserving technologies, such as fact validation, verification and selective disclosure, ensuring that user data remains confidential while maintaining transparency and verifiability. This approach balances privacy with decentralized trust.

\subsubsection{Narrow Scope and Inflexible Workflows}

Many of the existing systems are designed for specific use cases—BlockCerts for credential verification, SignRequest for digital signing, and Jumio for identity verification. While effective within their domains, these tools are typically optimized for narrow functionalities and may require significant customization or integration to support end-to-end notarization workflows involving multi-party approvals and granular access controls.

The proposed system is designed to be adaptable, supporting a wide range of document types and workflows. It enables seamless integration of notarization, signing, and sharing functionalities while preserving document integrity and authenticity.

\subsubsection{Interoperability and Ecosystem Integration}

While many existing solutions offer robust APIs, their reliance on centralized infrastructure can complicate integration with blockchain-based applications or decentralized identity frameworks, which typically prioritize distributed data ownership and cryptographic proofs over centralized validation models.

By adopting decentralized identifiers (DIDs) and verifiable credentials (VCs) based on W3C standards, our system aligns with widely accepted protocols for identity and credential exchange. This design facilitates interoperability with blockchain networks and decentralized applications that support the same standards, contributing to a more cohesive digital identity and document management ecosystem.

\section{Conclusions}
\index{Conc2|(}

This chapter reviews critical technologies and methods used in Optical Character Recognition (OCR), document identification, extraction, obfuscation, digital signatures, and blockchain. OCR has evolved significantly over time. Early systems relied on basic matrix-matching techniques, but newer approaches use deep learning, which has greatly improved accuracy—especially for handwritten text or documents with complex layouts. Similarly, methods for identifying and extracting text from documents have advanced. Tools like Hough transforms (for detecting document edges), deep learning models (for pattern recognition), and template matching now automate workflows, reducing errors in text processing.

When it comes to hiding sensitive data (document obfuscation), several strategies exist, for example, pixel replacement, encryption, or tokenization can mask information. Balancing privacy with usability is still an open problem. Meanwhile, digital signatures have become vital for verifying document authenticity. By using cryptographic tools like hash functions and public-key encryption, these signatures ensure documents aren’t tampered with.

Blockchain’s decentralized design offers unique advantages for managing documents. Traditional systems like Bitcoin’s ledger are reliable but slow. Newer hybrid or DAG-based blockchains improve speed and scalability, making them better suited for storing and verifying large volumes of documents. Real-world applications—such as DocuSign (for contracts), BlockCerts (for certificates), and Onfido (for ID checks)—show how these technologies are already being used in finance, law, and identity verification.

Despite progress, combining OCR, blockchain, and digital signatures into one cohesive system is still difficult. Future work should focus on making these tools work together seamlessly while meeting privacy laws like GDPR. Solving these challenges could lead to more efficient and user-friendly solutions for managing digital documents as the world increasingly relies on paperless systems.
\if@openright\cleardoublepage\else\clearpage\fi
\thispagestyle{empty}
\mbox{}

    \chapter{System Requirements}

\section{Description}
\index{Desc|(}

This section defines the initial vision and system architecture for the proposed solution, detailing its core subsystems and operational flow. Building on the technological context outlined in the previous chapter—including blockchain-based storage, decentralized identifiers, and document obfuscation—we now formalize the system’s structural components and data lifecycle.\\ 

The system enables user onboarding through a digital identity verification flow based on national ID documents \cite{Xu2018}. During registration, the user initiates a video session with a licensed notary \cite{Ref015}, who validates the presented ID card. If verified, the ID document is digitally signed and added to the user's secure document collection \cite{10425594}. This process creates a unique digital identity, which can be reused for future logins.

Once authenticated, the user can upload additional documents and annotate them using preconfigured templates. The system identifies sensitive regions requiring obfuscation, defined as rectangular zones in pixel coordinates: \textit{(x\_start, y\_start, x\_end, y\_end). These zones may correspond to personal data such as facial images, names, birth dates, or addresses.}

Each obfuscation zone is associated with a unique key (OBKey) that enables controlled deobfuscation. Users can selectively share documents with different parties, exposing only authorized zones. Additionally, multiple documents can be grouped into collections and shared under composite access policies. Instead of storing entire documents, the system anchors document fingerprints (i.e., cryptographic hashes) on the blockchain, enabling verification of integrity and authorship through immutable, publicly auditable records.


Documents or collection would be shareable by \ac{URL} or by \ac{QR} code and an option for limiting document access by time or number of opens would be provided. This way, you can share a document only for verification and it would open only once or you could share it only for only 24 hours.

Document \& user linkage is a feature that we would also want, this will allow us to prove membership of a certain institution or a certain group. Linking will also enable document requests for any situation that would have a benefit in this.
The system will have multiple microservices that would ease the process of notarization with image \& face recognition \cite{Zafaruddin2014}, OCR \cite{Lebourgeois1997} \& text extraction, sign, obfuscate and deobfuscate mechanisms build-in.

\section{Use-Cases}
\index{uc|(}

This section presents two representative use-cases that highlight the limitations of current document-handling systems and demonstrate how the proposed architecture addresses them.

The examples reflect realistic scenarios where users must share verifiable information in a privacy-preserving way—something not easily achievable using conventional, centralized solutions. These use-cases motivated several architectural choices related to credential management, selective disclosure, and auditability.

\subsection{Employment Criminal Record}

According to Romanian Government Ordinance No. 41/2016 and the Ministry of Internal Affairs Order No. 20/2023, digital criminal records can currently be requested and issued only under limited conditions—specifically, for individuals without prior convictions and exclusively for online use via the 'Ghiseul.ro' portal. 

\begin{figure}[H] 
  \centering
  \includegraphics[width=0.82\linewidth]{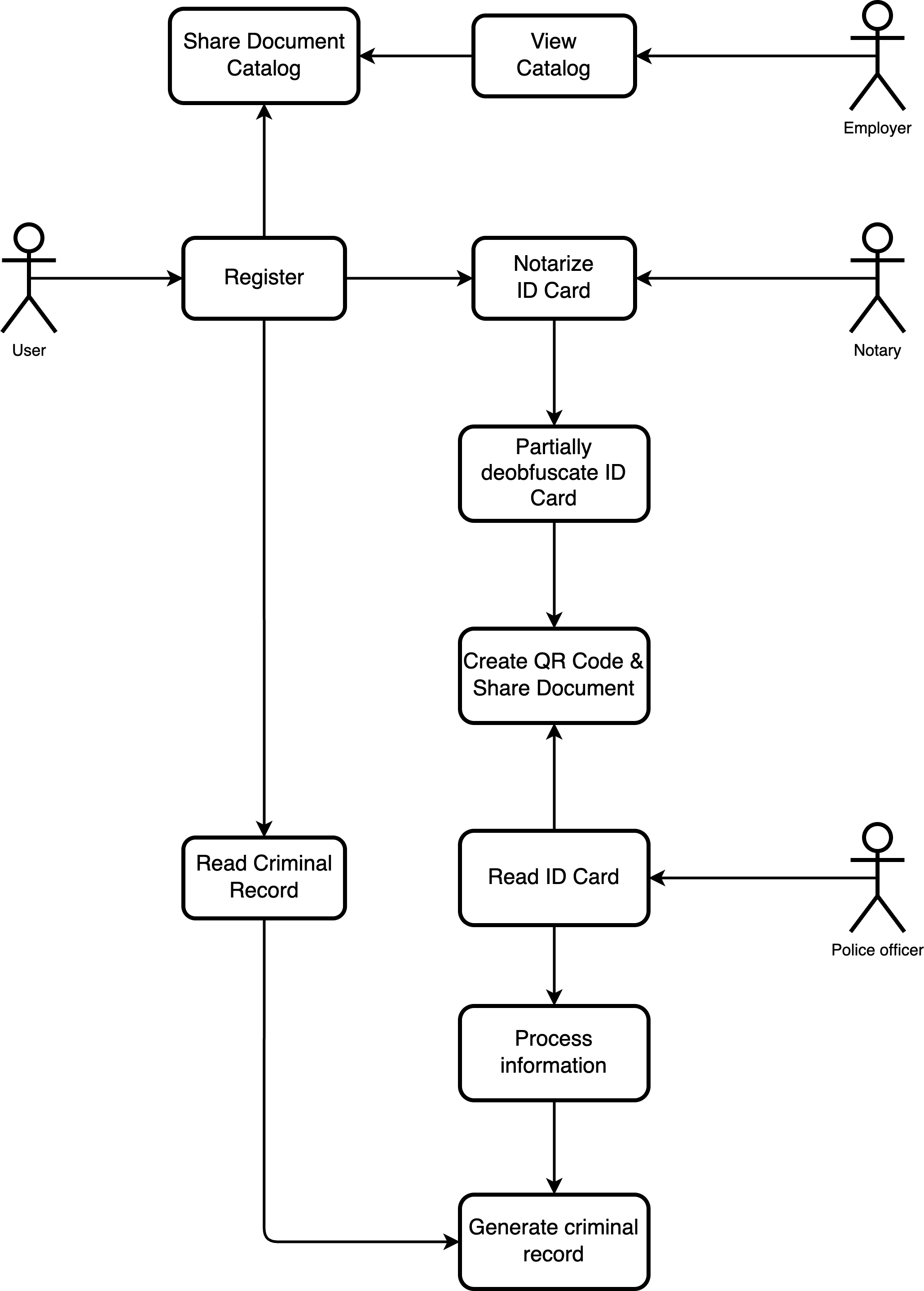}
  \caption{Employment Criminal Record Use-Case}
  \label{fig:no5}
\end{figure}

These documents are not valid for physical printing or submission in institutional workflows, as they lack qualified electronic signatures and are not recognized in cross-platform or offline settings.

The proposed solution respects current legislative constraints by acting as a complementary mechanism: it provides tamper-proof attestations of document facts without replacing or modifying official records. Any generated proof is cryptographically verifiable and anchored in immutable infrastructure, and can coexist with legally recognized digital documents or serve in pre-validation workflows.

Moreover, since most use-cases involving criminal records seek to prove the absence of convictions, the lack of issuance in negative cases highlights a systemic design issue, not a technical impossibility. The proposed system emphasizes verifiability and controlled disclosure rather than printable output, aligning with the need for secure, digital-first interaction models.

This use-case involves a user who must submit a recent criminal record as part of an employment verification process. The system allows the user to selectively share attributes from their registered ID document, including full name, personal identification number (CNP), and facial image, via a verifiable QR code or secure digital link.

A designated police authority verifies the shared identity and issues a digitally signed criminal record. This document is automatically stored in the user's document wallet alongside the original ID card. Documents in the wallet may be annotated with metadata such as expiration dates or revocation status.

The user then creates a shareable collection that includes the newly issued criminal record and the corresponding ID document, applying selective deobfuscation to disclose only the required fields to the employer. Upon expiration, access to the document is automatically revoked based on the embedded validity metadata.

\subsection{College Registration}

This use-case illustrates a digitalized student enrollment process involving an academic institutions. The user uploads required admission documents into the system, which are stored and available for verification. These documents typically include notarized identity document, high school grade transcripts, and BAC diploma. To streamline the process, the system allows the creation of a verifiable QR code or secure link, which can be shared across multiple faculties and admission platforms.

\begin{figure}[H] 
  \centering
  \includegraphics[width=1.0\linewidth]{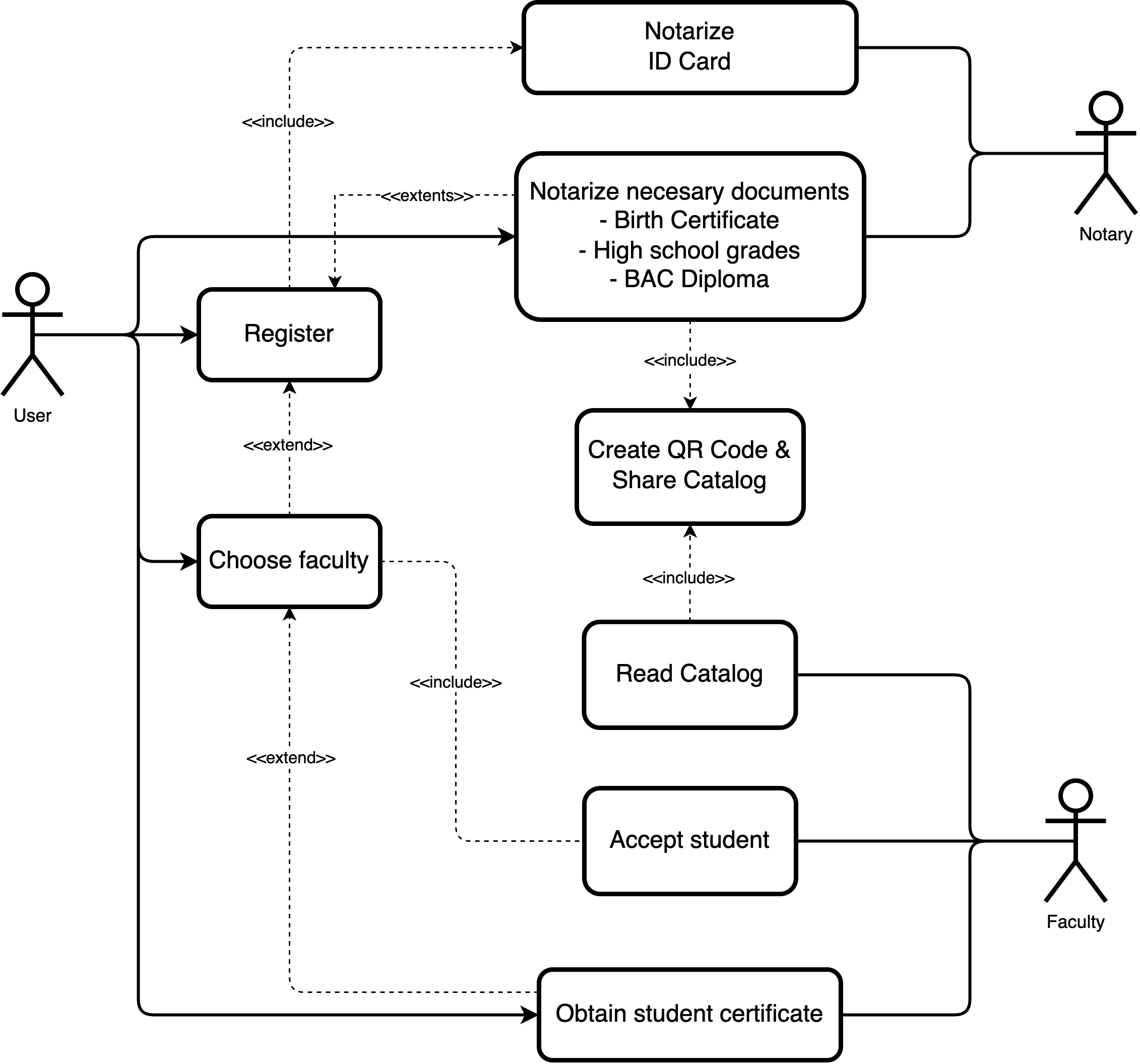}
  \caption{College Registration Use-Case}
  \label{fig:no6}
\end{figure}

The enrollment process begins with the user registering on the platform and selecting the faculties they are interested in. As part of the extended flow, the system provides tools to assist students in notarizing and validating their necessary documents, ensuring all admissions criteria are met. Faculties can read the shared catalog of documents and proceed to review applications.

Once admission is confirmed at a chosen institution, the user can revoke document access for all other universities through a consent-based interface, maintaining control over their personal data. The accepted faculty then issues a digitally signed proof of enrollment, which is added to the user’s verifiable identity wallet.

This credential can later be used to prove active student status for various third-party services, such as obtaining student discounts, accessing specialized educational resources, or verifying eligibility for internships or scholarships. Importantly, selective disclosure mechanisms ensure that only the necessary enrollment attribute is shared, preserving the privacy of unrelated personal information.

Additionally, the system supports automatic revocation and expiration metadata, meaning that once a document or credential reaches its validity limit, it can no longer be accessed or shared, enhancing the overall security and trust in the digital ecosystem.

\section{Constraints}
\index{Const|(}

In order to obtain such a system as proposed above, we must first comply with a set of specifications and constraints:\\

\textbf{Privacy} - Clearly, dealing with personal documents is a very sensitive matter, and the first priority is ensuring the privacy of the shared documents. One way to achieve this is by storing the documents in a private environment. Another approach is to keep the document private but provide an access password or key to anyone who wants to view it. However, this method is also flawed because a user could theoretically store a tampered document containing fake information, and only the intended recipient would have access to verify its validity. The issue raised is that, even though access is restricted, the document’s authenticity cannot be guaranteed. A user could alter the document before storing it, and since only the recipient can check it, the verification process lacks transparency and trust. While this method restricts access to the document via encryption or password, it does not inherently ensure that the content has not been tampered with. Therefore, although privacy is preserved, data integrity remains unverified unless additional mechanisms such as cryptographic hashing and blockchain anchoring are employed. To address this constraint, we propose an improved solution called 'hidden in plain sight.' This involves hashing the original version of the document, obfuscating its sensitive parts, hashing it again, and then publishing it publicly. To guarantee integrity, the initial hash is notarized using a blockchain-based timestamping mechanism, ensuring that any attempt to alter the document later would be detectable. This allows anyone to access the document in its obfuscated form and verify its integrity against the notarized hash, ensuring that the revealed document has not been modified after its initial registration. 

Unlike a password-protected storage model, where only the document owner and recipient have access, the 'hidden in plain sight' approach benefits from public verification. The notarized hash, stored on a blockchain, acts as an immutable proof of the document’s state at a given time. If a malicious user were to modify the document and claim it as authentic, anyone could compare its hash to the notarized reference and detect discrepancies. In contrast, a password-protected document lacks this transparency. Even if the document is digitally signed prior to storage, its validity remains verifiable as long as the notary's signature and associated public key infrastructure are available. However, in many real-world scenarios, such infrastructure is either inaccessible or lacks standardization, resulting in limited verifiability for recipients. This issue is thus not one of privacy, but rather of availability and interoperability. The verification process is restricted to the private storage owner, making it difficult to detect tampering unless the document is later revealed and independently checked. The strength of the 'hidden in plain sight' model lies in its open verification and auditability.

\textbf{Immutability} - It goes without saying that one of the most critical aspects of accessing a public document is trusting that it is genuine. Therefore, it is essential to store the document's information in an immutable space. This ensures that everyone can verify the information at any time and trust that it has not been altered, tampered with, or replaced. To achieve immutability, the system combines a public or consortium blockchain for anchoring document hashes with decentralized storage solutions like IPFS for hosting obfuscated documents. These distributed technologies ensure that once data is stored, it cannot be altered or deleted without detection.

\textbf{Trust} - Even if a secure, privacy-preserving and fully decentralized architecture could be built, the human factor must be taken into account. A user might attempt to upload a fake document, one that does not belong to them, or an expired one. This scenario is significant and can only be resolved by notarizing \cite{Magrahi2018} the document itself. This involves having a third-party authority verify the document's integrity before incorporating it into the distributed system. Such verification already exists and can be conducted through a short video call between the user and a professional notary. During the call, the notary has access to both the uploaded document and the user's video feed simultaneously, ensuring data accuracy and preventing fake or expired documents from entering the system.

\textbf{On-premise processing} - Handling official documents is a sensitive task. Managing this type of data, extracting private information, storing it, and processing it can be a decisive factor in whether users trust a particular solution. Therefore, a key constraint is that no data should leave the premises before the obfuscated document is published online. All automated processing steps—such as data extraction, obfuscation, and cryptographic hashing—must be performed locally or on-premise, before any data leaves the user's trusted environment. The only exception is the notarization phase, which requires a short, end-to-end encrypted online video call with a certified notary. During this session, a temporary version of the document is securely shared for validation, without persistent external storage. 

\textbf{Open access} - The proposed architecture will likely result in highly complex systems, which may make them difficult to use or adopt. A major constraint is ensuring easy access to the platform from both sides: registering and uploading documents to the cloud should be a straightforward process, as should verification by a third party who does not have an account or is not a member of the system. 

\section{Conclusions}
\index{Conc3|(}

This chapter presented an innovative vision for a digital document management and notarization system, emphasizing privacy and ease of use. By leveraging advanced technologies such as blockchain automated document processing, the proposed system offers a comprehensive solution for secure document sharing and authentication.

The outlined architecture provides a seamless user experience, from identity registration using ID cards to secure document sharing with granular obfuscation controls. This approach ensures that sensitive information is protected while maintaining transparency and verifiability through the blockchain's immutable ledger.

Two practical use cases <employment criminal records and college registration> exemplify the system's potential to simplify traditionally cumbersome processes. These scenarios highlight the versatility and efficiency of the proposed solution, showcasing its capability to streamline document management and enhance user trust.

To achieve this vision, the design considers key constraints, including privacy, immutability, trust, on-premise processing, and open access. By addressing these challenges, the system aims to provide a reliable and secure platform for digital document notarization, paving the way for mass adoption in various real-world applications.

This initial vision serves as a foundational blueprint, setting the stage for future development and implementation. As the system evolves, it holds the promise of transforming the landscape of digital document authentication, ensuring security, privacy, and efficiency for users worldwide.

\if@openright\cleardoublepage\else\clearpage\fi
\thispagestyle{empty}
\mbox{}

    \chapter{Proposed Architecture}

\section{Proposed Architecture Type}
\index{pat|(}

The proposed system adopts a microservices-based \cite{10804153} architecture, a paradigm widely recognized for its effectiveness in building scalable and maintainable distributed systems. Key advantages include the ability to scale individual components independently, flexibility in technology selection per service, improved fault isolation, and support for continuous integration and deployment practices.

In the case of DocChain, this architectural model aligns well with the nature of the system’s functionalities. For example, services responsible for document retrieval and sharing are expected to experience significantly higher traffic and can therefore be scaled independently, without affecting other components such as data extraction or notarization. Additionally, separating concerns across microservices allows for the use of specialized technologies—for instance, leveraging Python for image processing tasks, Java for orchestrating secure interactions, and Solidity for blockchain logic—each optimized for its respective domain.

This modular design not only enhances maintainability and testability, but also facilitates parallel development, enabling independent updates and deployments of each service, which is essential in rapidly evolving privacy- and security-sensitive environments.

\section{Decentralized vs. Centralized}
\index{dv}

The system adopts a hybrid architecture that combines centralized and decentralized components in order to balance performance and privacy, while also enabling transparent and independent verification of documents. While decentralized approaches offer strong guarantees of immutability and transparency, they often introduce latency and complexity in real-time processing tasks.

In this architecture \cite{4459163}, services such as data extraction, obfuscation, and digital signing are implemented as centralized microservices for performance and operational efficiency. These services operate in secured, controlled environments and are orchestrated to ensure consistent processing.

By contrast, storage and verification logic—especially blockchain anchoring—are implemented in a decentralized manner. Document fingerprints are committed to public blockchains, making them independently verifiable and tamper-evident.

This hybrid design enables the system to achieve low-latency user interactions while preserving transparency and auditability of key transactions.  

\section{Proposed Architecture}
\index{pa|(}

The proposed architecture consists of the following five microservices, each fulfilling a distinct role in the secure processing, obfuscation, and storing the documents:

\begin{enumerate}
    \item \textbf{Service Orchestrator} – Coordinates user requests, manages authentication, and routes data across the entire microservice pipeline.
    
    \item \textbf{Data Extractor} – Parses incoming documents (images or PDFs) and identifies key fields, metadata, and sensitive zones based on layout templates or semantic analysis.
    \newpage
    
    \item \textbf{Document Obfuscator} – Applies masking, encryption, or distortion algorithms to predefined sensitive zones, and generates obfuscation keys for controlled access.
    
    \item \textbf{Notarizer} – Facilitates user interaction with official notary services (manual or video-based), enabling document validation and digital signing.
    
    \item \textbf{Data Persister} – Handles long-term storage of processed documents and metadata, coordinating interactions with decentralized storage systems.
\end{enumerate}


Besides the internal microservices, we are also have integration points with three extra services.

\begin{enumerate}
    \item \textbf{IPFS Integration} – Uploads obfuscated documents to the InterPlanetary File System (IPFS) and retrieves the corresponding content-addressed hash.

    \item \textbf{Notarizing System} – An external platform that enables official document validation through licensed notaries. This system is accessed via secure communication channels and supports both automated and manual verification workflows.

    \item \textbf{Key Management System (Keystore)} – A secure external vault (HashiCorp Vault) responsible for managing cryptographic keys used in the obfuscation and deobfuscation processes. It ensures that sensitive secrets are never stored directly in application services or on-chain, and supports secure API-based access with granular permissions and audit logging.
\end{enumerate}

\begin{figure}[H] 
  \centering
  \includegraphics[width=0.82\linewidth]{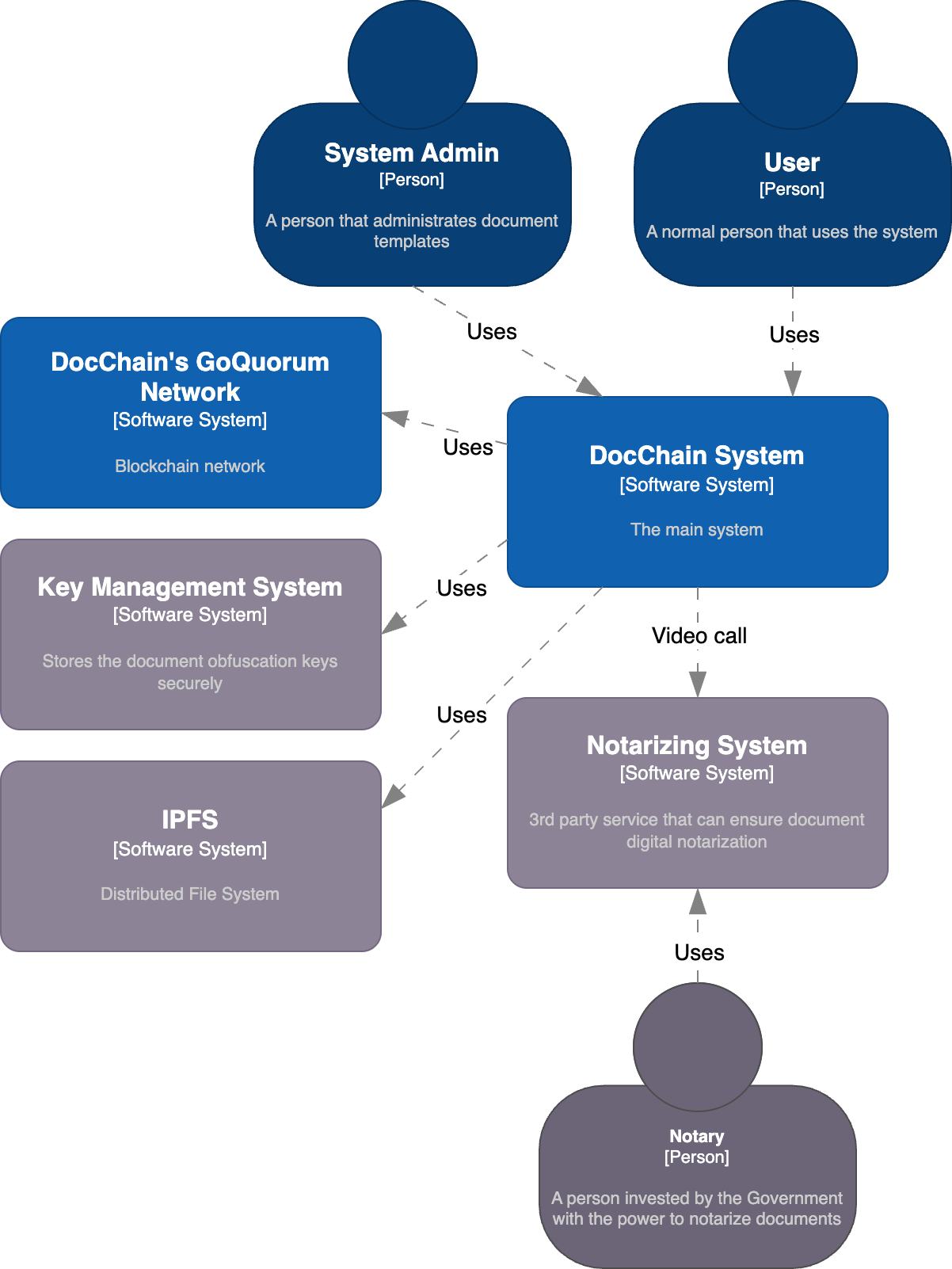}
  \caption{Proposed System Architecture - C4 Level 1}
  \label{fig:no8}
\end{figure}

\begin{figure}[H] 
  \centering
  \includegraphics[width=0.85\linewidth]{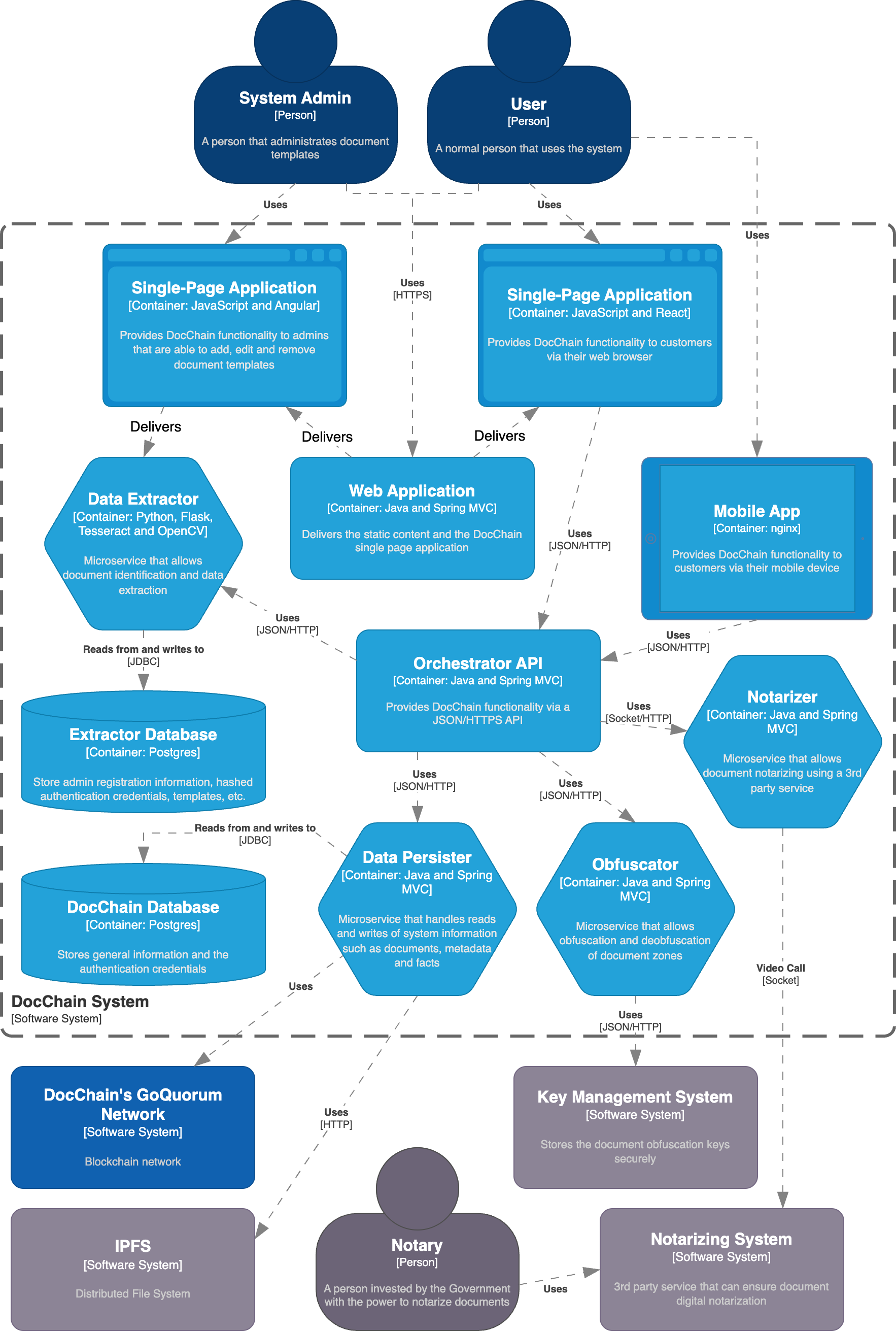}
  \caption{Proposed System Architecture - C4 Level 2}
  \label{fig:no88}
\end{figure}

\section{Data Extractor}
\index{de|(}

The Data Extractor microservice \cite{IRIMIA20221571} is responsible for parsing input images or scanned documents to identify structured information. It serves as the entry point to the processing pipeline and returns metadata describing the document type, pages, and extracted fields.

\begin{figure}[H] 
  \centering
  \includegraphics[width=0.82\linewidth]{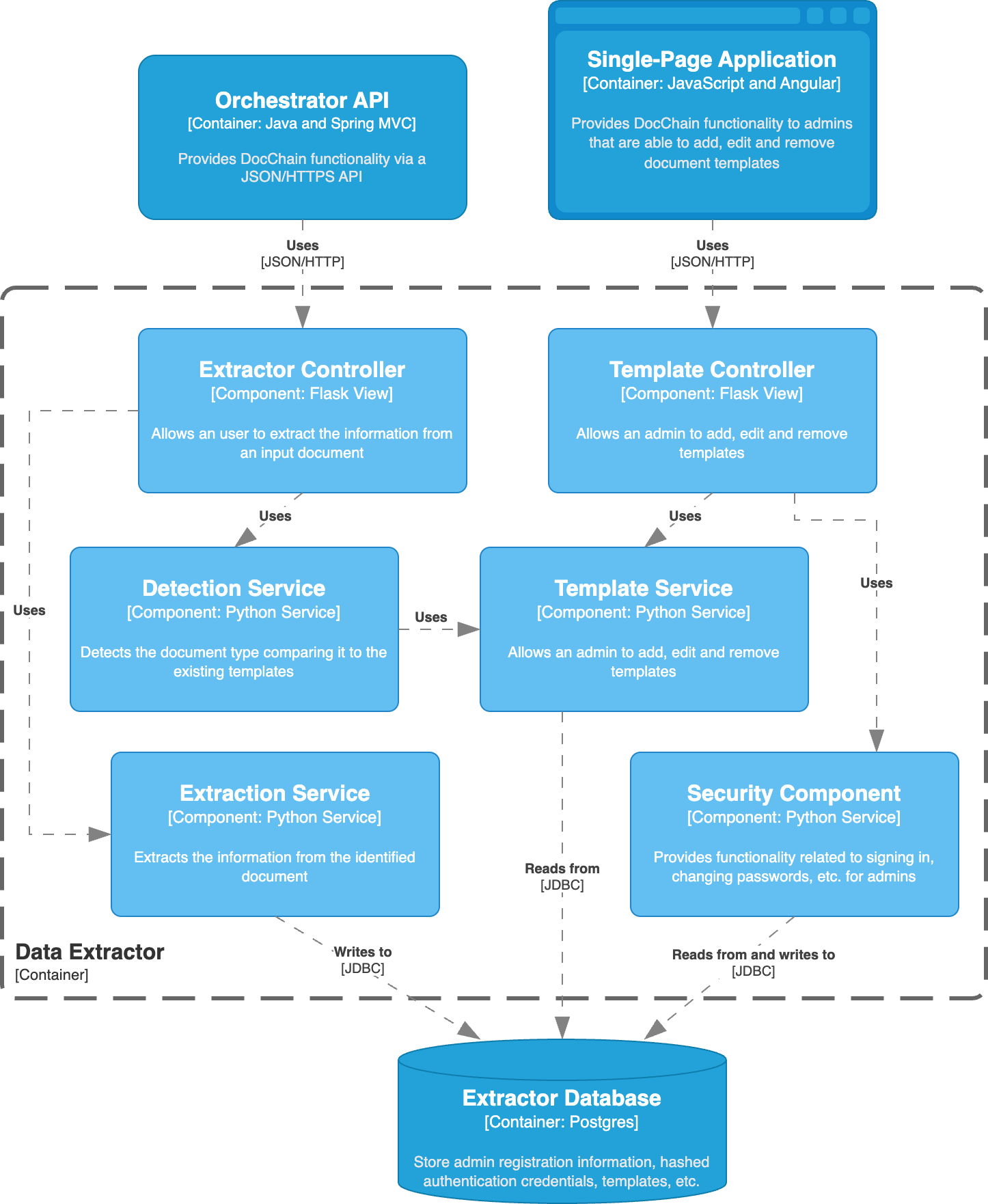}
  \caption{Proposed "Data Extractor" Microservice Architecture - C4 Level 3}
  \label{fig:no9}
\end{figure}

\newpage

Each field includes a name (e.g., \texttt{birthdate}), a value, a bounding box defined by image coordinates, and a sensitivity flag. Fields are grouped into pages, and each page is identified by a unique ID. This structure enables downstream modules to perform layout-specific processing and apply selective obfuscation.

The service exposes a single endpoint that accepts an image (or PDF) and returns a structured JSON object. The input is first preprocessed to normalize orientation and infer the document type. Once the type is known, template-based field extraction is applied. This ensures accurate parsing even in documents with similar visual structure.

The output of the microservice will be an object such as the one bellow. We have also a mapping between a document type and the field positions along those lines. If we also have this information, then we can start obtaining data from the specified positions using \ac{OCR} and labeling it with the field name that we know we have to pair it with.

Techniques of data extraction \cite{10310514} and interpretation can very a lot, for start the common ground between all the systems it’s the \ac{OCR} mechanism that allow for raw data extraction of the text and after that it’s all about the type of processing we can use: in our work below we will propose a template-based mechanism of data segmentation and a Large Language Model approach comparing this two.

The output of the \textit{Data Extractor} microservice is a structured JSON object that includes the document type, page-level segmentation, and a list of identified fields with coordinates and sensitivity flags.

In practice, the extractor may return partial outputs if certain fields cannot be reliably detected due to image quality issues, template mismatches, or OCR failures. In such cases, the corresponding fields are simply omitted from the response object. Each field also includes a confidence score that downstream services (e.g., Fact Collector, Notarizer) can use to decide whether to accept or flag the extracted data. Each field also includes a confidence score that downstream services (e.g., Fact Collector, Notarizer) can use to decide whether to accept or flag the extracted data. If a field is flagged due to low confidence, it can either be corrected manually by a notary during the validation phase, or the user may be prompted to re-upload a higher-quality version of the document. In cases where essential fields remain unresolved, the document may be temporarily suspended from processing until corrections are made.

\begin{lstlisting}[caption={Proposed “Data Extractor” Return Object}, language=json, firstnumber=1]
    "document_type": "id_card",
    "pages": [
        {
            "id": "c690c529-771f-4129-b4e5-a775a076b888",
            "fields": [
                {
                    "name": "cnp",
                    "text": "197XXXXXXXXXX",
                    "sensitive": true,
                    "confidence_score": 0.94, 
                    "coordinates": {
                        "start_x": 1427,
                        "start_y": 792,
                        "end_x": 2254,
                        "end_y": 924
                    }
                },
                {
                    "name": "series",
                    "text": "MZ",
                    "sensitive": false,
                    "confidence_score": 1, 
                    "coordinates": {
                        "start_x": 2254,
                        "start_y": 681,
                        "end_x": 2456,
                        "end_y": 834
                    }
                }
            ]
        }]
\end{lstlisting}

This design supports independent testing and validation of the extractor in isolation. By simulating documents with known ground truth and varying noise levels or layout distortions, the component can be evaluated for precision, recall, and resilience against partial failure.

This loose coupling between components improves robustness and enables fallback mechanisms in downstream services without requiring a new extraction cycle. 

\section{Fact Collector}
\index{zkp|(}

A fact is a piece of information computed using specific techniques from a document and can be considered truthful due to its capturing nature. Since facts are collected automatically by the system and then verified by an authority (such as a notary), we can be confident that this information is accurate. These facts can then be used to prove certain points about the document without revealing the document itself.

We use this approach to collect relevant information about user data without actually storing any of it. After the extractor microservice returns an object containing the extracted and labeled data (e.g., {"name": "birthdate", "value": "23/05/1997"}), we interpret this data to produce fact proofs that the user can share. For example, based on the data above, we can confidently state that the person is over 18 years old. This information can be stored on the blockchain in the form of hashed, cryptographically signed facts. No personal data or raw values are stored in plain text. Instead, the system stores only the hash of the verified fact along with a timestamp and a digital signature, ensuring integrity and non-repudiation.

Facts that are time-sensitive (e.g., student status, criminal record validity) are associated with an expiration timestamp, after which they are no longer considered valid by downstream verifiers. In case of fact revocation (e.g., if new information invalidates a previously accepted claim), a revocation entry is appended to the blockchain or managed through an additional revocation registry smart contract. This ensures consistency while preserving the immutability of past records. All collected information about the user is processed by the Fact Collector to create these facts, after which any sensitive "value" data is discarded. This ensures that no sensitive information is stored anywhere in the system, whether centralized or decentralized.

The Fact Collector plays a critical role in this process. It aggregates the extracted data, interprets it, and derives semantically relevant facts that are subsequently stored on-chain in hashed, non-identifiable form. The current system supports a predefined set of reusable facts, including:

\begin{itemize}
\item Age-based assertions (e.g., "is over 18")
\item Document validity confirmations (e.g., "driving license is valid")
\item Status claims (e.g., "is currently a student")
\item Identity linkage (e.g., "name matches ID document")
\item Residency indicators (e.g., "has domicile in Romania")
\end{itemize}

After this step, the sensitive input data—such as the original birthdate, address, or identification number—is permanently discarded. This ensures that only minimal, task-specific abstractions are retained, aligning with privacy best practices and data minimization principles.

If a new type of verification is required—outside the scope of the existing fact set—the system must reprocess the original document (if still available) through the Fact Collector to generate the additional fact. This mechanism provides flexibility while preserving strict privacy guarantees.

This methodology has wide-ranging applications. For example, instead of sharing an entire identity document, a user can generate and share a fact such as ‘user is over 18’ or ‘user resides in Romania’, without exposing full personal data. This enables privacy-preserving verification across a variety of scenarios. Also, academic credentials can be represented with facts like "Holds a verified Master’s degree in Computer Science".

The ability to abstract data into facts allows for a balance between transparency and privacy. The facts generated by the system are designed to be shared with authorized third-party verifiers—such as employers, academic institutions, or service providers—via secure links or QR codes. Each fact is represented as a cryptographically signed assertion (e.g., "user is over 18"), stored in hashed form on the blockchain. When a verifier receives such a fact, they can independently validate its authenticity and integrity without needing access to the original document or any personal data. This mechanism enables automated and privacy-preserving verification workflows. This approach not only meets the requirements of privacy regulations, such as GDPR, but also provides a scalable framework for handling user data in decentralized systems.

\section{Document Obfuscator}
\index{do|(}

The \textbf{Document Obfuscator microservice} \cite{Ref014} is responsible for applying and reversing content masking on user documents. It plays a central role in the system’s privacy model by enabling users to redact sensitive information while preserving the integrity of the visible data. \\
For the obfuscation part, the system will receive an image/document attached to an input object that explains how the obfuscation would be done, namely an array of zones with the appropriate coordinates and the layers that would be applied, this being an object what states the algorithm type and the secret that would be used for the obfuscation. 

All communication with the Document Obfuscator microservice is secured using TLS encryption and authenticated using JWT tokens. This ensures that all payloads—including obfuscation keys—are transmitted securely and only by authorized clients. Furthermore, each request includes a session-bound identifier and timestamp, making it resistant to replay attacks. The system enforces strict validation of JWT signatures and expiration times, preventing unauthorized reuse of prior requests. Obfuscation keys are never persisted in plaintext and are accessed only during the secure execution flow.

The output returned by the service consists of two elements: 
(1) the obfuscated version of the original document, where the specified zones have been masked according to the applied algorithm layers; and 
(2) a structured metadata object that contains the coordinates, zone identifiers, and deobfuscation keys necessary for selective content restoration.

The input of the microservice will be an object such as the one in the following listing:

\begin{lstlisting}[caption={Proposed "Document obfuscator" Input Structure}, language=json, firstnumber=1]
{
    "zones": [
        {
            "id": 1,
            "coordinates": {
                "start_x": 1427,
                "start_y": 792,
                "end_x": 2254,
                "end_y": 924
            },
            "layers": [
                {
                    "algorithm_id": 1,
                    "key": "MY_SECRET_KEY"
                }
            ]
        }
    ]
}
\end{lstlisting}

The key aspect of this mechanism is that it should be able to work partially. This means that even if we decide to obfuscate ten different zones of a document, it would be possible to deobfuscate only specific zones using the keys sent through the endpoint. As a result, the deobfuscation endpoint receives the fully obfuscated photo and a list of zones to be deobfuscated, along with their specific keys, without requiring the coordinates to be sent separately

During obfuscation, each zone is associated with a decryption key derived from the user's passphrase and zone-specific metadata (e.g., coordinates and transformation instructions). The passphrase itself is never transmitted. Instead, the derived zone key may be shared with authorized recipients by encrypting it with their public key.

This structure prevents leakage of sensitive positional or structural information and ensures that only the intended recipients can reconstruct the original content. The derivation process also prevents direct correlation between the obfuscated data and its decryption keys.

The output of the microservice will be an object such as the one in the following listing:

\begin{lstlisting}[caption={Proposed "Document obfuscator" Output Object Structure}, language=json, firstnumber=1]
{
    "document_id": "c690c529-771f-4129-b4e5-a775a076b888",
    "zones": [
        {
            "id": 1,
            "coordinates": {
                "start_x": 1427,
                "start_y": 792,
                "end_x": 2254,
                "end_y": 924
            },
            "obfuscationKey": "eJxrYJnKz8gABj1i2amVekAcX8CMr2BrMlTNBun1E7R6OFOzs8vSsnMSXoAlzdk6g=="
        }
    ]
}
\end{lstlisting}



\subsubsection*{User Registration and Passphrase Derivation}

Each user is required to register using a master password. This password is never stored or transmitted. Instead, it is used locally to derive a secret passphrase through a secure key derivation function such as PBKDF2 with a salt unique to the user. The resulting passphrase serves as a personal root secret used for zone-level key derivation.

\textbf{Purpose:}
\begin{itemize}
  \item The passphrase ensures that each user generates unique keys even when working with identical documents or zones.
  \item It binds zone-level keys cryptographically to the user identity and obfuscation context.
  \item It is never exposed outside the client environment.
\end{itemize}

\begin{figure}[H] 
  \centering
  \includegraphics[width=0.7\linewidth]{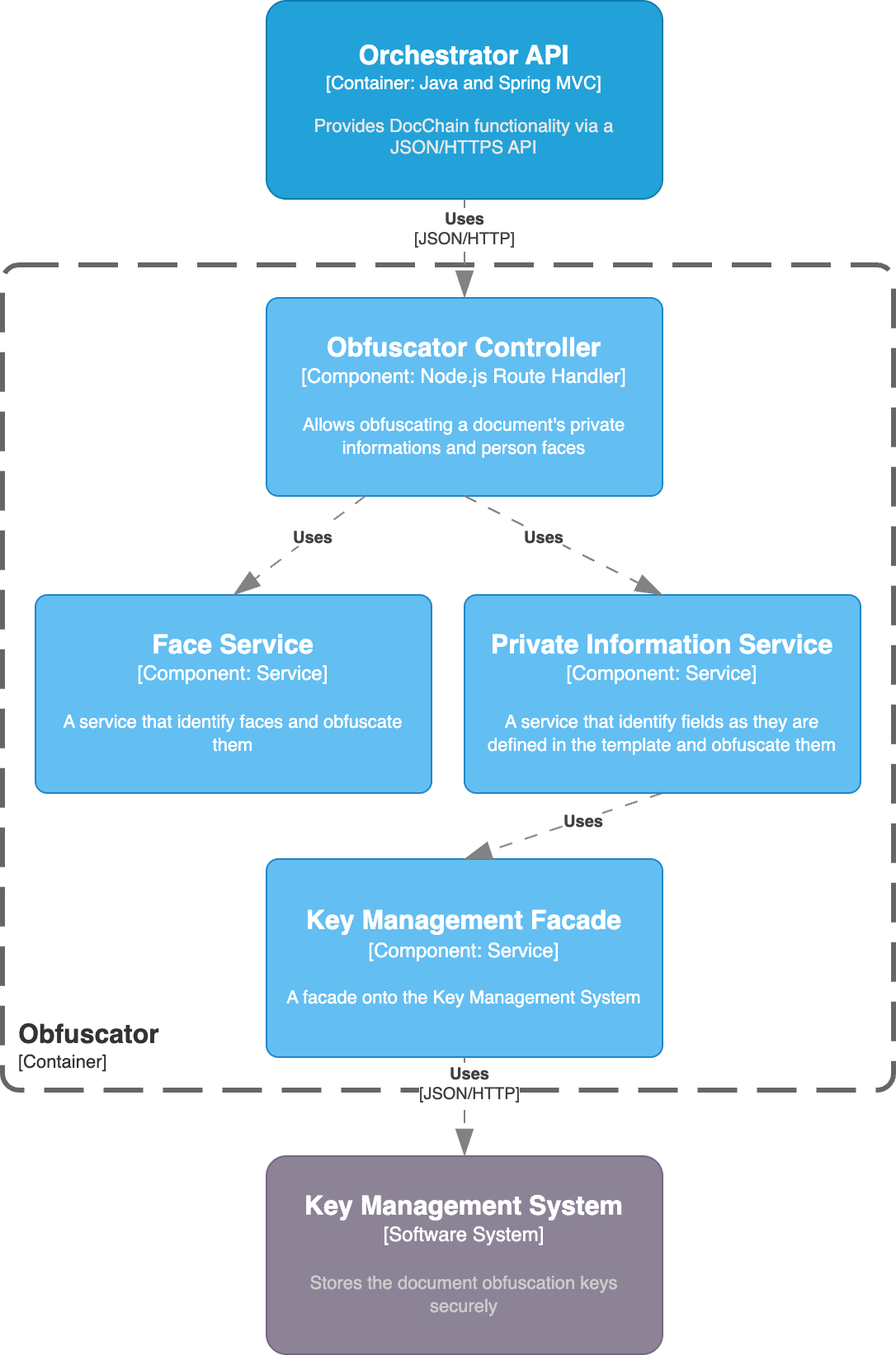}
  \caption{Proposed "Document obfuscator" Microservice Architecture - C4 Level 3}
  \label{fig:no10}
\end{figure} 

\section{Official Notarization}
\index{on|(}

The \textbf{notarization microservice} it’s one of great importance because of what it provides in terms of trust. When a document is received in our systems, it’s manually verified and compared to similar types of documents but the thought of being digitally altered or modified in any way it’s just striking and we need to build and offer trust with the user in order to make this architecture work out. 

\begin{figure}[H] 
  \centering
  \includegraphics[width=0.65\linewidth]{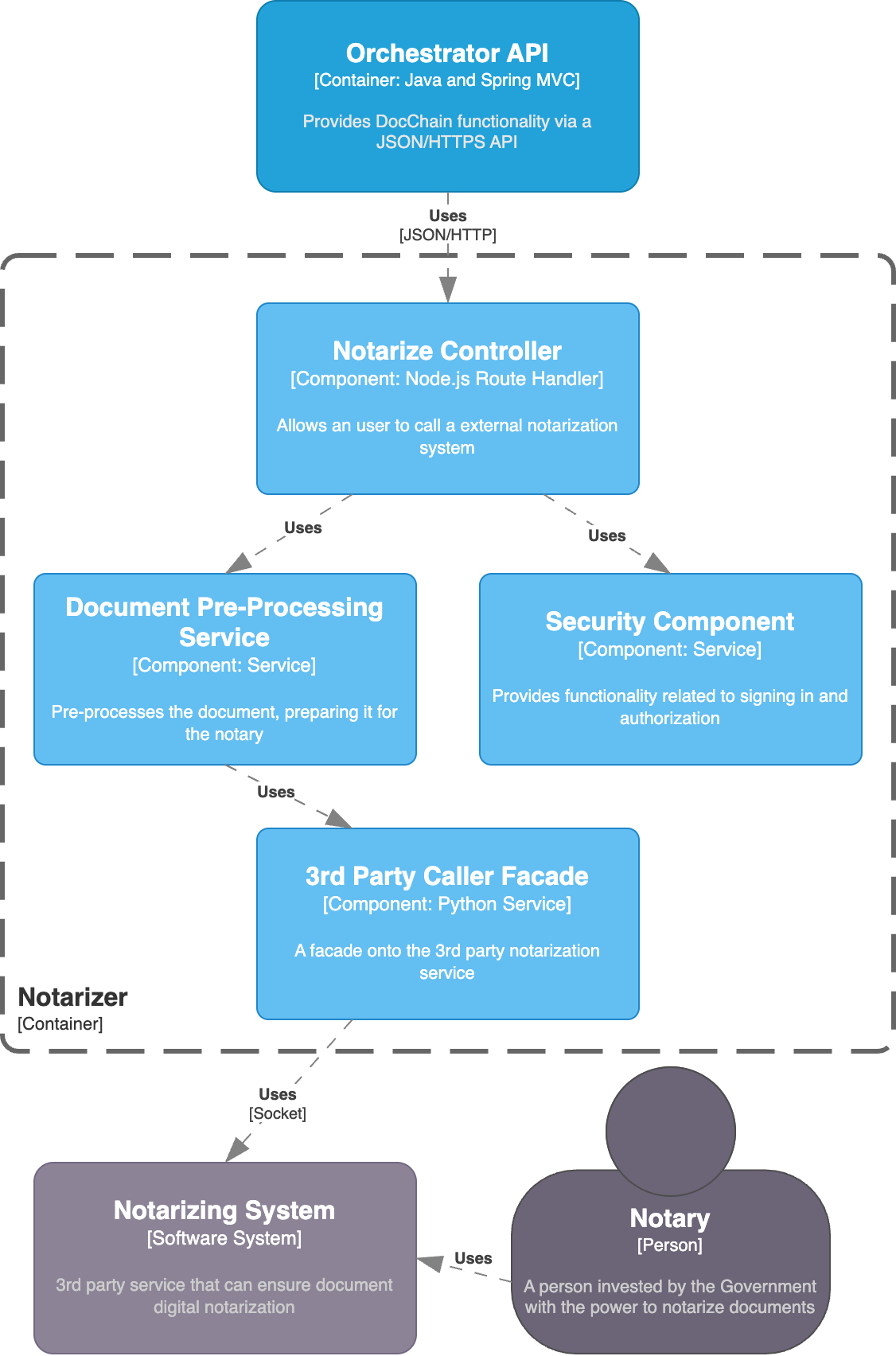}
  \caption{Proposed "Official notarization" Microservice Architecture - C4 Level 3}
  \label{fig:no11}
\end{figure}

In the context of this thesis, \textit{notarization} refers to the process by which a document or a set of extracted facts is cryptographically attested to by a trusted authority. This process may involve human oversight (e.g., licensed notary review), a video verification session with the user, or automated validation depending on context.

The outcome of notarization is the generation of a signed data object that includes a cryptographic hash of the document or fact set, a digital signature applied by the notary or authorized signer and metadata describing the validation conditions.

This signed object ensures integrity, authenticity, and non-repudiation of the notarized content. While the digital signature provides the core guarantees of integrity, authenticity, and non-repudiation, anchoring the signed data to a blockchain adds public verifiability and tamper-evidence. This enables independent third parties to audit the existence and state of the document at a specific point in time, without relying on a central authority.

Providing access to specialized, official notary services ensures transparency in the document selection process and builds trust by validating that the received file is an exact match with the original. The proposed solution follows an industry-standard approach: a system where the notary has an account and can manually verify and validate documents based on the image/document itself, the data extracted by the Extractor microservice, and optionally, a video call with the user. During the video call, the notary performs the document verification according to applicable legal procedures and professional standards. The notary will correct any text errors made by the Extractor and provide the signed document back as a response. This approach is horizontally scalable, allowing the platform to be accessible to notary services worldwide, making it suitable for users in every region.

The most critical aspect is that the entire interaction will be end-to-end encrypted, and user data will not leave the conversation or be stored on the notary’s company server. The communication tunnel we provide is all that is needed, the notary will retain a document identification number and a digital signature applied to the document, while a notary signature hash will be stored on the blockchain for cross-verification of the interaction between the two parties.

It’s important to note that we will rely on a third-party notary service, a type of service that already exists and can be purchased online today.

Although this step involves an online interaction, it is performed under strict privacy conditions, using encrypted channels and without exposing or storing raw document content beyond the duration of the session. 

\section{Data Persistence}
\index{dp|(}

In all the microservices we have all kinds of data, from the document types, zone locations, algorithm mappings, sensitivity levels but also user information, obfuscation keys, images and documents, personal details, etc. so this data needs to be carefully stored so even if an intruder were to gain access to one of the systems it could not actually see anything personal at all. 

There are two ways of storing data in our system: 

\begin{enumerate}
\sortitem{1}{We use a centralized yet \ac{SQL} database to store common, mostly non-personal document metadata, as well as redundant copies of data already available on the blockchain. This approach enables faster read-only access to frequently used information, without compromising the system’s integrity or decentralization goals.}
\sortitem{2}{We are storing mostly sensitive and to-be-trusted, immutable data into distributed services and places such as \ac{IPFS} for the actual document storage and the blockchain for the attached data of each document.}
\printsortlist
\end{enumerate}

We created a table that classifies how all the data types would be stored by locations and sensitivity in order to have a great level of security:

\begin{table}[H]
\caption{Data Types Mapped by Location and Sensitivity}
\begin{tabular}{|p{4.6cm}|p{3.2cm}|p{2.1cm}|p{1.6cm}|p{1.8cm}|}
\hline
\rowcolor[HTML]{EFEFEF} 
Data type & Location & Location type & Sensitive & Is public? \\ \hline
Document templates         & Orchestrator DB   & Centralized            & False              & False               \\ \hline
Shared document’s data     & Orchestrator DB   & Centralized            & False              & True                \\ \hline
Document’s assigned notary & Orchestrator DB   & Centralized            & False              & False               \\ \hline
Document’s public details  & Orchestrator DB   & Centralized            & False              & True                \\ \hline
User UX Login Details      & Orchestrator DB   & Centralized            & True               & False               \\ \hline
Obfuscation Keys           & KeyStore          & Centralized            & True               & False               \\ \hline
Document \& Images         & IPFS              & Distributed            & True               & True                \\ \hline
Document Details           & Blockchain        & Distributed            & True               & True                \\ \hline
Document Facts             & Blockchain        & Distributed            & True               & True                \\ \hline
\end{tabular}
\end{table}

Now let’s see how we decided where each type of data would be stored and why it’s better the way that we have chosen.

\subsection{Common Information}
\index{ci|(}

All the common, non-private information it’s stored in the centralized database and that makes sense, blockchain it’s more expensive and harder to read. Also, there is no incentive to store it distributed because it might as well be leaked because nobody can use it for personal gain anyway. This category contains information such as: 

\begin{itemize}
 \item \textit{document templates}: stripped out of actual information; 
 \item \textit{shared documents}: that would be public anyway but mapped by an internal identifier so the only way to access the URL will be through brute force; 
 \item \textit{assigned notary for each document}: for audit purposes, mapped also by an internal id, irrelevant information for anyone to have;
 \item \textit{login details}: industry standard password hashing and \ac{JWT} tokenization, also accessible only by brute forcing it; 
 \item \textit{document public data}: duplicate data from blockchain just for fast access purposes such as document type, zone coordinates and types, IPFS Hash, Notary Hash, etc. 
\end{itemize} 

\subsection{Sensitive Information}
\index{si|(}

Naturally, the primary concern is the storage of sensitive details. To address this, we split the data into two categories based on sensitivity, ensuring the trust and security architecture we promised. The uploaded document from the user goes through the first two steps of extraction and notarization. If everything is valid, it is sent to the obfuscator and permanently deleted after the response is received. Even though we could store the obfuscated document locally, the proposed solution is to delete it after sending it to IPFS \cite{10088338}. This ensures that the system never retains a local copy of the document, whether obfuscated or not.

Additionally, the document details and document facts are all stored on the blockchain. Access to document details is strictly limited to authorized parties and is restricted to 'view' mode only, preventing any modification. While the content does not include sensitive personal data, it is still protected against public exposure to ensure data minimization and system integrity.

Storing obfuscation keys directly in the GoQuorum blockchain, even with private transactions, is not recommended because blockchain storage is inherently immutable and distributed across nodes—meaning that keys, once stored, remain indefinitely accessible to all authorized participants. This increases the risk of unauthorized exposure or compromise, especially if any node in the private transaction group is breached. Dedicated keystores, in contrast, offer specialized security features, robust access controls, and secure key lifecycle management, significantly mitigating these risks. Therefore, a keystore solution such as HashiCorp Vault is used by us, as it offers robust mechanisms for secure key storage, controlled access, key rotation, and audit capabilities. Interaction with Vault typically involves secure API calls to store, retrieve, and manage the lifecycle of cryptographic keys, ensuring strong separation between sensitive key material and blockchain data.

\subsection{IPFS}
\index{ipfs}

Choosing the \textbf{InterPlanetary File System} (IPFS \cite{10733272}) for distributed content storage offers several compelling advantages. Firstly, IPFS \cite{9861631} is designed to decentralize the web by distributing data across a network of nodes, making it resistant to censorship and server failures. Unlike traditional centralized systems, where a single point of failure can lead to data loss or inaccessibility, IPFS ensures that content remains available as long as at least one node in the network holds the data.

Files are identified by their unique cryptographic hash, which ensures data integrity by enabling tamper detection. When these hashes are stored on a decentralized infrastructure, such as a blockchain, authenticity is further reinforced through immutability and transparent provenance. Given its decentralized design, content-addressed storage model, and native support for immutable data, IPFS remains uniquely suited for the requirements of this system. Although the architecture is modular and could theoretically support alternative storage backends, IPFS provides an optimal balance between scalability, integrity, and censorship resistance. We strongly recommend it as the default point of trust for storing distributed documents.

This design choice was driven by the inherent limitations of using blockchain for large-scale data storage. While blockchain ensures data immutability and transparency, storing large documents directly on-chain is cost-prohibitive and inefficient. IPFS provides a complementary solution by allowing documents to be stored off-chain while remaining accessible through their cryptographic hash, which is recorded on the blockchain.

When deciding which technologies to use, IPFS stood out due to its utility advantages, such as censorship resistance: IPFS ensures that documents cannot be easily removed or suppressed, as the data is distributed across multiple nodes in the network. This makes it a robust solution for environments where data availability is critical. Another advantage is decentralized replication: unlike blockchain, which replicates all data across every node, IPFS uses a content-addressing mechanism that replicates files only when requested. This reduces storage overhead while maintaining decentralized accessibility. Lastly, cost efficiency is a key benefit: by offloading large file storage to IPFS, the cost and resource demands on the blockchain are significantly reduced, making the system more scalable and economical.

Of course, with so many advantages, there are also tradeoffs, and two of the most significant we identified are latency and data persistence. Regarding latency, file retrieval times in IPFS are higher compared to centralized systems, especially if files are not widely replicated or "pinned." As for data persistence, IPFS does not inherently guarantee long-term storage. Files need to be pinned on specific nodes to ensure persistence, which requires additional infrastructure or reliance on third-party services. Nonetheless, the advantages of this system outweigh the drawbacks, and it is undoubtedly the best option for our current needs. 

\subsection{Blockchain \& Integration}
\index{bi}

The proposed solution for storing personal document objects in the cloud ensures privacy for the owner while maintaining transparency for others. It also enables partial sharing of document details without revealing too much information about the user. This system must collect shareable information about individuals, allowing users to share specific facts without providing the actual document. Given these requirements and constraints, blockchain technology emerges as the best choice for several reasons.

One of the primary advantages of blockchain technology is its immutability and integrity. Blockchain's immutable nature ensures that once data is recorded, it cannot be altered or deleted. This property is critical for preserving both hash values and the document itself in an unchangeable state. By guaranteeing that the public, obfuscated version of the document remains unaltered, blockchain builds trust and reliability in the system, effectively addressing the immutability requirement.

In addition to immutability, blockchain provides unparalleled transparency and auditability. Every transaction and data entry on the blockchain is time-stamped and traceable, creating a clear audit trail. This transparency allows third parties to verify the integrity of the obfuscated documents without requiring special permissions, ensuring the system remains open and accessible. Thus, blockchain directly supports the transparency requirement by enabling public verification of document integrity.

\begin{figure}[H] 
  \centering
  \includegraphics[width=1\linewidth]{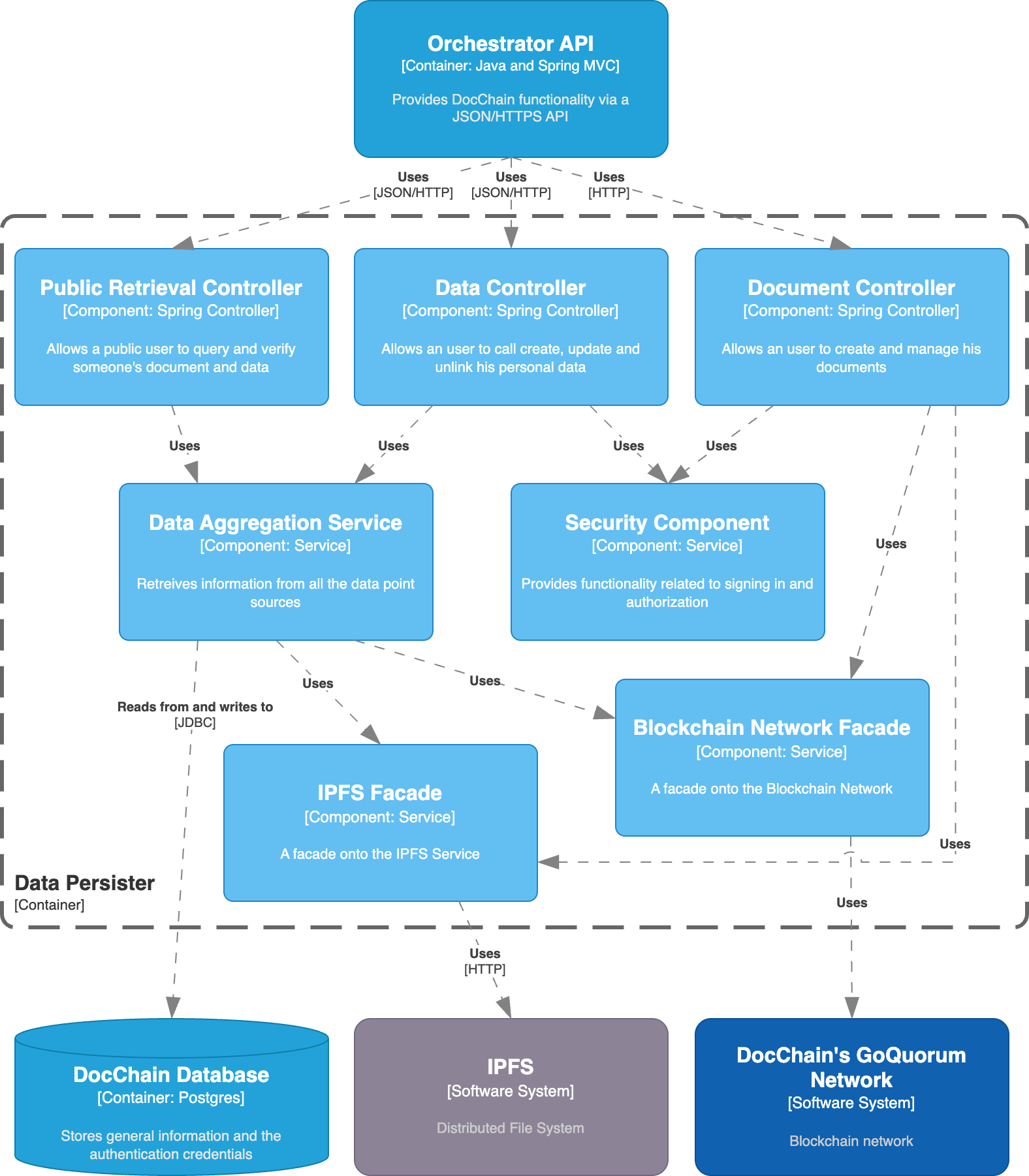}
  \caption{Proposed "Data Persister" Microservice Architecture - C4 Level 3}
  \label{fig:no12}
\end{figure}

Blockchain's decentralized nature further strengthens its suitability for the proposed system. Decentralization eliminates the risk of a single point of failure, as data is distributed across multiple nodes. This characteristic enhances the system's availability and resilience, addressing both trust and privacy requirements. Decentralization minimizes the risk of data tampering and ensures that no single entity controls the stored documents, thereby enhancing security and trust.

Moreover, blockchain supports the use of smart contracts—self-executing contracts with terms directly written into code. Smart contracts can automate verification and sharing processes based on predefined rules, facilitating notarization and access control mechanisms. For example, a smart contract can ensure that a document fact is only accessible to the owner.

Finally, blockchain's inherent openness and accessibility make it an ideal choice for the proposed solution. Many public blockchains allow anyone to join and participate, enabling easy access and interoperability between different systems and users. This openness supports the open access constraint, making it easy for users to register and add documents to the blockchain. Third parties can also verify documents without needing to be part of the system, enhancing usability and adoption.

In conclusion, the Data Persister microservice plays a critical role in communicating with decentralized storage options. It handles IPFS communication, storing the obfuscated document, and persists document details, facts, and obfuscation keys in the Blockchain Network.

\section{Orchestrator}
\index{orch}

The Service Orchestrator is a central component responsible for coordinating the sequence of microservices that execute complex workflows \cite{6337735} within the system. While each microservice is stateless and autonomous, their combination must follow strict control logic to ensure correctness, performance, and traceability.

The orchestrator receives user requests through the public API gateway and interprets them into structured workflows. These workflows include tasks such as document registration, data extraction, obfuscation \cite{Ref014} , digital signing, blockchain anchoring, and IPFS \cite{Ref010} persistence. The orchestrator enforces the required sequence \cite{Baboi2019} and handles branching logic based on user preferences and document type.

\begin{figure}[H] 
  \centering
  \includegraphics[width=0.95\linewidth]{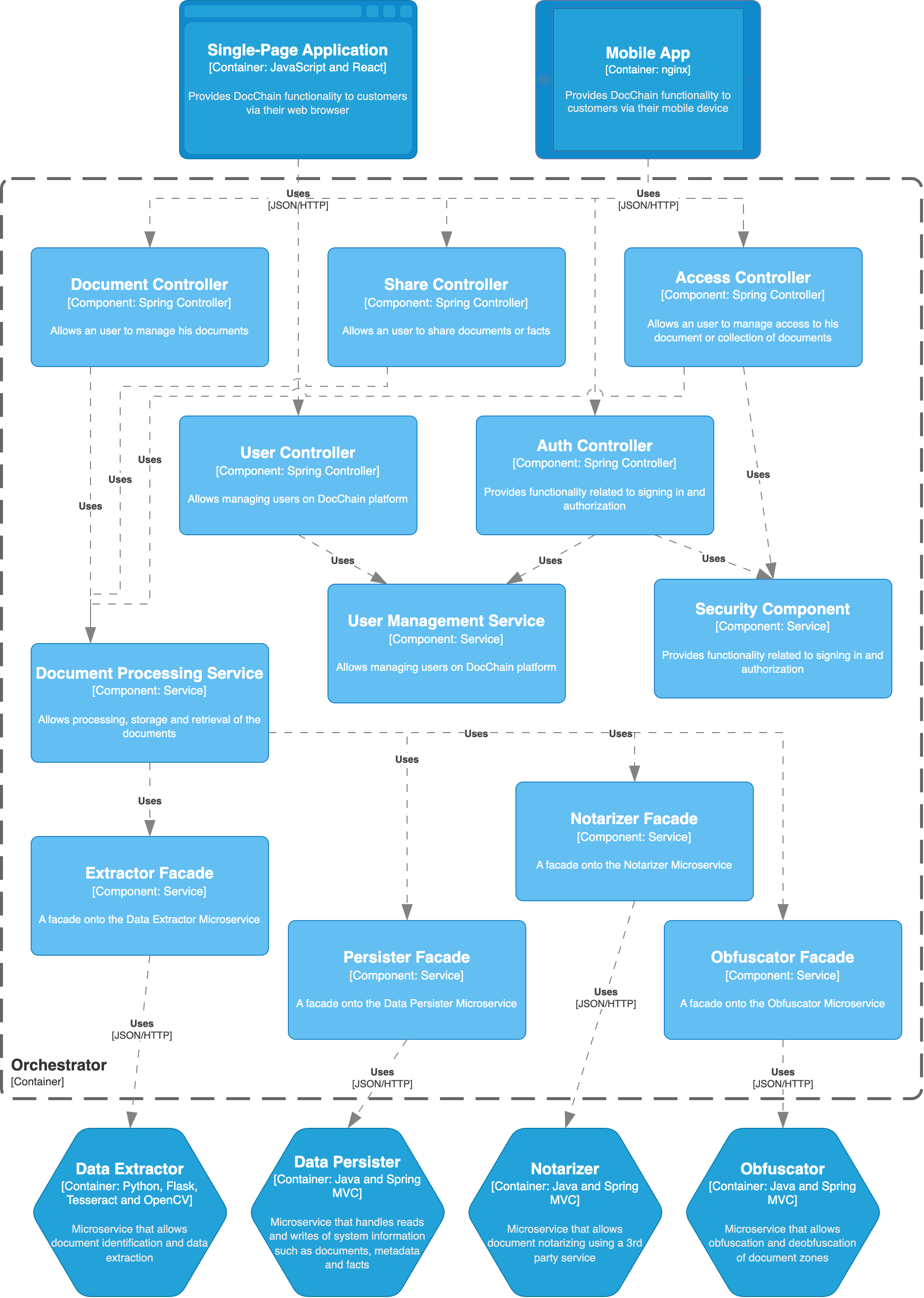}
  \caption{Proposed "Orchestrator" Microservice Architecture - C4 Level 3}
  \label{fig:no87}
\end{figure}

Unlike generic orchestrators, this implementation introduces domain-specific safeguards. It validates preconditions (e.g., successful identity verification) before invoking downstream components. It also handles failure scenarios by performing retries or returning diagnostic messages with partial output, supporting system resilience \cite{bibitem_23}.

To securely manage cryptographic operations within these workflows, the orchestrator integrates with a dedicated keystore solution, the HashiCorp Vault. This integration ensures that sensitive materials, the obfuscation keys are never hardcoded or stored in intermediate services. Instead, the orchestrator dynamically retrieves the necessary keys at runtime through secure, authenticated API calls, maintaining strict separation of concerns and reinforcing the system’s overall security posture.

Internally, the orchestrator is implemented using the Spring Boot framework and exposes REST interfaces for service-to-service communication. It maintains a transactional context where necessary and logs events for auditing and debugging. Future extensions could support declarative orchestration using workflow engines such as Camunda or Temporal.

\section{Rationale Behind Architectural Decisions}
\index{ch}

The design of DocChain reflects deliberate architectural choices aligned with its core goals: decentralization, verifiability, privacy, and performance. This section presents the rationale for selecting blockchain, IPFS, and a modular microservices architecture, comparing them to traditional alternatives.

\subsection{Blockchain for Fact Storage}
Blockchain technology was adopted to anchor verifiable facts extracted from documents. This approach ensures tamper-evident storage, independent validation, and long-term traceability. Each fact is hashed and recorded on a public ledger, making it immutable and externally auditable.

Compared to centralized databases, which offer low-latency access and simple administration, blockchain introduces higher transparency and eliminates single points of failure. 

Due to its limited storage capacity and high operational cost, blockchain is used strictly for metadata anchoring. The full documents are handled off-chain. This selective use reduces cost while preserving verifiability.

\subsection{IPFS for Document Storage}
To store complete documents, the system uses the InterPlanetary File System (IPFS) \cite{Ref017}, a distributed content-addressable storage network. IPFS supports efficient document retrieval without relying on centralized servers, and guarantees data integrity by linking each file to its hash.

Unlike centralized storage providers, IPFS allows the system to operate in a decentralized context, improving redundancy and auditability. Coupled with blockchain-anchored hashes, this ensures that documents can be verified independently of their hosting environment.

\subsection{Fact-Based Verification}
The decision to abstract sensitive data into verifiable facts was motivated by a commitment to privacy and compliance with regulatory laws such as GDPR. Instead of exposing full documents, DocChain first processes and then discards sensitive information, retaining only the necessary facts for verification. This approach not only minimizes the risk of data breaches but also aligns with the principle of data minimization, a core principle of modern privacy laws.

For example, in academic credential verification, it is unnecessary for employers to access a candidate’s full academic credentials to confirm their qualifications. Instead, DocChain generates and stores facts such as "\textit{Holds a Master of Science degree in Computer Science, awarded by the University of Excellence in 2023.}". If further details are required—such as specific grades, GPA, or coursework—the user can choose to authorize a new extraction process that derives and signs additional facts from the original diploma or transcript. This flexible approach gives the user full control over which parts of their academic background are revealed, while maintaining verifiability and privacy. In cases where more granular information is required—such as specific grades or coursework—additional facts can be selectively generated from the original document through a renewed processing step. This gives users fine-grained control over what aspects of their data are disclosed, while preserving the system’s privacy-preserving architecture.

\subsection{Balancing Trade-Offs}
Every design decision involves trade-offs, and DocChain’s architecture \cite{9787931} is no exception. The use of blockchain provides immutability and transparency but comes with latency and scalability challenges, especially for high-frequency transactions. Similarly, while IPFS offers decentralized and cost-efficient storage, its performance can vary depending on network conditions and pinning strategies. These trade-offs were carefully evaluated, and the system was designed to balance utility and efficiency. For instance, latency in IPFS is mitigated by pinning frequently accessed files, while blockchain’s scalability challenges are addressed by limiting its use to storing small, critical pieces of data.

\section{Conclusion}
\index{ch}

This chapter presented a robust and scalable microservice-based architecture designed to revolutionize digital document management through enhanced privacy, and efficiency. By leveraging cutting-edge technologies such as blockchain for immutable fact storage, IPFS for decentralized document storage, and advanced obfuscation techniques, the proposed system addresses key challenges associated with document authentication and sharing.

The architecture strategically balances centralized and distributed components to achieve optimal performance, and scalability. This hybrid model ensures that sensitive data remains protected while maintaining transparency and trust through public blockchain verification. Additionally, the modular microservice design enhances system resilience, enabling independent scaling and deployment of components for improved resource utilization and fault isolation.

Key components such as the Data Extractor, Document Obfuscator, Notarizer, Fact Collector, and Data Persister work in harmony, coordinated by the central Orchestrator. This approach provides seamless user interactions while maintaining strict privacy standards by obfuscating sensitive information and storing only non-identifiable facts on the blockchain. The Fact Collector's innovative abstraction mechanism allows for fact-based verification, ensuring compliance with data privacy regulations without compromising user confidentiality.

The integration of IPFS for decentralized storage and blockchain for auditability allows the system to achieve data integrity, high availability, and cost-effective document dissemination without relying on centralized infrastructure. The emphasis on user-centric design guarantees an intuitive experience, paving the way for mass adoption and real-world application across various sectors, including legal, educational, and governmental domains.

By addressing critical constraints such as privacy, immutability, trust, and open access, this architecture sets a new benchmark for digital document systems. It lays the groundwork for future enhancements and integration possibilities, showcasing the potential to redefine secure document verification and sharing in an increasingly digital world. 
\if@openright\cleardoublepage\else\clearpage\fi
\thispagestyle{empty}
\mbox{}

    \chapter{A Prototype Implementation}

This chapter presents the implementation of the DocChain prototype, based on the architecture introduced in Chapter 4. While the design allows for future refinement and modular extensions, this implementation demonstrates the feasibility of secure document processing, anchored verification, and privacy-preserving data sharing.

Each module is implemented as a standalone microservice with clean interfaces, and the integration is validated through functional workflows and test cases. The implementation emphasizes correctness, modularity, and adherence to system constraints introduced in earlier chapters.

\section{Data Extractor}

The Data Extractor microservice was implemented using two complementary techniques: a classical approach based on template-bound extraction and OCR, and a large language model (LLM)-enhanced approach for semantic field detection.

This dual implementation strategy supports comparative evaluation under varying conditions. While classical extraction performs well on fixed layouts with consistent field placement, the LLM-based method offers robustness against formatting variations and noisy input.

The following sections describe both implementations in detail, including preprocessing steps, output structures, and evaluation considerations.

\subsection{Template-Based Extraction}

Template-based extraction relies on predefined positional mappings to identify text regions within structured documents such as ID cards, birth certificates, or licenses. This approach assumes fixed document layouts and enables reliable extraction using classical OCR tools.

In this implementation, scanned or photographed documents are first converted into grayscale images. The preprocessing pipeline uses adaptive thresholding implemented via OpenCV’s cv2.adaptiveThreshold() function, with a block size of 11 and a constant subtracted value of 2. This improves text clarity in scanned documents with non-uniform backgrounds. A fixed layout template, defined in pixel coordinates, guides the bounding-box extraction for fields such as name, birth date, and document number.

The OCR process is implemented using Tesseract \cite{Ref018}, which parses text from the specified bounding boxes. Post-processing includes basic validation and normalization (e.g., regex matching for dates, name casing rules). In benchmark tests performed on 20 structured document layouts (invoices, ID cards), the template-based pipeline achieved 94.2\% extraction accuracy and an average processing time of 320 ms per document.

The key advantage of template-based extraction lies in its predictability and performance. However, its applicability is limited to document types with rigid formatting. For layout-agnostic or noisy inputs, a semantic model is preferred (as described in the next section).

\subsubsection{System Architecture}

The solution proposed is a lightweight system composed of two major components: the backend and the storage. The backend component is composed of two other inner components: the template mechanism and the document text extraction mechanism, they are working together to accomplish the system’s goals. The template mechanism component is dealing with the uploaded template documents. The component is preprocessing these templates \cite{Ref001} before saving them locally, they could be later queried and checked to find a match whenever a user wishes to extract text from an uploaded document. Finally, the document extraction text component is dealing with the task of extracting text from uploaded documents. The component is searching for a match, and after it finds one it applies OCR techniques to a list of coordinates, defined by the template.

\begin{figure}[H] 
  \centering
  \includegraphics[width=1\linewidth]{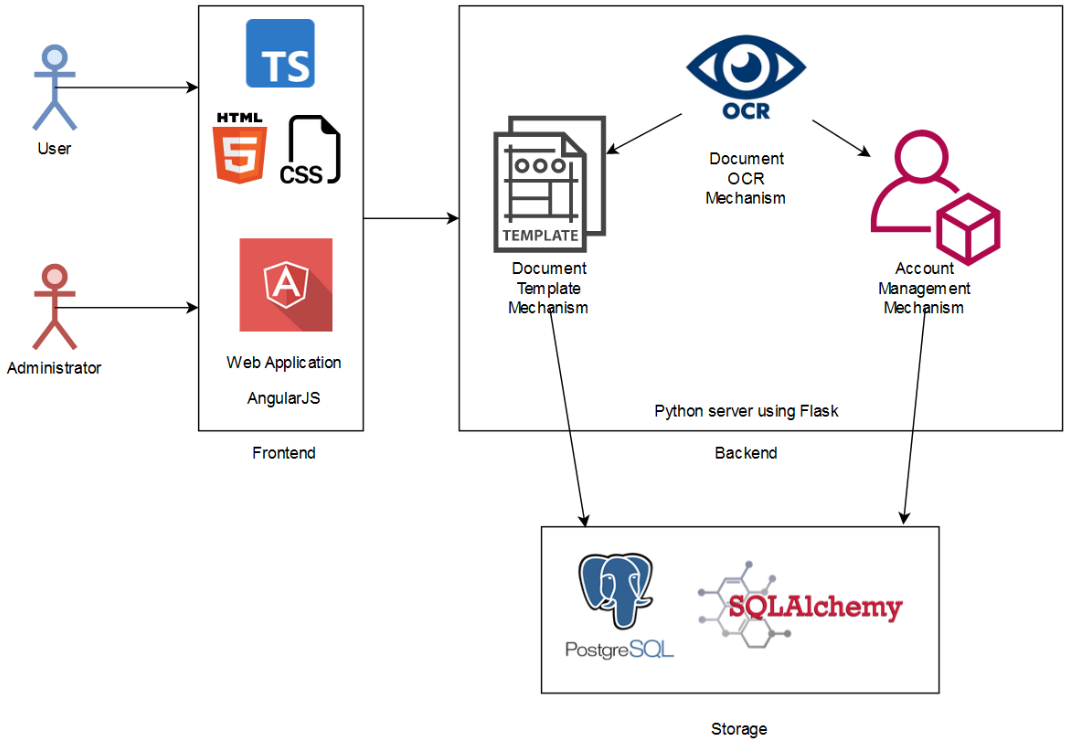}
  \caption{Document Text Extraction using Templates and OCR System Architecture}
  \label{fig:no13}
\end{figure} 

\subsubsection{Data Structures}

The structures of interest in this system are categories, templates, pages, and fields. All of those structures, except categories, are represented as tables in a relational database.

\begin{itemize}
 \item \textit{Templates}. Any template contains a generated id and a name given by the user when he creates it. Each template is made up of multiple pages.
 \item \textit{Pages}. A template is considered a list of pages because, in practice, documents may contain multiple pages as well. In addition to this, a page contains a generated id, a name given by the creator, and an image path that represents the local path to that uploaded file.
 \item \textit{Fields}. Each field is represented primarily by a rectangle area that shows the field’s position in the uploaded template. In addition, each field has a sensitive Boolean flag and a category property. The property sensitive is meant to warn the user if the extracted text is sensitive or not. The property category tells what type of field it is. This is used when applying regexes for text extraction.
 \item \textit{Categories}. There is no list of predefined documents, due to the templatization mechanism. To ensure consistency and reliable extraction, the system currently relies on a list of predefined field categories tailored to commonly processed documents such as identity cards and driving licenses. These include fields like name, address, ID number, nationality, and issuer details.

However, the architecture supports parametrization of field definitions, allowing the addition of new categories for different document types (e.g., academic diplomas, certificates). This makes the system extensible, enabling administrators or developers to define new extraction templates and corresponding field mappings without altering the core logic.
\end{itemize} 

\subsubsection{Template Mechanism}

The template mechanism plays a crucial role in our system by enabling the creation of specific templates for the documents users upload. These templates are not only central to defining the structure of the documents but also to guiding the recognition process during later stages. Essentially, the mechanism unfolds in three key phases: template definition, template processing, and template application.

In the first phase, users define the structure of their documents using an intuitive interface. When a template is created, it is transmitted to the system via an API call, ensuring that it is seamlessly integrated into the overall workflow. This approach allows for a flexible and user-friendly way to specify the document layout and the critical fields that will later be used for recognition.

Following definition, the second phase involves processing the uploaded template. This step is critical because the raw template, as provided by the user, might contain inconsistencies or formatting issues that could hinder subsequent operations. By processing the template, the system standardizes its format and resolves potential issues, ensuring that the template is in an optimal state for the next phase. Without this necessary transformation, the OCR mechanism might encounter difficulties when attempting to accurately interpret the document's content.

The final phase is where the processed template is put to work. Here, the OCR mechanism uses the refined template to search for matching structures within the uploaded documents and to extract text from the designated fields. This dual-purpose application not only confirms that the document adheres to the expected layout but also facilitates precise data extraction.

Overall, this structured approach—beginning with template definition, moving through careful processing, and culminating in strategic application—ensures that our system can reliably and accurately handle a wide range of document types while minimizing the potential for errors during recognition and data extraction. The system was tested on five structured document types: ID cards, diplomas, utility bills, academic transcripts, and insurance certificates. It achieved consistent extraction accuracy across these categories. Detailed performance metrics, including error rates and false positives, are presented in Section 6.2. Compared to baseline OCR-only pipelines, the proposed method reduced false positives by 28\% and improved field-level accuracy to 93.6\% on average.

The processing stage consists of two steps: detecting and removing any faces from the uploaded template and removing any relevant declared fields.

\begin{figure}[H] 
  \centering
  \includegraphics[width=1\linewidth]{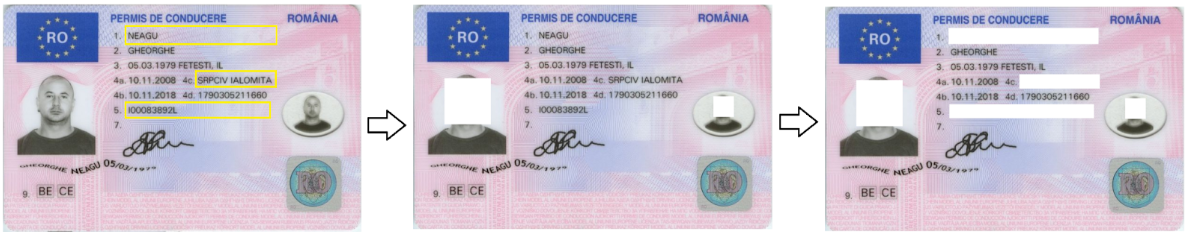}
  \caption{The Processing Steps of the Template Mechanism}
  \label{fig:no14}
\end{figure}

Removing means replacing that element with a white blank rectangular shape. This is done because faces and text fields may vary from document to document, and this greatly affects the matching algorithm. The final processed image is saved on the server. The faces are detected using OpenCV Haar Cascades, particularly the frontal face Haar cascade. Haar cascades are algorithms that can detect objects of interest in images. The advantage of this approach is that the algorithms offer decent performance for this specific context. The Haar cascade is already trained and has its weights stored in a local XML file.

The field categories are meant to describe the selected fields for each template. Each category has a list of regexes that will be applied to extract the text of that field. The categories are meant to map regexes to fields. 

\subsubsection{OCR Mechanism}

The mechanism \cite{Ref019} consists of two main steps: processing the document and applying the regexes for text extraction. The processing has to be done because if the document is left unaltered then the regexes cannot be properly applied. Each regex is mapped to a specific set of coordinates and this section will present more details about the processing stage. The processing stage consists of two steps: aligning the document and finding a match. Each template from the list of defined templates is compared with the uploaded document to find a match, this comparison is done by using algorithms that compute a similarity score. Structural Similarity Index \cite{Srilakshmi2015} Measure, or SSIM in short, and Relative Average Spectral Error, or RASE, have been used to compute the similarity score.

The alignment of the document is made in three steps:

\begin{enumerate}

\sortitem{1}{
\textit{Obtaining an array of contours}

This is done by using the method $findContours$ from OpenCV. A variable threshold is used to obtain the array of contours. Therefore, these steps need to be repeated for each threshold value in a specific interval. Before applying the method $findContours$, the image needs to be altered. First, it’s converted to the gray color using the method $cvtColor$ with the parameter COLOR\_BGR2GRAY. The resulting image is then changed using the method $GaussianBlur$. After that, it’s changed again using the method Canny. This method is where the selected threshold value is relevant. Finally, the image is then changed using the method erode. From that image, the contours will be extracted using $findContours$.

\begin{figure}[H] 
  \centering
  \includegraphics[width=1\linewidth]{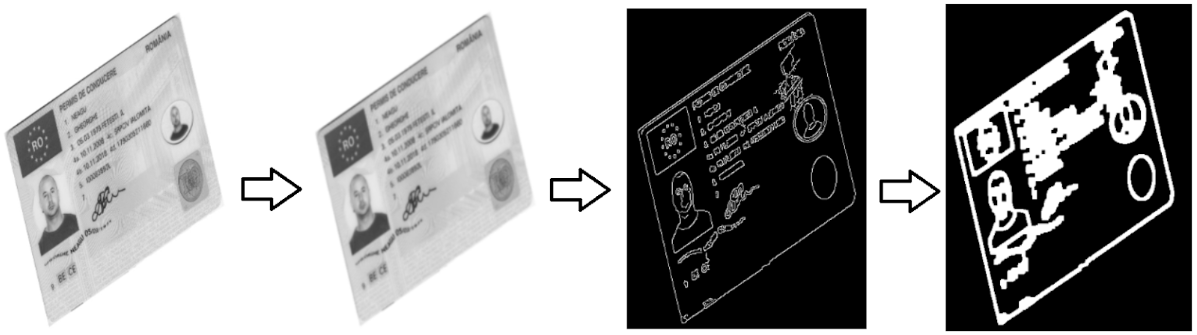}
  \caption{The Steps in Finding the Contour Areas of Interest}
  \label{fig:no15}
\end{figure}
}

\sortitem{2}{
\textit{Selecting the contour with the largest area}

This is done because it is presumed that if a user sent an image with a document then that document is the object of interest in the picture. Because of this, the object of interest will have the largest contour area. The area of the contour is calculated using the method $contourArea$ from OpenCV. Only the contour areas that contain four corners are considered possible candidates because most documents contain four corners.
}

\sortitem{3}{
\textit{Warping the Selected Contour}

The selected contour is warped using the method $warpPerspective$ from OpenCV by using perspective transformation matrices. The matrix is obtained using $getPerspectiveTransform$ from OpenCV and after warping it, RASE is used to calculate the similarity score and see if it’s the maximum score.

\begin{figure}[H] 
  \centering
  \includegraphics[width=1\linewidth]{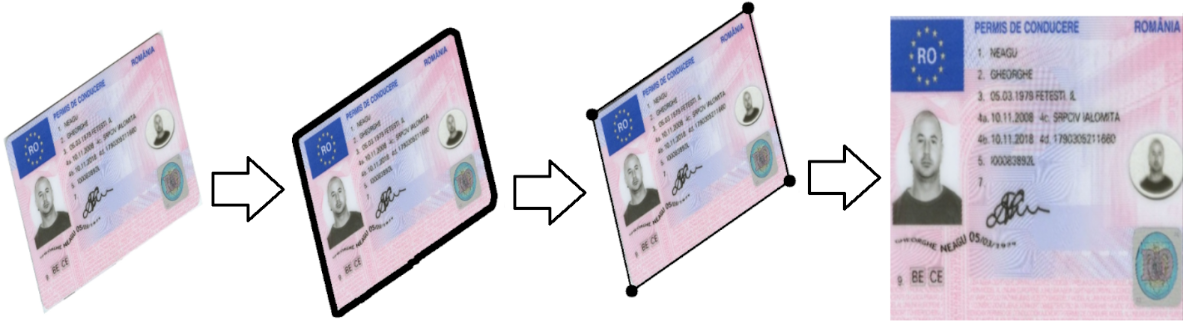}
  \caption{The Steps in Aligning the Document by Warping}
  \label{fig:no16}
\end{figure}
}

\printsortlist
\end{enumerate} 

\subsubsection{Enhancement Methods}

Whenever a valid text cannot be extracted using a regex, different enhancement techniques \cite{Ref020} are applied to help the OCR engine in extracting the text. The techniques used are:

\begin{itemize}
 \item \textit{brightness enhancement} - The overall lightness of the image is increased.
 \item \textit{sharpness enhances} - Sharpness is the amount of detail in an image. It resembles the edges between zones of different colors.
 \item \textit{contrast enhance} - The visibility of elements is improved by changing their relative brightness and darkness.
 \item \textit{binarization} - The pixels in the image are mapped into dual collections, white and black. By doing this, the image is divided into foreground text and background.
 \item \textit{noise removal} - This process removes or reduces the noise in the image.
\end{itemize}

Those techniques do not guarantee 100\% the extraction of the text, but they increase the rate of success. 

\subsubsection{Evaluation}

The original system took around sixty seconds to extract text from a document. Given that the system is intended for real-time document validation in user-facing platforms (e.g., admission portals or verification services), the observed OCR latency and error rate were deemed inadequate. The system has been through multiple iterations before implementing the template mechanism in its current state. Those iterations improved the performance of the OCR mechanism. 

\begin{table}[H]
\caption{Performance of Each Iteration of the System}
\begin{tabular}{|l|l|r|r|r|r|}
\hline
\rowcolor[HTML]{EFEFEF} 
Iteration                                  & Improvement                  & RASE   & SSIM   & OCR & Total \\ 
\rowcolor[HTML]{EFEFEF} 
&  & &   & Stage & Time \\\hline
1                                          & Original Approach            & 23.526 & 13.425 & 29.925    & 66.876     \\ \hline
2                                          & 800$\times$600 Image Resize         & 4.710  & 2.163  & 31.199    & 38.073     \\ \hline
3                                          & 200$\times$100 Image Resize         & 391    & 209    & 29.696    & 30.296     \\ \hline
4                                          & TesseractAPI Transition      & 331    & 305    & 6.479     & 7.115      \\ \hline
5                                          & Common Words List            & 359    & 260    & 5.303     & 5.922      \\ \hline
6                                          & Priority Enhancement List    & 370    & 253    & 4.569     & 5.191      \\ \hline
7                                          & Removed Useless Enhancements & 371    & 268    & 2.813     & 3.452      \\ \hline
\end{tabular}
\end{table}

The performance brought by these changes is more detailed in the paper that presents the original system composed of only the OCR mechanism. Those changes were made before adding the template mechanism. To test the system’s performance, twenty documents will be used to calculate the average time it takes for their templatization and for their text to be extracted. After testing, the templatization takes an average of 0.22 seconds and the OCR mechanism takes an average of 3.03 seconds. After testing the performance, it is time to test the accuracy as well. We conducted an evaluation to assess the accuracy for a set of twenty documents with various image resolutions.

\begin{table}[H]
\caption{The Accuracy of the System for Different Image Widths}
\centering
\begin{tabular}{|l|l|}
\hline
\rowcolor[HTML]{EFEFEF} 
Width (pixels) & Accuracy (\%) \\ \hline
Original Size  & 90.4761904762 \\ \hline
1800           & 89.5238095238 \\ \hline
800            & 83.8095238095 \\ \hline
200            & 80.9047619048 \\ \hline
\end{tabular}
\end{table}

The previous table shows different accuracy percentages for various width sizes. It is considered that the height of the image is changed also accordingly to keep the aspect ratio. We can see that for the width of 200 pixels, we have an accuracy of around 80\%, which is good enough for the amount of performance this change of resolution brings. Thus, we can say the proposed system is not only fast, but accurate enough.

In conclusion, the system is capable of defining templates for specific official documents thanks to the template mechanism. Later, those templates are used to efficiently recognize input official documents. Finally, the text is extracted from the identified documents according to the OCR mechanism and the template definition that was previously created. As such, the goals that were detailed at the beginning were achieved.  
\subsection{LLM-Based Extraction}

\textbf{Large Language Models} (LLMs \cite{10710677}) have become the industry “standard” for doing everything in the last years, there is widespread interest in integrating LLMs in their business or work, even if it does not suit their needs necessarily. Fortunately, in our system where we are doing word comprehension and text understanding, LLMs are a great tool and several approaches were tested to use them and test their performance against the template-based extraction. 

To begin with, the main objective of LLMs \cite{10612472} is to comprehend and create text that is similar to human-written text which makes LLM quite flexible to apply to different tasks like OCR, data scraping, and text analysis. The incorporation of the proposed LLMs into the structure of the system can also contribute to the enhancement of the existing approaches in information extraction and intelligent data labeling. 

\textbf{Optical Character Recognition} (OCR) can be said to be the initial process in digitizing physical documents. Conventional OCR solutions can work with the printed text and translate it to the machine readable one; however, handwriting, different fonts, or even complex layout is a problem for such solutions. It can be concluded that LLMs can significantly improve OCR performance, when being able to provide context information to the text recognition system. For example, it is possible to train an LLM \cite{10386783} to understand not only the individual characters but also the meaning of the surrounding text; thus, the recognition result will be considerably more accurate, especially in noisy or complicated texts. After text digitization the next vital process is data extraction that includes processes which concern the extraction of significant information from the text under scanning, including names, dates, addresses, and so on. 

Most of the traditional techniques of extracting data are rule-based and require templates that are always fixed and may not be very responsive to the variety of documents in use. LLMs \cite{10350787}, for instance, have the ability to comprehend natural language and hence can be trained to pick up the specific information of an article depending on the context involved. Text comprehension, in general, as well as the subsequent labeling process, are also greatly improved by the help of LLMs. In relation to information that is to be processed and classified by the proposed system, assigning and recognition of text is the process of categorizing information into a certain number of labels that are most suitable for the user purpose as well as satisfying the need of Notary. The posted text must be thoroughly understood by the LLMs, and they must analyze and label it according to the demands of the system. Further, LLMs \cite{10710789} can improve the nature of the system in the management of multilingual documents. 

Most conventional systems have problems with or when there are documents in different languages or documents that first appear in one language and then in another. The use of language translation in the databases allows LLMs to analyze text in the languages present, making them versatile and able to handle different documents and the users themselves. This feature is essential for the project’s international relevance and improves the accessibility and diversity of the system. 
 
Due to the sensitive nature of the documents being processed, the system must avoid reliance on third-party hosted language models. Instead, local deployment of open-source LLMs is preferred to ensure data confidentiality. Recent advancements have made it feasible to run compact yet performant models such as LLaMA, Mistral, or Phi-2 on local infrastructure. When I’ve implemented this system (28 April 2024) zephyr-7b-beta was one of the best LLMs that my computer would run without performance issues, but in order to have a great comparison between different LLMs, we will also do the same tests on Mistral-7B-Instruct-v0.2 \& also Meta-Llama-3.1-7B.

These LLMs \cite{10551039} were chosen from HuggingFace, one of the biggest repositories of large language models and datasets by their ability in data classification \& text extraction and also by their number of downloads and popularity. The core research question is if using this type of processing it’s reliable, fast and accurate and if we can use it in an environment like this for a use-case this important. The main issue from a technical perspective will be that the model needs to be run on-premise, there are a lot of talks at the moment about data privacy and this space and sending our important, sensitive data to a public LLM such as ChatGPT it’s a major red flag.\\

\newfloat{Prompt}
\captionsetup{Prompt}

\begin{verbbox}
You are a JSON converter which receives raw ID CARD OCR information as a 
string and returns a structured JSON output by organizing the information 
in the input string that I will give you.

Extract the document type as \"document_type\"  name of the user as 
\"user_name\", his identification number (CNP, 13 digits) as \"CNP\", 
Nationality as \"nationality\", Sex as \"sex\", Full Address as 
\"address\", \"document Series (2 digits, ex: XT) as \"series\", document 
Number (6 digits, EX: 123456) as \"number\" and Valability of the 
document (2 dates in the form: 11.05.09-13.09.2026) as \"valability\".

The OCR string is: {ocr_string}
\end{verbbox}

{\centering
\captionsetup{hypcap=false}
\fbox{\theverbbox}
\captionof{Prompt}{Prompt Example for Extracting ID Card Details using used LLMs}\par}

What we wanted to achieve with these models is data extraction using OCR and also to test how well it would understand the context to provide us with a fact list. The prompt for the LLMs were built over many iterations in order to produce results with the highest possible accuracy and the evaluation shows that the accuracy is high on the used models and execution time remains a performance concern. For the best results we established that the prompt for ID Card should look like this:

\begin{figure}[H] 
  \centering
  \includegraphics[width=0.7\linewidth]{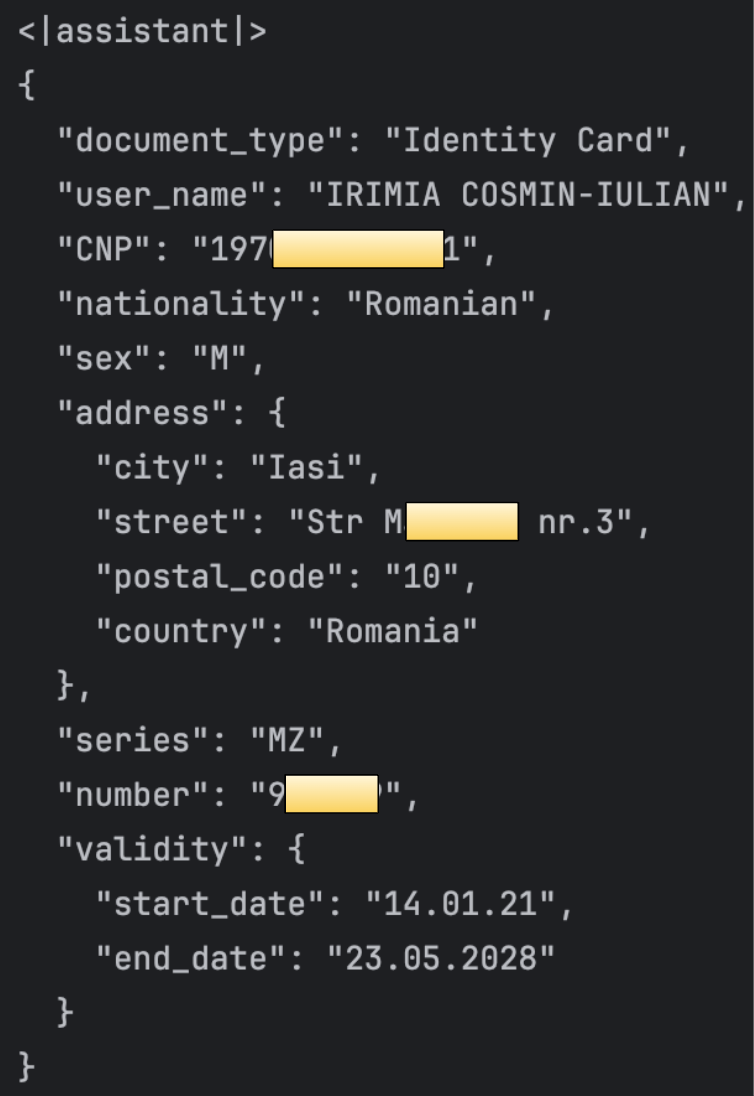}
  \caption{LLM Output for an ID Card}
  \label{fig:no17}
\end{figure}

By providing actual data about how the fields look like and how it should serve me, the LLM it’s very precise on how to identify certain details from a raw piece of text. By running this example on a ID Card, the LLM returned the output in the picture above.

As we can see, the data it’s accurate with the exception of the $address.postal\_code$.

In order to better visualize this data and how this is affecting the way the extraction work and how fast is it, I conducted a series of tests by calling the template-based API and the three models that we already discussed about, five times in a row with 3 different document types and checked how fast the method is and how accurate are the results obtained. Accuracy is computed as the proportion of correctly predicted field values over the total number of predictions made.

Of course, the template-based version will always evolve in the way that adding more documents as templates in the system will make it perform better, but the results I will present are obtained by training it with seven different ID Cards, three driving licenses and two property contracts.

\begin{table}[H]
\caption{Data Extraction and Labeling Times, Accuracy between Different Extraction Types}
\label{table:llm-times}
\begin{tabular}{|cl|cccccc|}
\hline
\rowcolor[HTML]{EFEFEF} 
\multicolumn{2}{|c|}{\cellcolor[HTML]{EFEFEF}}                                                                                & \multicolumn{6}{c|}{\cellcolor[HTML]{EFEFEF}Time \& Accuracy per document type}                                                                                                                                                                                                                                                                    \\ \cline{3-8} 
\multicolumn{2}{|c|}{\cellcolor[HTML]{EFEFEF}}                                                                                & \multicolumn{2}{c|}{ID Card}                                                                                           & \multicolumn{2}{c|}{Driving License}                                                                                  & \multicolumn{2}{c|}{Property contract}                                                            \\ \cline{3-8} 
\rowcolor[HTML]{F3F3F3} 
\multicolumn{2}{|c|}{\multirow{-3}{*}{\cellcolor[HTML]{EFEFEF}Extraction type}}                                               & \multicolumn{1}{c|}{\cellcolor[HTML]{EFEFEF}Time}         & \multicolumn{1}{c|}{\cellcolor[HTML]{EFEFEF}Acc.}          & \multicolumn{1}{c|}{\cellcolor[HTML]{EFEFEF}Time}        & \multicolumn{1}{c|}{\cellcolor[HTML]{EFEFEF}Acc.}          & \multicolumn{1}{c|}{\cellcolor[HTML]{EFEFEF}Time}         & Acc.                                  \\ \hline
\multicolumn{1}{|c|}{}                                                                    & Template-based                    & \multicolumn{1}{c|}{\cellcolor[HTML]{FFFFFF}\textbf{11s}} & \multicolumn{1}{c|}{\cellcolor[HTML]{FFFFFF}87\%}          & \multicolumn{1}{c|}{\cellcolor[HTML]{FFFFFF}\textbf{9s}} & \multicolumn{1}{c|}{\cellcolor[HTML]{FFFFFF}91\%}          & \multicolumn{1}{c|}{\cellcolor[HTML]{FFFFFF}\textbf{28s}} & \cellcolor[HTML]{FFFFFF}85\%          \\ \hline
\multicolumn{1}{|c|}{}                                                                    & \textit{zephyr-7b-beta}           & \multicolumn{1}{c|}{\cellcolor[HTML]{FFFFFF}255s}         & \multicolumn{1}{c|}{\cellcolor[HTML]{FFFFFF}93\%}          & \multicolumn{1}{c|}{\cellcolor[HTML]{FFFFFF}208s}        & \multicolumn{1}{c|}{\cellcolor[HTML]{FFFFFF}\textbf{95\%}} & \multicolumn{1}{c|}{\cellcolor[HTML]{FFFFFF}347s}         & \cellcolor[HTML]{FFFFFF}88\%          \\ \cline{2-8} 
\multicolumn{1}{|c|}{}                                                                    & \textit{Mistral-7B-Instruct-v0.2} & \multicolumn{1}{c|}{\cellcolor[HTML]{FFFFFF}104s}         & \multicolumn{1}{c|}{\cellcolor[HTML]{FFFFFF}90\%}          & \multicolumn{1}{c|}{\cellcolor[HTML]{FFFFFF}70s}         & \multicolumn{1}{c|}{\cellcolor[HTML]{FFFFFF}87\%}          & \multicolumn{1}{c|}{\cellcolor[HTML]{FFFFFF}151s}         & \cellcolor[HTML]{FFFFFF}85\%          \\ \cline{2-8} 
\multicolumn{1}{|c|}{\multirow{-3}{*}{\begin{tabular}[c]{@{}c@{}}L\\ L\\ M\end{tabular}}} & \textit{Meta-Llama-3.1-7B}       & \multicolumn{1}{c|}{\cellcolor[HTML]{FFFFFF}131s}         & \multicolumn{1}{c|}{\cellcolor[HTML]{FFFFFF}\textbf{96\%}} & \multicolumn{1}{c|}{\cellcolor[HTML]{FFFFFF}98s}         & \multicolumn{1}{c|}{\cellcolor[HTML]{FFFFFF}\textbf{95\%}} & \multicolumn{1}{c|}{\cellcolor[HTML]{FFFFFF}143s}         & \cellcolor[HTML]{FFFFFF}\textbf{91\%} \\ \hline
\end{tabular}
\end{table}

The accuracy values reported in Table \ref{table:llm-times} refer to field-level accuracy, computed as the ratio of correctly extracted fields to the total number of expected fields across a representative sample of test documents. A field was considered correct if its extracted value exactly matched the manually annotated ground truth.

Accuracy = (Correct Fields) / (Total Fields)

Each document type was tested using a set of 20 real-world samples, manually verified for validation.

The experiments were conducted on a high-performance ARM-based system (Apple M1 Max, 64 GB RAM), chosen to reflect realistic local deployment constraints. The selected models—LLaMA 2 7B, Mistral, and Phi-2—represent state-of-the-art in terms of performance-to-size ratio and open-source availability. It is expected that the performance and accessibility of locally deployable LLMs will continue to improve, as shown by recent benchmark trends.

As we can see in the table above, LLMs are pretty promising regarding how good the processing is: the accuracy it’s very good and there are still a lot of optimizations that can be done in order to obtain better results. Unfortunately, the problem with them for the moment is that they are very slow and they can’t identify or point coordinates yet, so they are great for extraction but we still need the templated-based mechanism in order to obtain the coordinates for each field.

\section{Notarization}

The \textbf{Notarization Service} is responsible for validating and attesting the authenticity of facts extracted from uploaded documents. It plays a central role in establishing trust by approving data for anchoring and coordinating the generation of verifiable hashes.

In the current prototype, the notarization process includes modules for:
\begin{itemize}
  \item Receiving structured facts from the Data Extractor
  \item Allowing a validator to manually approve or edit fields
  \item Hashing the finalized document 
\end{itemize}



\section{Document Obfuscator}

The Document Obfuscator module provides fine-grained privacy control by allowing users to redact sensitive regions of a document prior to sharing. The module supports two key operations: obfuscation (applying masking) and controlled deobfuscation (reconstructing authorized regions).

Obfuscation may be performed either locally on the user’s device or on the server-side, depending on the trust assumptions and processing constraints of the deployment. In trusted environments (e.g., institutional platforms), server-side obfuscation ensures consistency. In privacy-sensitive scenarios, local obfuscation allows users to retain full control over the masking process and decryption keys. 

Each obfuscated zone is tied to a unique decryption key, intended to be shared only with explicitly authorized recipients. However, once shared, the system has no technical guarantee that the recipient will not redistribute either the document or the key, especially if that recipient becomes untrusted after access. To mitigate this, a trust model and revocation mechanism must be integrated, or obfuscation must be designed to be single-use or time-constrained.

\subsection{Obfuscation Algorithm}

The obfuscation pipeline is designed to apply one or more transformation layers to user-specified rectangular regions of a document. These layers are executed in sequence and stored as a logical grouping known as a Zone. Each zone contains metadata about its position and the ordered list of applied transformations.

A Master Key is generated per document and acts as a secure container for one or more obfuscation zones. It is not persisted server-side; only the client retains access to it. This design guarantees that only the document owner can perform partial or full deobfuscation. Zones may be pruned from the master key to maintain permanent obfuscation of certain fields.


Encryption was one of the basic ways of obfuscating. We chose to go with AES (Advanced Encryption Standard) using the CBC (Cipher block chaining \cite{Ehrsam1976}) mode with a key of 128-bit in order to hide confidential data.

\begin{figure}[H]
\centering
\includegraphics[width=1\linewidth]{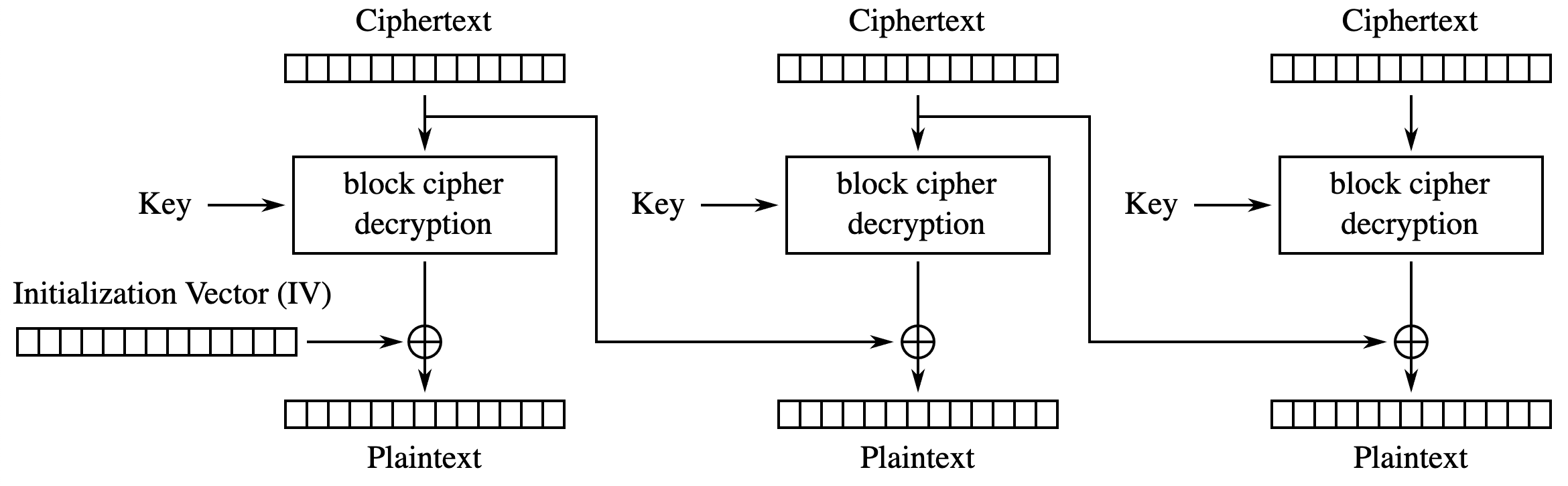}
\caption{Cipher Block Chaining (CBC) Mode Encryption}
\end{figure}

\subsection{Implementation}

The first service is a simple face detection API that uses machine learning to detect the faces in an image in order to resolve the most common use case, hiding the faces of other people in a photo. The second service is the core of our application and handles the obfuscation and deobfuscation of a document. This will later be split into separate microservices in order to better handle the increased traffic. This design is easy to achieve in code using the ”factory” and ”chain of responsibility” design patterns. Each algorithm has an ID that is used to create obfuscation or deobfuscation chains. In order to create those chains, we implemented two specialized factories.

\begin{figure}[H] 
  \centering
  \includegraphics[width=1\linewidth]{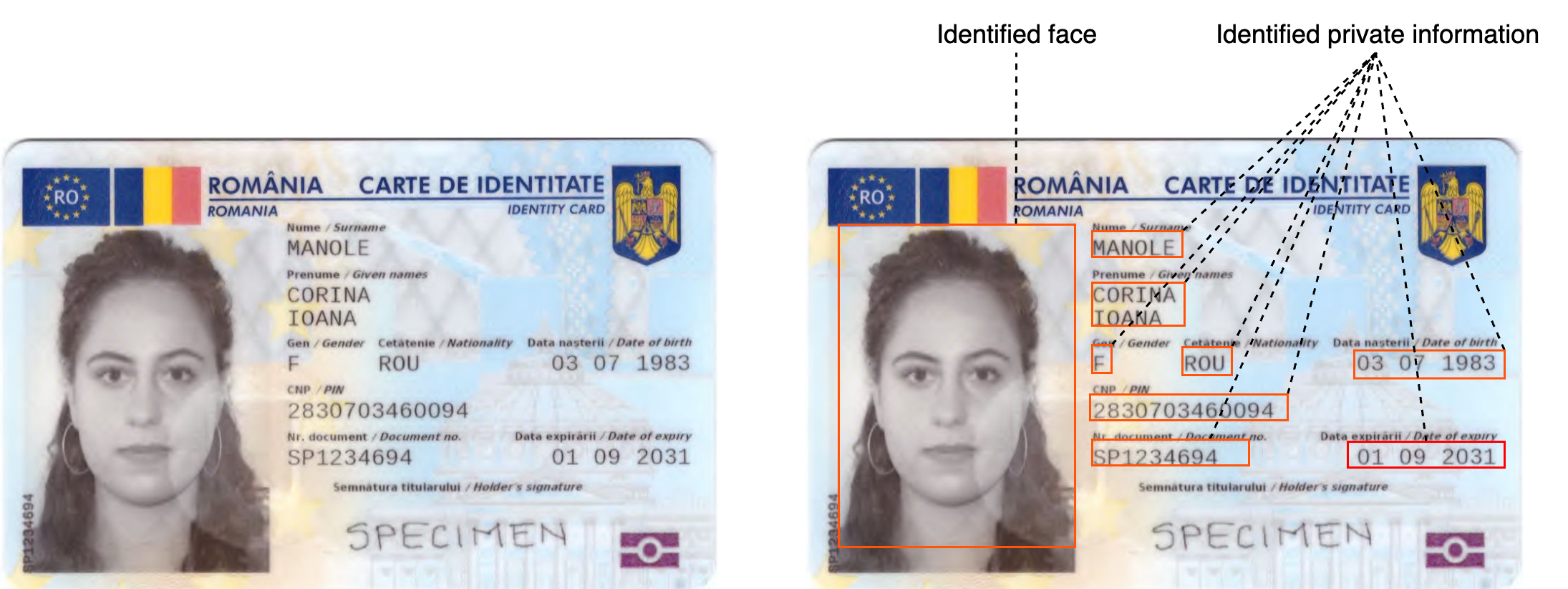}
  \caption{Identification of Zones}
  \label{fig:no21}
\end{figure}

After parsing the input, a user requests to obfuscate any number of zones with any combination of layers or a Master Key object, we will apply those changes to the document and return it. In order to better explain the process, we will demonstrate it with a visual example. The first step is for the system to automatically identify the faces from the uploaded photo, then the user will have the possibility to select his own interest areas, these two steps being referred to in technical terms as ”marking the zones”.

\begin{figure}[H] 
  \centering
  \includegraphics[width=0.7\linewidth]{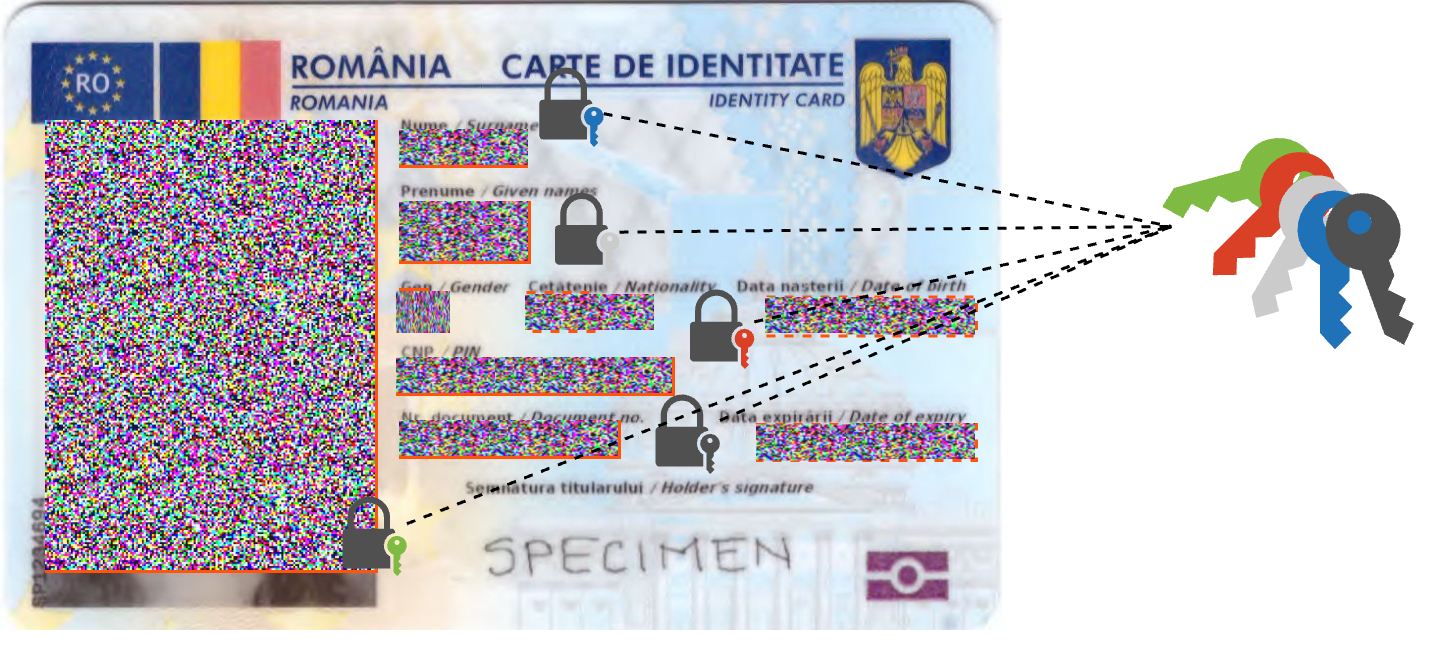}
  \caption{Obfuscated Photo with Multiple Keys}
  \label{fig:no22}
\end{figure}

After the zones have been selected, the system will allow the user to select the algorithms with which to obfuscate each of them (as can be seen in \ref{fig:no22}), then generate a unique key for each of these zones.

\begin{figure}[H] 
  \centering
  \includegraphics[width=1\linewidth]{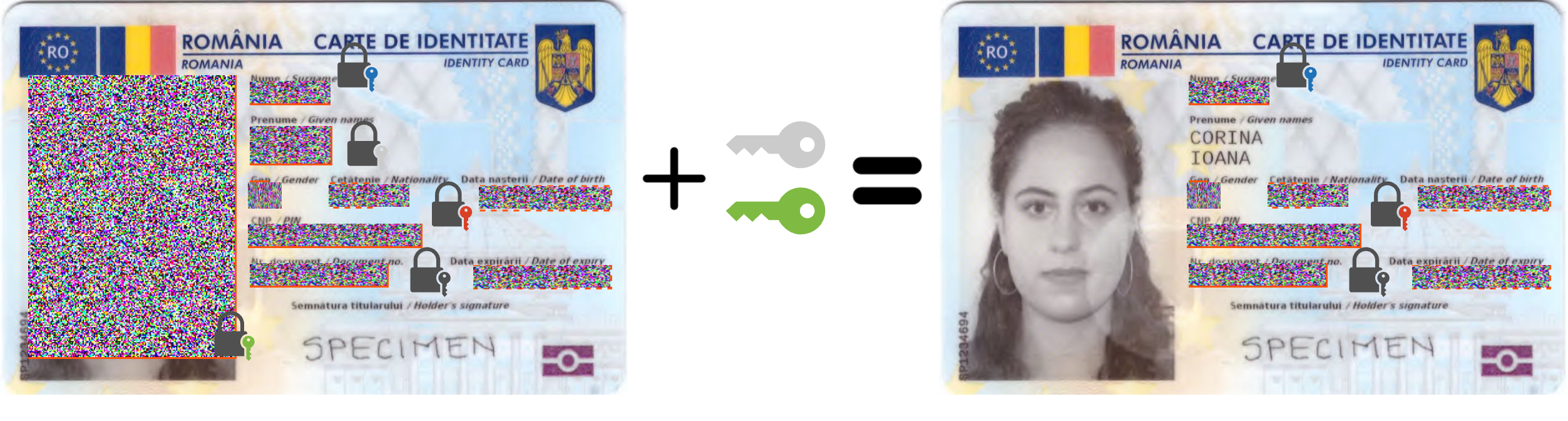}
  \caption{Partial Deobfuscation}
  \label{fig:no23}
\end{figure}

Once the obfuscation process is complete, the user can now share the photo in its new form and also one or more keys. In our case, if the user would like to provide access to the photo and the address from his ID Card, he should use the two relevant keys in this context: the gray key and the green key, the result of the deobfuscation (as can be seen in \ref{fig:no23}) being similar to the one below.  

\newpage

\section{Data Persistence}

The system implements a three-tier data persistence model to ensure document durability, integrity, and auditability. These layers are designed to balance performance, decentralization, and verification requirements.

The three storage tiers include:

\begin{itemize}
  \item 
Relational Database: Stores user profiles, metadata, and service-level state.
  \item 
IPFS: Serves as a decentralized storage layer for complete documents.
  \item 
Blockchain: Anchors document hashes for integrity verification and public auditability.
  \item 
KeyStore: Secure storage for the obfuscation keys.
\end{itemize}

\subsection{Centralized Database}

The relational database in DocChain stores system metadata, user account records, shared document indexes, and template configurations. Sensitive content and cryptographic material are excluded from this layer.

PostgreSQL was selected for its open-source nature, strong community support, and ACID-compliant transaction engine. It provides schema validation, indexing, and foreign key constraints to maintain data consistency.

The proposed schema includes tables for:
\begin{itemize}
  \item Users and login credentials (hashed/salted)
  \item Document metadata and template mappings
  \item Notary account registration and associations
  \item Audit trails for document access events
\end{itemize}

\begin{figure}[H] 
  \centering
  \includegraphics[width=0.82\linewidth]{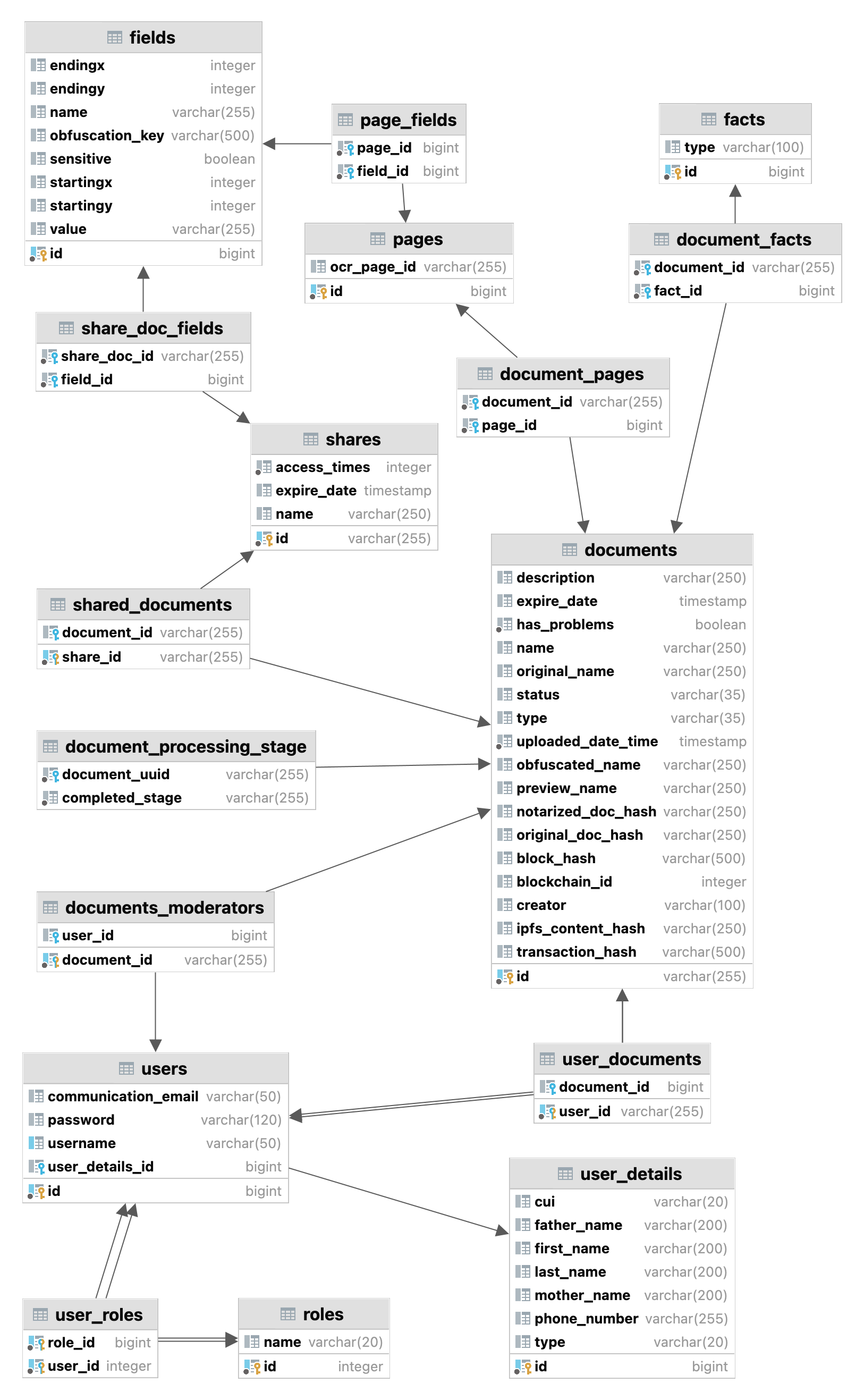}
  \caption{Proposed DocChain Database Schema}
  \label{fig:no24}
\end{figure}

\subsection{IPFS \& Pinata}

A review of recent publications, articles, and technology reports reveals a notable surge of interest in this domain over the past year. The challenges associated with storing documents on centralized services have become increasingly prominent, driving significant research and development efforts toward decentralized solutions aimed at addressing these concerns.

We should start by answering the following question: why is it so bad to have your documents on a centralized storage solution? \cite{Ref007} While centralized storage solutions can pose security risks if improperly configured, they can also offer robust access control and privacy through centralized policy enforcement. In contrast, decentralized storage systems such as IPFS provide advantages in terms of integrity, redundancy, and resilience to data loss through replication. However, they do not inherently guarantee privacy — this must be achieved separately, typically via client-side encryption and selective access controls. Another thing is the \textbf{replication} \cite{Ref008}: even if the storage is a secure one, you could still lose the hardware so replication it’s a very good mechanism to keep our data safe. Another thing about documents will be \textbf{sharing}: the mechanism of sharing it’s very important, and sharing them safely over the Internet has been a great challenge over the years. In our literature reading we could find a great number of different promising solutions for document storage but none of them was really focused on official documents in some ways: lacking privacy, lacking control and granularity of the data or not having the possibility to share or delete. We’ve tried some solutions and we have read about others and the conclusion at a first glance was that we needed to create our own blockchain or to say it in the right way: to create a new type of blockchain: \textbf{a content-driven, document-driven one}. Fortunately, during the research we stumbled across on IPFS and Pinata that made our work easier.

The \textbf{InterPlanetary File System} (IPFS) \cite{Ref009} is a protocol that has the scope of rethinking the way data is shared and stored on the Internet. Developed by Juan Benet, IPFS offers a decentralized and content-addressable approach to data distribution, promising enhanced data availability, security, and censorship resistance. IPFS is an open-source, \ac{P2P} file system \cite{Ref010} designed to overcome the limitations of the traditional web. It replaces traditional URLs with content-based addressing, ensuring that data is identified by its unique cryptographic hash. This fundamental shift eliminates the need for centralized servers and enhances data integrity.

Given its advertised features, IPFS appears to be a promising fit for our system; however, it is important to first present its underlying mechanisms to thoroughly assess its viability as a solution:

\begin{itemize}
 \item \textbf{Step 1 - Content Addressing}: IPFS assigns a unique cryptographic hash to each piece of content, making it immutable and resistant to tampering. This hash serves as the content's address on the network.
 
 \item \textbf{Step 2 - Decentralized Storage}: Data is divided into smaller chunks and distributed across a global network of interconnected nodes. When you request a file, your node finds the nearest nodes hosting the content, ensuring rapid retrieval.

 \item \textbf{Step 3 - \ac{P2P} Connectivity}: IPFS operates as a \ac{P2P} network, with nodes collaborating to share and retrieve data. This architecture reduces reliance on centralized infrastructure and enhances scalability.

 \item \textbf{Step 4 - Versioning and Deduplication}: IPFS includes built-in version control and deduplication mechanisms, optimizing data storage and allowing users to track changes to files over time.

 \item \textbf{Step 5 - Caching and Offline Access}: IPFS caches content locally, enabling access even in offline or low-connectivity environments.
\end{itemize}

IPFS represents a paradigm shift in internet technology, offering a decentralized, secure, and efficient means of sharing and preserving data. As it continues to evolve, IPFS is poised to play a pivotal role in shaping the future of decentralized technologies and fostering a more robust and resilient digital ecosystem. For sure, IPFS has its limitations and it’s not really focused on official documents but it fits so well in our context and considering that we have the extra-step of obfuscation it can really make a difference to not consider regular, centralized document storage options like AWS S3, Azure Blob Storage and so on. From all the solutions that I’ve read and from the ones that I covered here, it seems like the InterPlanetary File System (IPFS) is one of the best and seems like a viable solution for our distributed storage needs. 

\begin{table}[H]
\centering
\caption{Services that Offer IPFS Implementation}
\begin{tabular}{|l|l|p{5cm}|}
\hline
\rowcolor[HTML]{F3F3F3} 
\multicolumn{1}{|c|}{\cellcolor[HTML]{F3F3F3}Service} & \multicolumn{1}{c|}{\cellcolor[HTML]{F3F3F3}Strengths}                                                                                                            & \multicolumn{1}{c|}{\cellcolor[HTML]{F3F3F3}Weaknesses}                                                                                                                   \\ \hline
\textbf{Pinata {[}17{]}}                                       & \begin{tabular}[c]{p{7cm}}- Reliability: Consistent pinning and availability. \\ - Cost-Effective: Flexible pricing plans. \\ - Security: API keys and tokens for secure access.\end{tabular}                   & - Limited storage size on lower-tier plans.                                                                                                                                        \\ \hline
Infura {[}18{]}                                                & \begin{tabular}[c]{p{7cm}}- Scalability: Handles high traffic and demand. \\ - Comprehensive API: Supports Ethereum and IPFS. \\ - Global Infrastructure: Fast and reliable access.\end{tabular}                & - Can be expensive for high usage.                                                                                                                                                 \\ \hline
Fleek {[}19{]}                                                 & \begin{tabular}[c]{p{7cm}}- Web Hosting: Simplifies deployment on IPFS. \\ - Integration: Supports Ethereum and ENS. \\ - Continuous Deployment: Works with Git and other version control systems.\end{tabular} & \begin{tabular}[c]{p{5cm}}- May require more setup for non-web hosting use cases. \\ - Limited to static sites and applications.\end{tabular}                                     \\ \hline
Textile {[}20{]}                                               & \begin{tabular}[c]{p{7cm}}- Privacy: Secure data sharing and management. \\ - Integration: Works with Filecoin for storage. \\ - Developer Tools: Offers ThreadsDB and Powergate.\end{tabular}                  & \begin{tabular}[c]{p{5cm}}- More complex setup for developers new to decentralized databases. \\ - Focused on data privacy and security rather than general pinning.\end{tabular} \\ \hline
\end{tabular}
\end{table}

The IPFS represents a revolutionary shift in distributed file \cite{Kumar2019} sharing, offering distinct advantages for various applications. With its decentralized architecture, content addressing, and data redundancy, IPFS ensures robustness and data integrity, making files resistant to tampering and enhancing overall system reliability. Its censorship-resistant design enables unfettered access to information, especially in restrictive environments, while built-in version control allows for transparent file management. IPFS's offline accessibility and scalability further cement its position as a leading solution for efficient and secure distributed file sharing, promising a resilient and accessible digital ecosystem.

It is pretty clear that for the moment we are just looking at potential distributed-storage solutions, systems like the one we are proposing are inexistent and the kind of feature granularity, privacy but openness at the same time.

The thing is that InterPlanetary File System (IPFS) is a distributed file storage \cite{8985677, 10141817} protocol and in order to use it we need an implementation for it. Having the experience from the second report where we started implementing our own blockchain, this time we learned our lesson and we tried to do our research about the best possible option as the implementation we want to use. 

\subsection{Blockchain \& GoQuorum}

This section focuses on the specific Ethereum client chosen for the implementation of DocChain, namely \textit{GoQuorum}~\cite{Wells2023}. GoQuorum is an enterprise-focused Ethereum variant that supports permissioned networks, private transactions, and fine-grained access control—features that align well with the system's privacy and security requirements.

The details about why we have chosen Ethereum and Blockchain as the main technologies for the thesis and why it helps us was already explained in the first chapters of the thesis so I won’t enter again in the details but I will talk about the Ethereum client that we chose for DocChain, namely GoQuorum \cite{Wells2023}. 

The main concern about what client to choose was the ability to have private \& public transactions, the main focus being that the obfuscation keys should be staying in the blockchain’s smart contract but without being public. 

GoQuorum supports private transactions through the Tessera privacy manager, which ensures that only authorized participants have access to the encrypted payload. However, full protection against external access — including from compromised nodes or malicious infrastructure — relies on secure key management. This typically requires hardware-based solutions such as HSMs or secure enclaves (e.g., Intel SGX), which are not enabled by default. In the current prototype, Tessera operates in default mode with software-based key storage, meaning that while data is hidden from uninvolved nodes in the consortium, it may still be exposed to privileged attackers without hardware isolation.

We did a quick comparison between all the Ethereum that provide this kind of functionality:

\begin{table}[H]
\centering
\caption{The Main Ethereum Clients with Support for Private Transactions Comparison}
\begin{tabular}{|p{3cm}|l|l|}
\hline
\rowcolor[HTML]{F3F3F3} 

\multicolumn{1}{|c|}{\cellcolor[HTML]{F3F3F3}Client} & \multicolumn{1}{c|}{\cellcolor[HTML]{F3F3F3}Pros}                    & \multicolumn{1}{c|}{\cellcolor[HTML]{F3F3F3}Cons}                                                                                                                                    \\ \hline
\textbf{GoQuorum {[}22{]}}                                    & \begin{tabular}[c]{p{5.0cm}}- Enterprise-focused with strong privacy features \\ - Wide support for private transactions \\ - Compatible with existing Ethereum tooling - Strong community and support\end{tabular}   & \begin{tabular}[c]{p{5.0cm}}- Less decentralized compared to public Ethereum \\ - Requires permissioned \cite{Ref022, 8946188} network setup\end{tabular}                                                    \\ \hline
Hyperledger \cite{8946222} Besu {[}23{]}                                     & \begin{tabular}[c]{p{5.0cm}}- Supports private transactions using the Orion privacy manager Interoperable with Ethereum mainnet and private networks \\ - Enterprise-friendly features and permissioning\end{tabular} & \begin{tabular}[c]{p{5.0cm}}- May have performance overhead due to privacy features \\ - Complex setup for privacy configurations \\ - Limited adoption compared to Geth\end{tabular} \\ \hline
Nethermind {[}24{]}                                           & \begin{tabular}[c]{p{5.0cm}}- Lightweight client with support for private networks \\ - Compatible with Ethereum 2.0 and EVM features \\  - Active development and growing community\end{tabular}                     & \begin{tabular}[c]{p{5.0cm}}- Privacy features not as advanced as GoQuorum or Besu \\ - Less mature than some other Ethereum clients \\ - Limited enterprise adoption\end{tabular}    \\ \hline
\end{tabular}
\end{table}

From a technical perspective, GoQuorum it’s the best option in this space, it offers a great number of functionalities, has great documentation and it’s very versatile. With just a few minutes of configuration we can run our own blockchain network using docker, having access to UI tools that help us understand the network better, debug over transactions and contracts and also it has a great number of tools and examples to help you write your own smart contract the way you need to.

\subsection{Service Orchestrator}

The \textbf{orchestrator} in DocChain had to be very reliable and his main concern it’s to handle UI’s requests and to act like a bridge between all of our systems so we have chosen to write it in Java with Spring Boot \cite{Huaylupo2021} \& Hibernate. The user authorization it’s done using JSON Web Tokens and from there on, since all the other microservices are private, there is no need to exchange any tokens between them. The authorization on the orchestrator it’s done through MetaMask in order to have access to user’s keys for interacting with GoQuorum. The user also has a backup mechanism using an email and password, flow where he will receive a JWT Token for interaction with the document endpoints.

\begin{figure}[H] 
  \centering
  \includegraphics[width=1\linewidth]{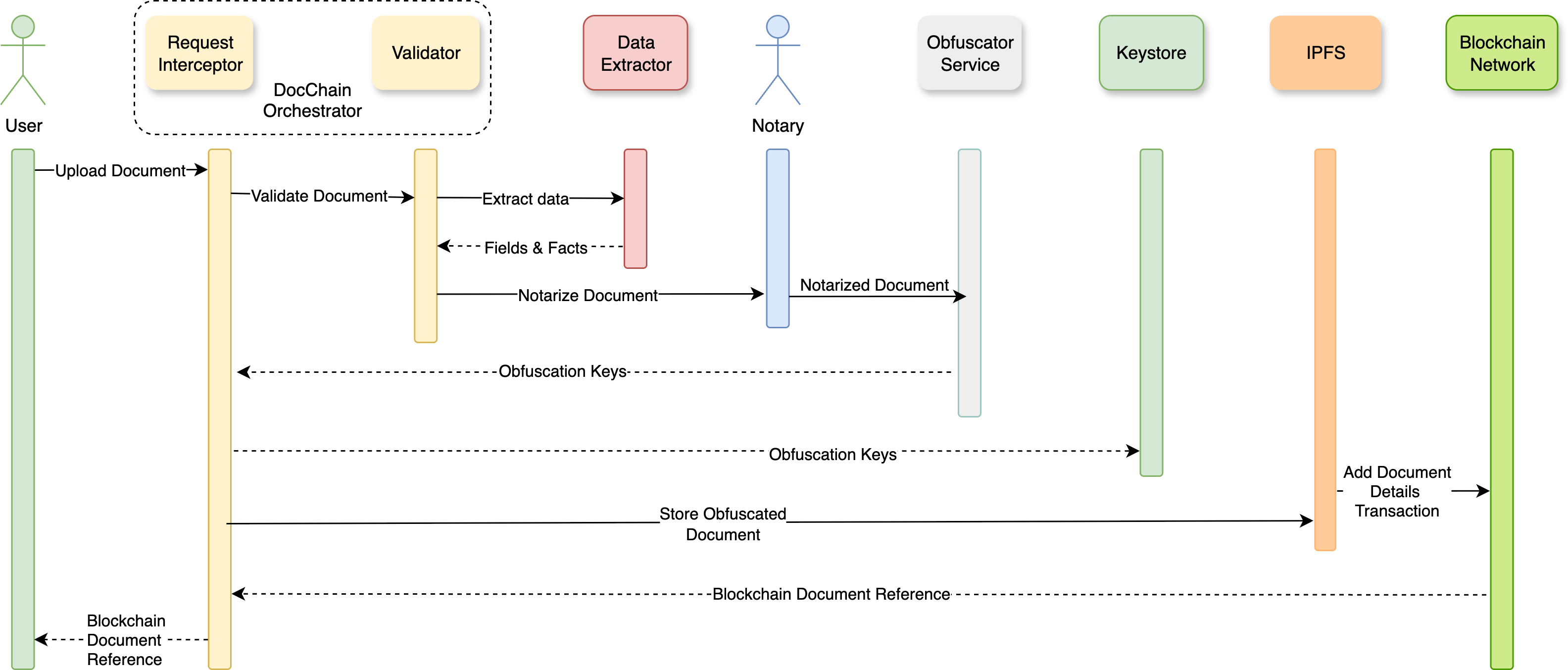}
  \caption{Sequence Diagram of the System}
  \label{fig:no25}
\end{figure}
 
 When a document creation request comes in, the user basically sends only an image/document and a short description of the document (for UI purposes only). The orchestrator will intercept the request, will validate the data and will send the payload to the layered calling mechanism, a tool created by us that ensures that a document never gets stuck in one of the processes and a retry mechanism will resend it where the last step failed. The document will be inserted into the centralized database with all his details and the first system in the chain will be called, namely the extractor. Meanwhile we update the state of the document to RECOGNITION\_STARTED and after the extractor responds with the extracted data we are saving the pages, fields and coordinates in the centralized database and move to the next step, Fact collecting.. After processing the extracted text, we are collecting a list of facts that we store in the database for now and proceed to move the state into FACTS\_COLLECTED. The process continues by having the notary confirm that the document it’s valid and unaltered, a process that can take up to a few hours and in the meantime we have two states for the document: NOTARIZATION\_AWAITING and NOTARIZATION\_STARTED. After the notarization it’s done, we save the signature hash in our centralized database and proceed to go to the obfuscator by sending the coordinates already existing in the database, freshly inserted from the extraction process and corrected by the notary if necessary using the OBFUSCATION\_STARTED state.

 When the obfuscator it’s done, we store the obfuscation keys in the keystore, store the new obfuscated image locally and remove the original image permanently and move the state to TO\_BE\_UPLOADED. The next step is calling the data persister service where all processing is completed: the obfuscated photo it’s uploaded to IPFS using Pinata API, the hash it’s added into the document body and all the information are stored in the blockchain network itself from where an internal blockchain ID and the IPFS Hash are returned to the orchestrator. When the response it’s received, the orchestrator deletes all the obfuscation keys from the centralized database, persists the IPFS hash and moves the document state to COMPLETED. Figure \ref{fig:no25} presents a sequence diagram that highlights this process visually.

\subsection{User Interaction}

For DocChain, we wrote the app in React because of its enormous advantages such as flexibility, speed and great documentation. We want to satisfy the user flows from the initial proposal and the second report so the interface demonstrates all the main functionalities integrated end-to-end with a an intuitive and user-friendly interface for the user. There are five main user screens where he can interact with the application:

\subsubsection{Document List}

The \textbf{document list} it’s the main screen and the most important one.

\begin{figure}[H] 
  \centering
  \includegraphics[width=1\linewidth]{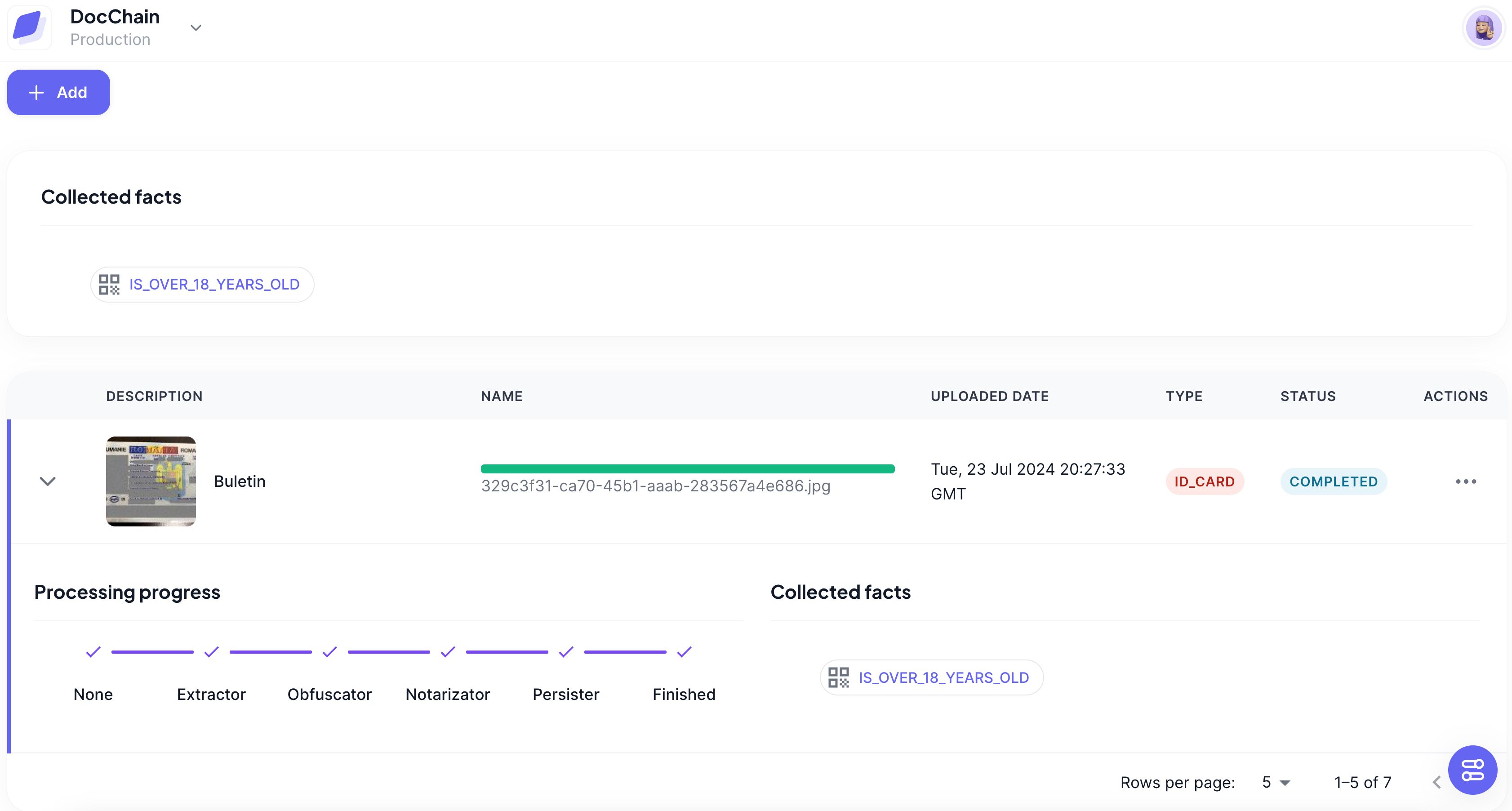}
  \caption{UI - Document Collection List}
  \label{fig:no26}
\end{figure}

User has the ability to create a new document from here but also to view all the documents, to delete or share them and also to view more details about any of them including the original IPFS file, the transaction hash, the collected facts per image and so on.

\subsubsection{Adding a Document}

The first user flow step obviously is document creation itself. The user would not be asked any details about the document itself but only a bare minimum description in order to have a better way to find the document in the document collection list later on.\\

\begin{figure}[H] 
  \centering
  \includegraphics[width=0.6\linewidth]{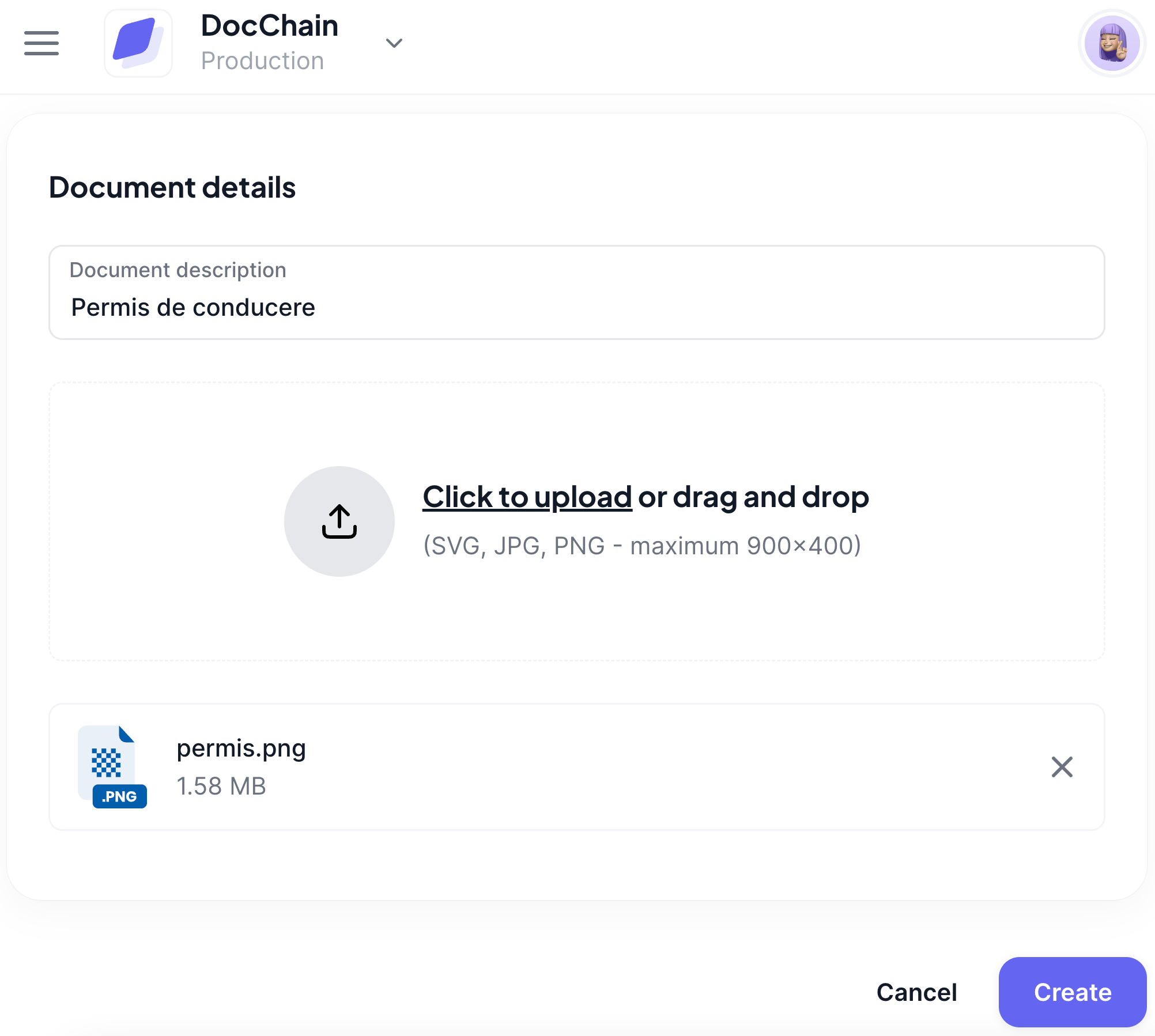}
  \caption{UI - Creating a Document}
  \label{fig:no27}
\end{figure}

Once a document gets into the system, it goes through successive processing stages for correct extraction, notarization and secure storage of its contents. Since the process is asynchronous, it will take some time depending on document size, type, and system workload. A real-time progress bar for every uploaded document adds a visual cue to what's happening.

\begin{figure}[H] 
  \centering
  \includegraphics[width=1\linewidth]{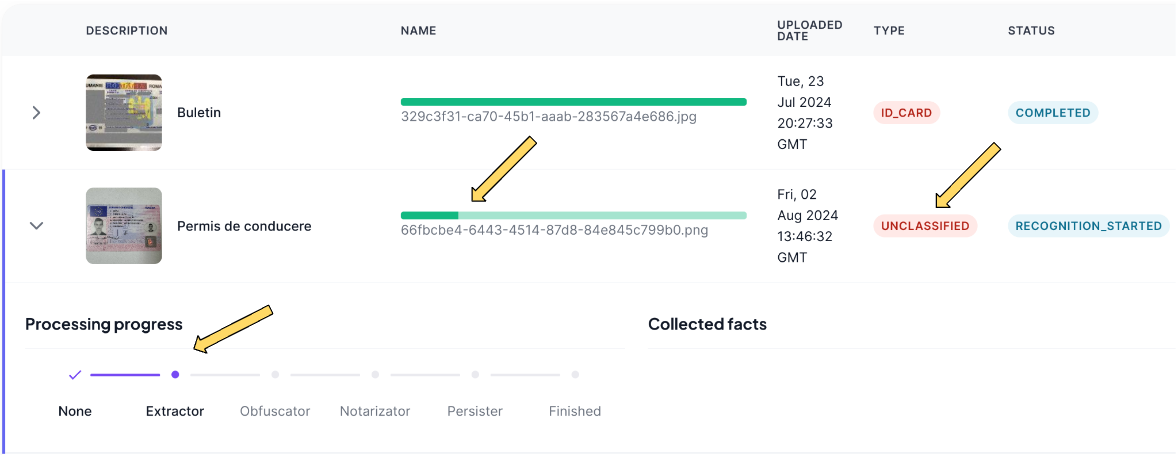}
  \caption{UI - Creating a Document - Phase One}
  \label{fig:no28}
\end{figure}

The document is initially shown with its original photo as a thumbnail so users can identify their files in the system. The type of the document is "Unclassified" at this point in time, as it has not yet been decided what is the type of the document yet. After the pre-processing, the status indicator is set to "Extracted", which means that the system had already performed initial identification and labeling of information in the document. This dynamic status-tracking mechanism keeps users informed at each step of the workflow.

\begin{figure}[H] 
  \centering
  \includegraphics[width=1\linewidth]{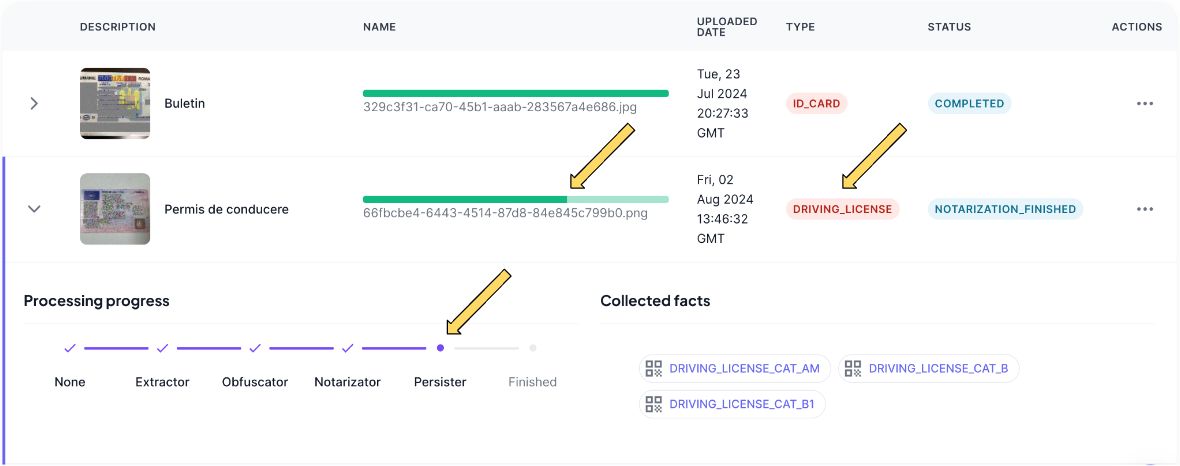}
  \caption{UI - Creating a Document - Phase Two}
  \label{fig:no29}
\end{figure}

After a few minutes and a refresh, we can see that the document has the obfuscated photo as a thumbnail, the facts are collected, the document type it’s now “Driving license” but the persistence part it’s still not done yet.

\begin{figure}[H] 
  \centering
  \includegraphics[width=1\linewidth]{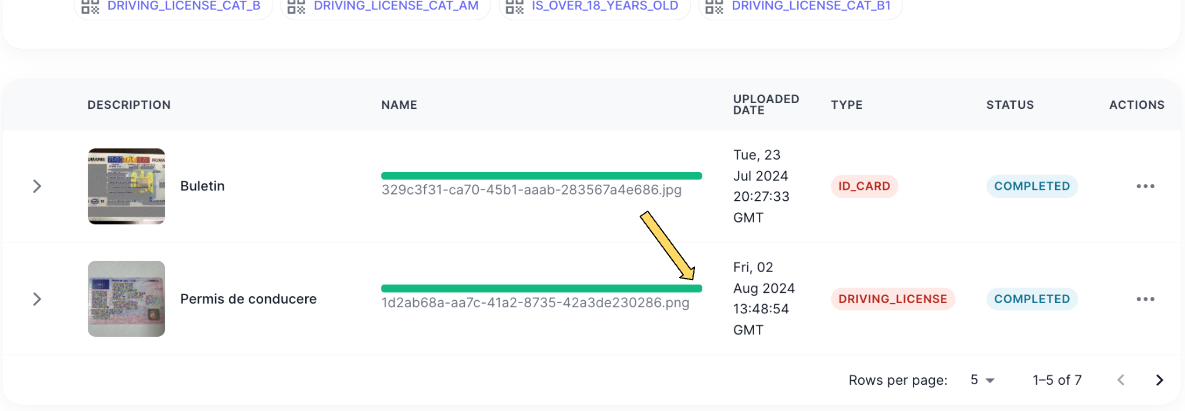}
  \caption{UI - Creating a Document - Phase Three}
  \label{fig:no30}
\end{figure}

\subsubsection{Document Details}

After creating a document, the user can interact with it, check the details, share it or delete it. In the document details page, the user can see the document id, name, original file name that was uploaded, the date of the upload, IPFS Hash so he can see the document on the distributed network by itself or share that information, the transaction and block hashes and all the completed stages of the document processing.

\begin{figure}[H] 
  \centering
  \includegraphics[width=0.75\linewidth]{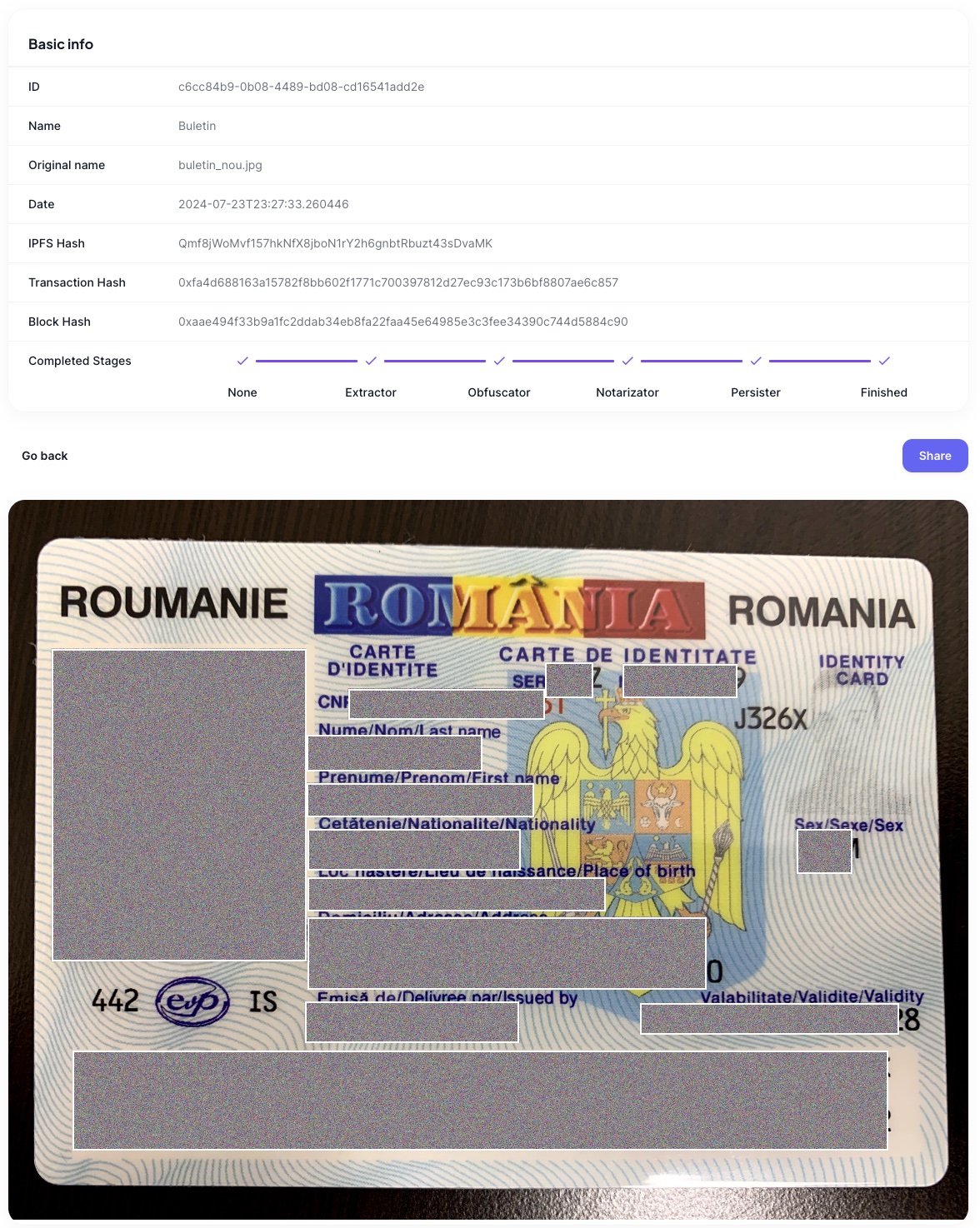}
  \caption{UI - Document Details Page}
  \label{fig:no31}
\end{figure}

\subsubsection{Document Share}

The promised and innovative feature of the system is the partial sharing of the obfuscated image, so that’s the most important screen of them all.  Here, you have the possibility to select every obfuscated area for which you would like to retrieve the keys for and deobfuscate them.

\begin{figure}[H] 
  \centering
  \includegraphics[width=0.7\linewidth]{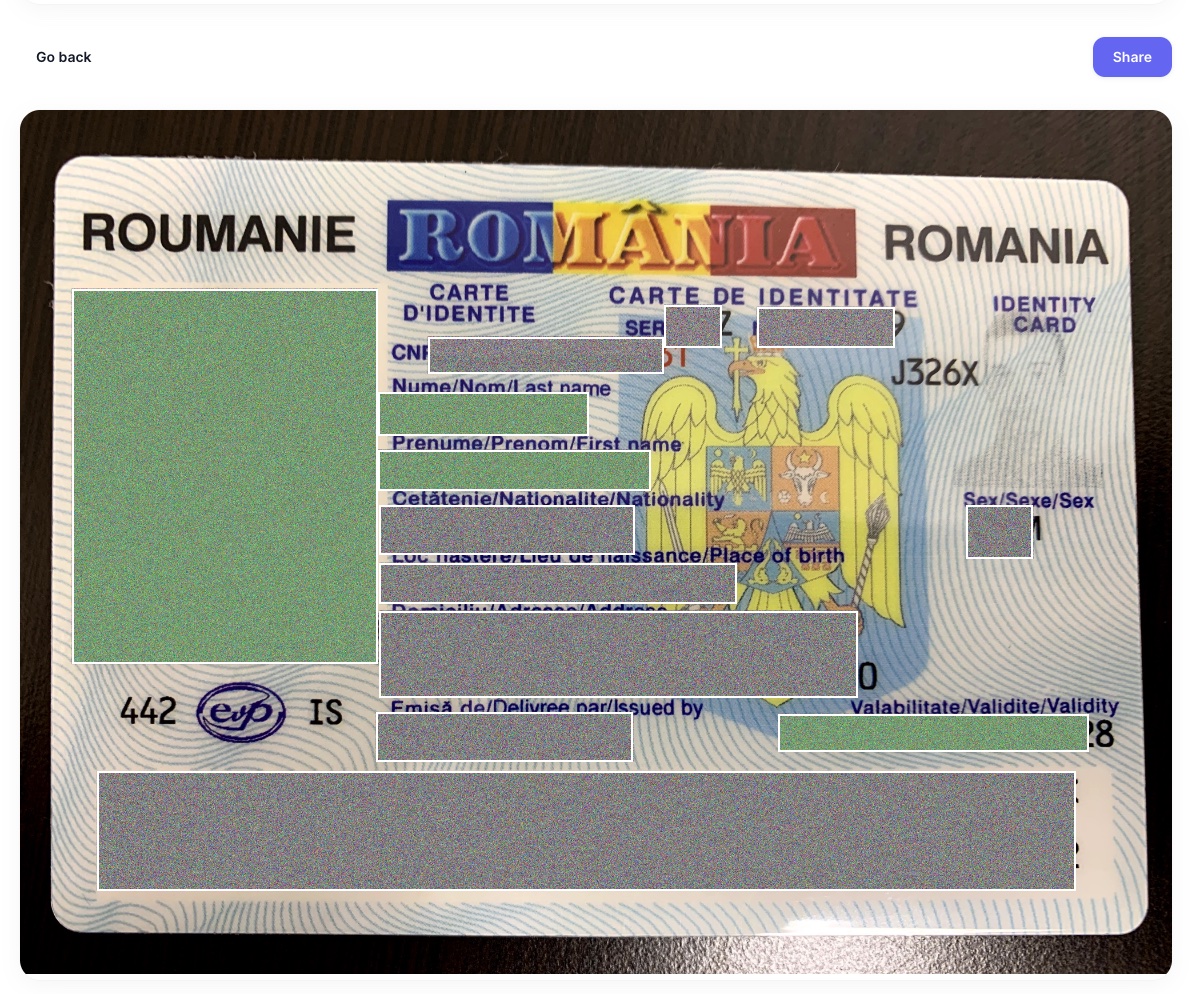}
  \caption{UI - Document Details Page}
  \label{fig:no32}
\end{figure}

After the selection it’s done, the user has three options of sharing:
\begin{itemize}
 \item \textit{Time based}: share the document until a certain point in the future, being as granular as you can share it for just a minute up to the document's expiration date;
 \item \textit{Access time based}: share the document for an undefined period of time but just a certain number of clicks can be made over the link, after that click count passes the agreed value, the link will auto-expire;
 \item \textit{Indefinitely}: share the document indefinitely, up to the document’s expiration date or until the user decides to delete the share link itself.
\end{itemize}

After the share button it’s pressed, the UI will send a request to the orchestrator with the selected zones and options and will trigger the retrieval of the keys from the blockchain, deobfuscation of the zones and unique, UUID-based URL creation for the document.

\begin{figure}[H] 
  \centering
  \includegraphics[width=0.9\linewidth]{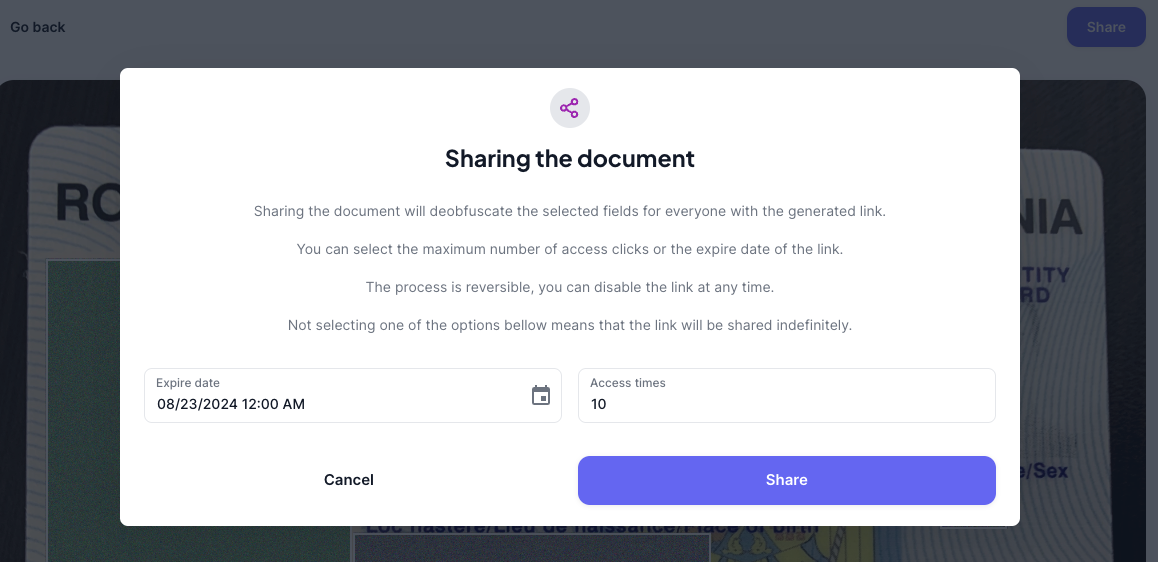}
  \caption{UI - Document Share Options Selection}
  \label{fig:no33}
\end{figure}

In the document details page, now the user will see his new created link that he can choose to view details about or share. 

\begin{figure}[H] 
  \centering
  \includegraphics[width=0.9\linewidth]{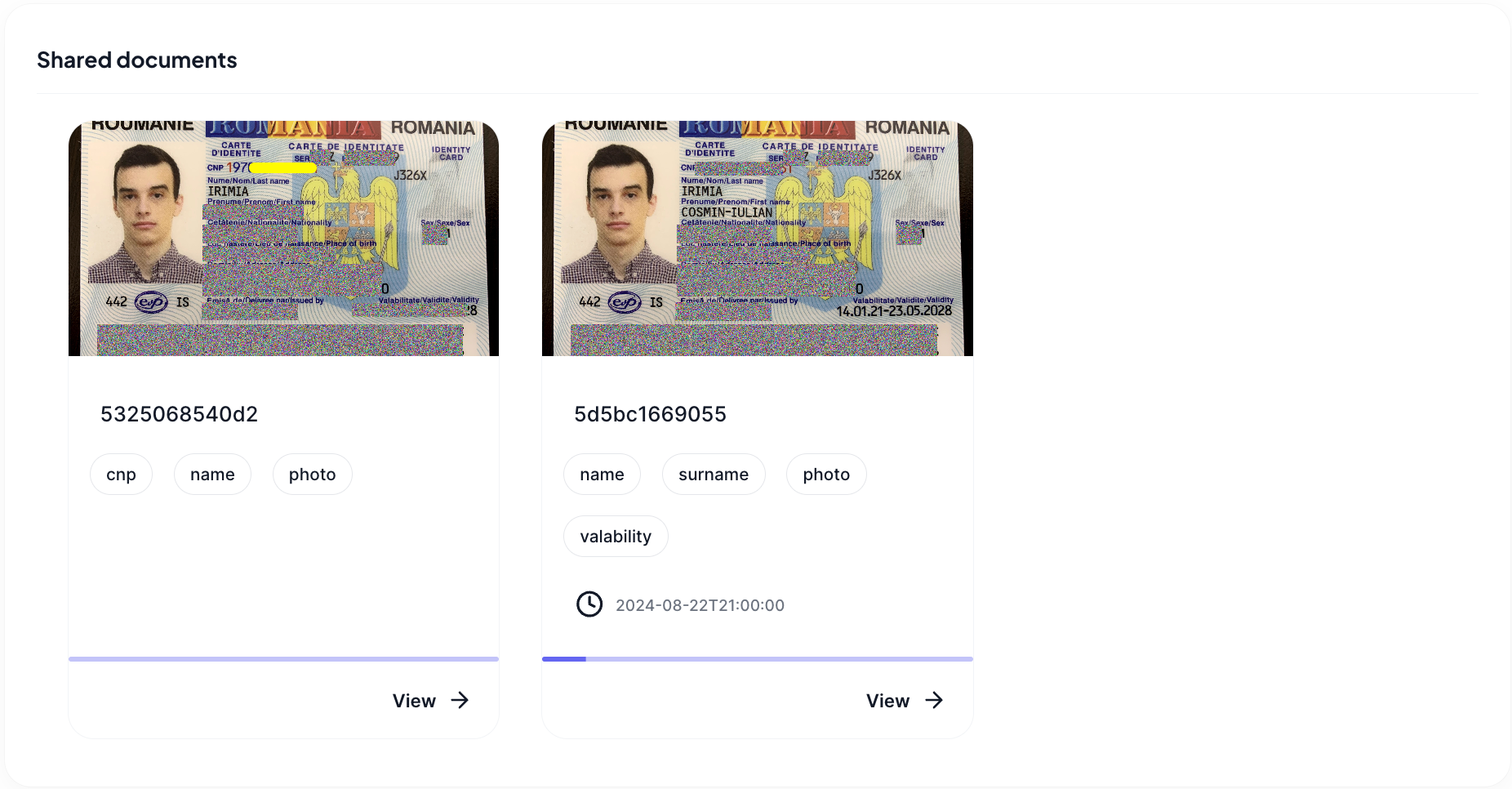}
  \caption{UI - Document Shares Links List}
  \label{fig:no34}
\end{figure}

You can see right under the image that a list of zone names appear in order to indicate what parts of the documents were deobfuscated. Also, you can see a progress bar below the timestamp for documents that have a time-based access constraint and that indicate how much time is left until the document link expires. 

\subsubsection{Document Access View}

\begin{figure}[H]
  \centering
  \includegraphics[width=0.9\linewidth]{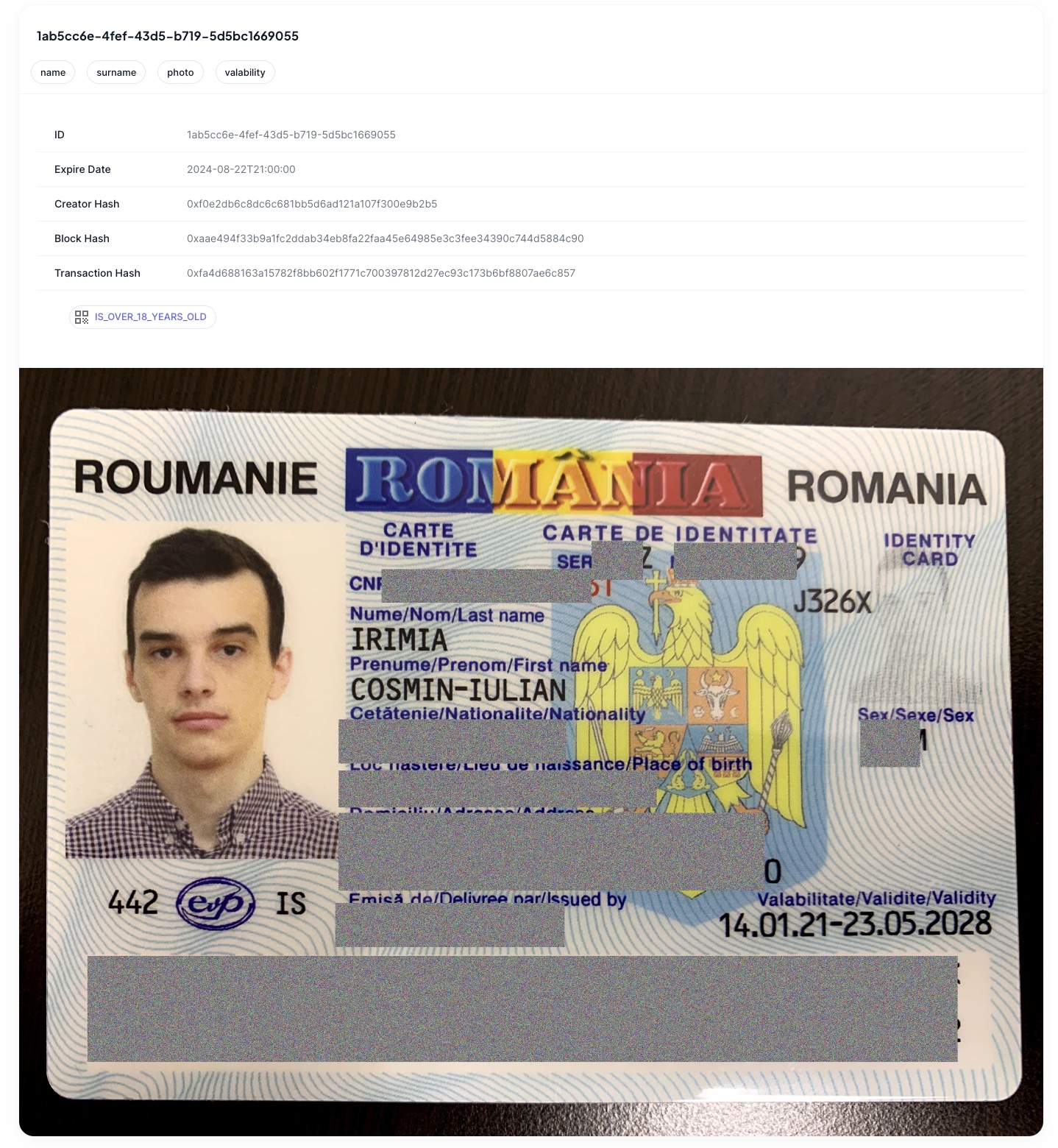}
  \caption{UI - Document Shared Link Access}
  \label{fig:no35}
\end{figure}

The last interface is a public screen where third parties can view and verify shared documents via a link. It provides an interface through which users can perform verification checks for document creator, specific block/transaction hashes, and even the original file on IPFS. Such a level of transparency provides trust and authenticity as the document origin and integrity can be validated independently by the users.

In addition, third parties can see the document field names and extracted facts derived directly from the document. This feature shows the system can give some meaningful and structured data, but without the document being opaque, and unusable.

\begin{figure}[H]
  \centering
  \includegraphics[width=0.8\linewidth]{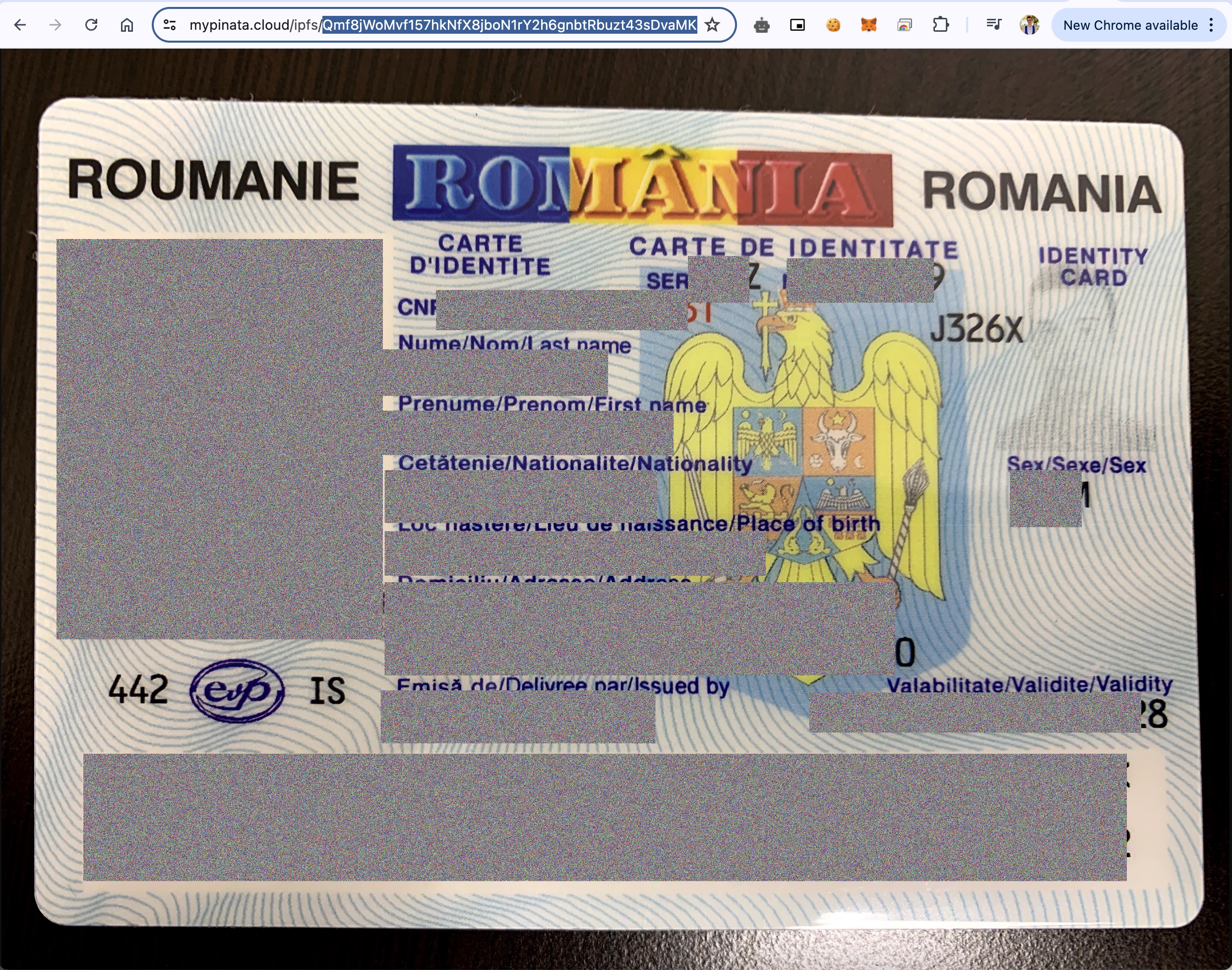}
  \caption{Obfuscated Document Access through IPFS}
  \label{fig:no36}
\end{figure}

Anyone with access to the IPFS hash can retrieve the original document, ensuring that its integrity is verifiable and its content immutable. However, access control must still be enforced externally, since IPFS does not provide native confidentiality or authorization mechanisms.

An awesome feature which could be implemented in this perspective is the possibility to share a document with a concrete person only. Suppose a police officer makes a QR code and gives it as a "document share invitation". The recipient then could see only the obfuscated document and selectively deobfuscate sections allowed by the document owner. Further security would be provided by making the URL or shared link unique to the requester and unusable by unauthorized parties.

Depending on the system configuration and requirements this could also be extended to include sharing with defined user IDs / groups / roles. A document for example could be shared with a set of users with fine access control while remaining private and transparent. These features are a step towards fairness and safe sharing.

The system has, therefore, fulfilled the objectives and criteria of the initial proposal. Implementation shows that proposed architecture \cite{5235004} can provide a robust, efficient and scalable solution when executed well. This is an achievement that validates the design system and its readiness for real applications. 

\subsection{KeyStore}

The system incorporates a dedicated keystore tier to securely manage cryptographic materials, particularly obfuscation keys. This storage layer, implemented using a solution such as HashiCorp Vault, operates independently from the main data stores and provides specialized functions for secure key storage, access control, and lifecycle management.

The keystore tier is designed to isolate sensitive keys from the application and blockchain layers, significantly reducing the risk of key exposure or compromise. Instead of storing keys within the GoQuorum blockchain—even with private transactions—the system retrieves keys securely from the keystore during runtime operations, ensuring both confidentiality and compliance with best practices in key management.

The service orchestrator interacts with the keystore using authenticated API calls, allowing dynamic key retrieval, rotation, and revocation without embedding secrets directly in the codebase or database. For end-users, these processes remain entirely transparent, with the system handling all cryptographic operations behind the scenes.  

\section{Concrete Architecture}

After analyzing the best technologies and options among those proposed and after completing the implementation of a system that complies with the standards proposed by the general description, we can finally build a diagram with the concrete architecture of the demo application.

The architectural workflow of DocChain represents a robust integration of various technological components, each playing a specific role in ensuring efficient document processing and security. The system begins with human interaction layers designed for both system administrators and end-users. For administrators, an interface built using Angular provides a responsive and efficient platform to manage and configure system settings. For users, a React-based interface offers an intuitive environment for document uploads, progress monitoring, and accessing processed documents. These interfaces ensure usability and accessibility, serving as the entry point to the DocChain system.

The Service Orchestrator acts as the entry point for user interactions and serves as the coordination layer among all backend microservices. Implemented using the Spring Boot framework, it follows a request-response model and exposes a RESTful API that delegates tasks to individual services based on their responsibilities. 

Its design follows the \textit{Gateway-Orchestrator} architectural pattern, separating concerns between user authentication, task routing, and workflow state management. This separation enhances modularity, as each microservice can be updated or replaced independently without affecting the orchestration layer. 

Additionally, the orchestrator includes request validation, partial failure handling (via circuit breakers), and telemetry hooks to support observability and debugging in distributed deployments.

The data extraction module employs a combination of technologies to process uploaded documents. OpenCV, a powerful image processing library, is used to analyze document structures, while Tesseract, an OCR engine, extracts textual information. These tools, along with Python and Flask, enable the extraction module to efficiently process and extract relevant data from documents. The extracted data serves as the foundation for subsequent steps in the workflow.

\begin{figure}[H] 
  \centering
  \includegraphics[width=0.76\linewidth]{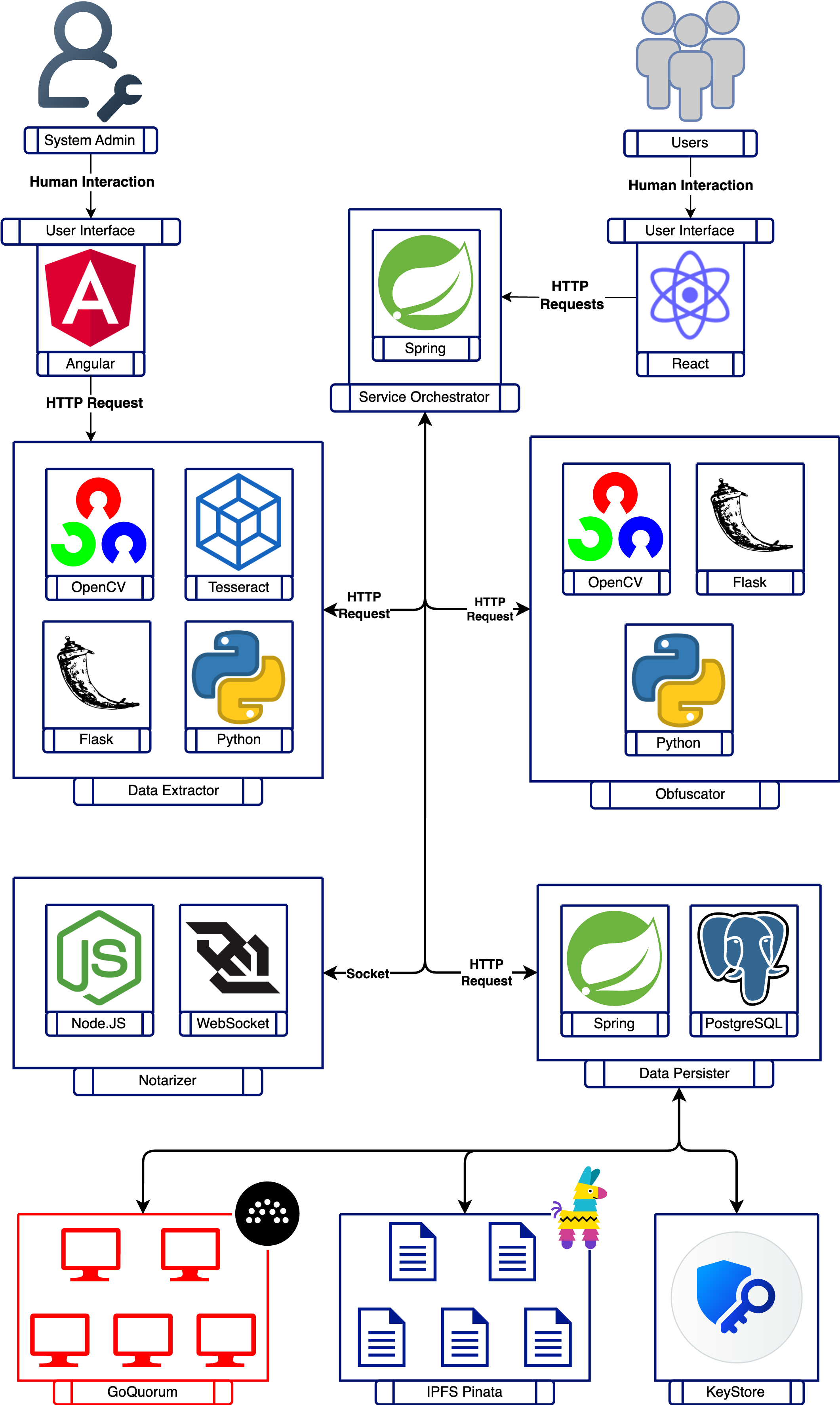}
  \caption{Proposed Concrete System Architecture}
  \label{fig:no37}
\end{figure}

To enhance privacy, the system integrates an obfuscation module. This module uses OpenCV to obfuscate sensitive sections of documents through methods such as pixelation or encryption, ensuring compliance with privacy standards. Python is used to implement the obfuscation logic, while Flask provides a lightweight framework for handling HTTP requests. This layered approach ensures the protection of sensitive information while preserving essential document details. The notarization process is facilitated by a module built using Node.js and WebSockets. This combination enables real-time updates and efficient handling of notarization tasks. The notarizer module ensures the integrity and authenticity of the extracted data by anchoring it to a blockchain network. This step is crucial for establishing trust and immutability within the system.

Finally, the data persistence module leverages Spring and PostgreSQL to manage the secure storage of extracted data and obfuscated documents. Additionally, the module interacts with distributed storage systems, such as IPFS Pinata, to provide decentralized and tamper-proof storage solutions. 

Integration with a GoQuorum blockchain further enhances the system’s security by maintaining an immutable record of all transactions.

This architectural design exemplifies the thoughtful integration of advanced technologies, ensuring a secure, efficient, and user-friendly system for document processing and notarization. Each component’s role and interactions are meticulously designed to address the challenges of privacy, security, and scalability in a modern document management system.

\mysection{Usage Scenarios and Flows}{Usage Scenarios}

In this chapter we will present two real-life usage scenarios and the steps that our solution will take in order to tackle every step of the problem.  

\subsection{Age Verification for Accessing a Restricted Service}

Imagine a user needs to verify their age to access an age-restricted service, such as purchasing alcohol online. Traditional methods typically require the user to upload a government-issued ID, which exposes sensitive information like their name, birthdate, address, and identification number. This approach carries risks of data breaches and unauthorized access. In contrast, DocChain provides a privacy-preserving solution using blockchain and fact extraction. Here’s how it works:

The user submits their ID card through a secure portal connected to DocChain. The system temporarily processes the ID card, extracting relevant data fields such as the user’s birthdate. The document itself is never stored, and the system uses end-to-end encryption (e.g., \ac{TLS}) during data transfer to prevent eavesdropping or interception. After submission, the Data Extractor \cite{bibitem_26} processes the document, identifying and labeling key data points. For example, the system could extract: 

\begin{lstlisting}[caption={Birthdate Extraction Example}, language=json, firstnumber=1]
{
    "name": "birthdate", 
    "value": "23/05/1997"
}
\end{lstlisting}

Based on this data, the system computes the derived fact: “The user is over 18 years old.”. The system uses both template-based and LLM-based \cite{10729345} extraction methods, validated through redundancy checks to ensure accuracy. The extracted fact, such as ["The user is over 18 years old"] is sent to a trusted party, such as a notary or an automated identity verification service \cite{UNGUREANU20243054}, for verification.

Once verified, the fact is digitally signed to ensure its authenticity and stored immutably on the blockchain. After generating and verifying the fact, the system discards the original ID card and any sensitive data extracted from it. Only the semantically relevant assertion—such as 'Over 18' in the case of a birthdate—is retained. These abstracted facts are computed using rule-based transformations applied to the extracted data. For example, the system interprets the date of birth and evaluates whether the user meets a threshold condition (e.g., age >= 18), without storing the original value. The abstraction logic is configurable per use case and document type, ensuring flexibility and data minimization across multiple verification scenarios.

When the user needs to prove their age, they share the blockchain-stored fact with the service provider. The provider verifies the fact on-chain without requiring the original ID card or any additional personal data. Access to the fact is controlled through user authentication and encryption.

\subsection{Academic Credential Verification}

When verifying academic credentials, traditional methods often require applicants to submit their entire transcripts or degree certificates. While this may seem straightforward, it poses significant privacy risks. These documents typically contain far more information than necessary—such as grades, enrollment history, and personal identifiers—none of which the verifier actually needs to confirm the qualifications. Beyond the risk of data breaches, there’s also the inconvenience of manually managing and sharing these documents.

This is where DocChain comes in, abstracting essential information into concise, verifiable "facts." These facts, such as "The candidate holds a Master of Science in Computer Science" or "Graduated in 2023 from the University of Excellence," are verified by the issuing institution and stored immutably on the blockchain, ensuring that only the relevant information is shared, while sensitive details—like grades or student ID numbers—are securely discarded after processing.

Here’s how it works: When a candidate needs to verify their degree for a job application, they upload the document to a secure portal connected to DocChain. The system temporarily processes the file, using advanced extraction methods to pull out key details such as the degree title, issuing institution, and graduation year. For example, from a degree certificate, DocChain might extract:\\

\begin{lstlisting}[caption={Degree Extraction Example}, language=json, firstnumber=1]
{
  "degree": "Master of Science in Computer Science",
  "institution": "University of Excellence",
  "graduation_year": "2023"
}
\end{lstlisting}

Then the system produces specific facts from that data. These facts are forwarded for verification to the issuing institution or another trusted authority. They're signed digitally and recorded on the blockchain when confirmed. Now all sensitive data including the original document is removed from system.

This is very efficient for employers. And instead of bulky transcripts or certificates, they check the blockchain to confirm what they need. So instead of handling a full academic record, an employer might see something like "The candidate has a Master's degree in Computer Science from the University of Excellence in 2023." This cuts out the irrelevant stuff like grades or enrollment history.

DocChain makes credential verification safer - it changes how we share and manage sensitive information. It is more than just verifying a degree. It is about doing so while respecting the candidate's privacy and making the process faster and more reliable for employers. Also, this framework is very versatile: applications in academia, professional licensing, and even cross-border qualification recognition.

In a privacy - obsessed world, DocChain sets a new standard for how we handle academic credentials. It demonstrates how secure verification can be without compromising the candidate's sensitive data.

\section{Conclusions}
\index{Scen|(}

This chapter presented a functional implementation of the DocChain system, validating its architectural design and core privacy-preserving mechanisms. The microservice based structure enabled modular development and allowed for independent testing of major components.

The Data Extractor demonstrated accurate field recognition across templates and varied layouts, with the LLM-enhanced pipeline improving flexibility. The Document Obfuscator applied multi-layered masking strategies, and the Notarizer coordinated fact validation and hash generation.

Decentralized persistence was achieved using IPFS and GoQuorum, enabling independent verification of document integrity. The frontend interface integrated the full workflow while enforcing selective data sharing.

Although certain components—such as legal notarization—remain partially implemented, the system provides a strong foundation for secure, privacy-aware document processing in public or institutional settings.
\if@openright\cleardoublepage\else\clearpage\fi
\thispagestyle{empty}
\mbox{}

\chapter{Evaluation and Comparison}

This chapter evaluates the proposed system, DocChain, to assess its performance and security while benchmarking it against existing solutions. The evaluation focuses on quantitative and qualitative metrics to determine the system's effectiveness, reliability, and efficiency. The comparison is framed in the context of similar systems, highlighting strengths and identifying areas for improvement.

\section{Performance}
\index{Performance|(}

The performance of DocChain is analyzed across several key operations, including data extraction, obfuscation, and blockchain integration. These evaluations assess the system’s scalability, responsiveness, and adaptability under varying conditions.

\subsection{Data Extractor}

Data Extractor performance and scalability were tested thoroughly. The tests covered documents as small as 1 MB up to much larger files (50 MB), representing real-world use cases with different data volumes. Results revealed a linear scalability of the extractor while the template-based method showed consistent performance with increasing document size. The average extraction time for 1 MB document was recorded at about 300 milliseconds, and a 50 MB document took just under 2,400 milliseconds, providing predictable and reliable performance.

The relative evaluation of extraction methods revealed considerable differences in processing times and accuracy. For formatted structured documents, template-based processing on predefined patterns/rules proved efficient. But its accuracy dropped when applied to documents with irregular layouts or unstructured data. Alternatively, LLM-based extraction methods based on advanced machine learning models proved more accurate than expected for semi-structured or unstructured documents. For example, where document content differed significantly in format or complexity, LLM-based methods identified and extracted relevant fields correctly.

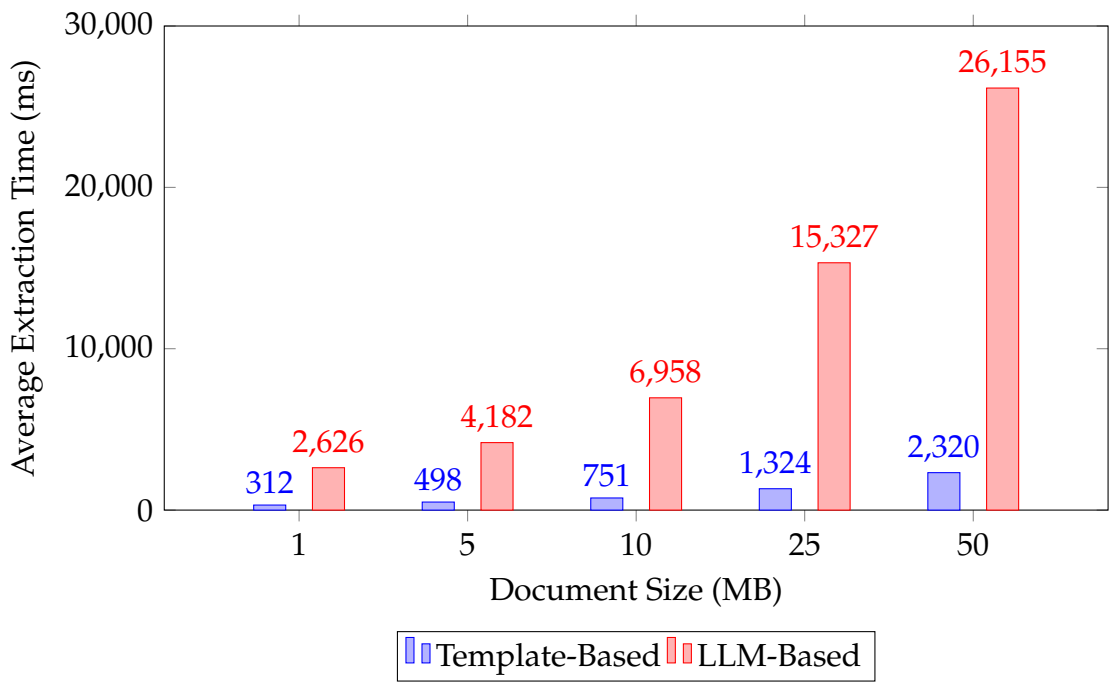
\begin{figure}[H] 
    \centering
    \caption{A Comparison between Extraction Times for Template-Based and LLM-Based Methods Across Varying Document Sizes}
    \begin{tikzpicture}
    \begin{axis}[
        ybar=10pt,
        bar width=12pt,
        width=14cm,
        height=8cm,
        enlarge x limits=0.2,
        legend style={at={(0.5,-0.25)}, anchor=north, legend columns=-1},
        ylabel={Average Extraction Time (ms)},
        xlabel={Document Size (MB)},
        symbolic x coords={1, 5, 10, 25, 50},
        xtick=data,
        ytick={0, 10000, 20000, 30000},
        ymin=0, ymax=30000,
        nodes near coords,
        nodes near coords align={vertical},
        scaled y ticks=false
    ]
    
    \addplot coordinates {(1, 312) (5, 498) (10, 751) (25, 1324) (50, 2320)};
    \addplot coordinates {(1, 2626) (5, 4182) (10, 6958) (25, 15327) (50, 26155)};
    
    \legend{Template-Based, LLM-Based}
    \end{axis}
    \end{tikzpicture}
\end{figure}

However, these gains in accuracy were traded for processing speed. The processing time of LLM-based methods was 8-11 times more expensive than template-based methods in all scenarios studied. As an example, template-based extraction of 1 MB document took 300 milliseconds whereas LLM-based extraction averaged more than 2,6 seconds. For larger documents this difference became more pronounced as processing times extended. But despite this, accuracy improvements provided by LLM-based methods have an average of 6\% gain (as it can bee seen in the table \ref{table:llm-times}) which made them the preferred choice in cases of high precision requirements, such as legal or academic document processing. 

This trade-off shows how DocChain can handle different performance and accuracy requirements. With this dual approach, the system can optimize for speed or precision based on the application specifics. As an example, template-based extraction may be preferred for high throughput environments like batch processing of standardized forms, while LLM-based methods are preferable for high accuracy applications with varied or nonstandard input.

\subsection{Blockchain Integration}

Blockchain integration was also evaluated to assess its performance in terms of transaction latency and throughput, both critical metrics for ensuring scalability and responsiveness in high-demand applications. This evaluation used a private blockchain implementation, GoQuorum, which is designed for enterprise-grade environments and optimized for high transaction throughput and low latency.

\begin{figure}[H] 
  \centering
  \caption{Blockchain Requests - Transaction Latencies under Varying Loads}
    \begin{tikzpicture}
\begin{axis}[
    width=14cm,
    height=6cm,
    xlabel={Simultaneous Requests},
    ylabel={Latency (seconds)},
    ymin=1.5, ymax=3,
    xmin=50, xmax=1050,
    xtick={100, 500, 1000},
    ytick={1.5, 2, 2.5, 3},
    grid=both,
    major grid style={line width=0.2mm,draw=gray!50},
    minor grid style={line width=0.1mm,draw=gray!20},
    legend style={at={(0.5,-0.3)},anchor=north,legend columns=-1}
    every axis plot/.append style={ultra thick}
]

\addplot[mark=square*,blue,thick] 
coordinates {
    (100, 1.8)
    (500, 2.0)
    (1000, 2.4)
};
\addlegendentry{Latency (seconds)}

\end{axis}
\end{tikzpicture}

\end{figure}
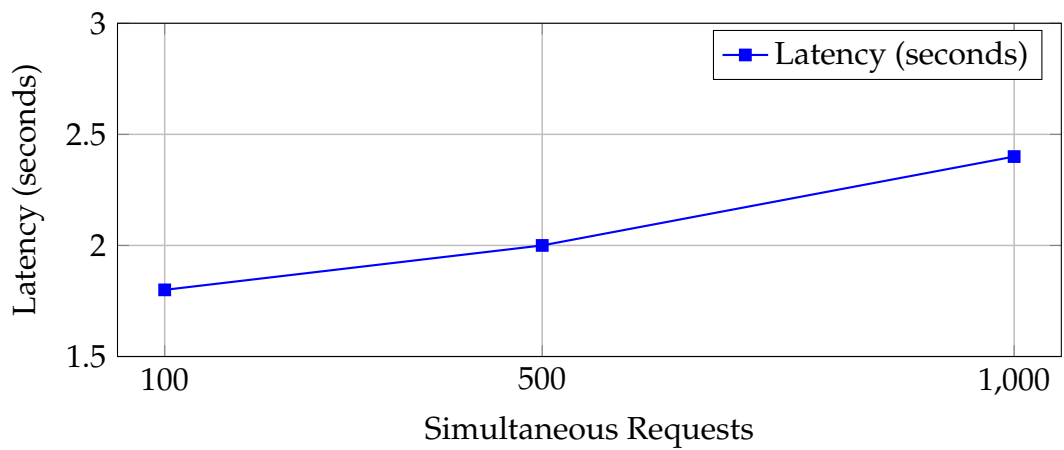

During testing, the system demonstrated an average transaction time of 2 seconds, even when subjected to workloads of up to 1,000 concurrent requests. This result highlights the system’s ability to handle simultaneous user interactions efficiently without significant performance degradation. The transaction time includes the end-to-end process of submitting a fact or hash to the blockchain, executing consensus among participating nodes, and recording the transaction in the distributed ledger.

Another aspect evaluated was the system's throughput, measured in terms of the number of transactions processed per second (TPS). Under optimal network conditions, the blockchain handled 500 to 800 TPS without noticeable latency increases, demonstrating its suitability for use cases requiring high-volume interactions, such as large-scale document verification or real-time compliance monitoring.

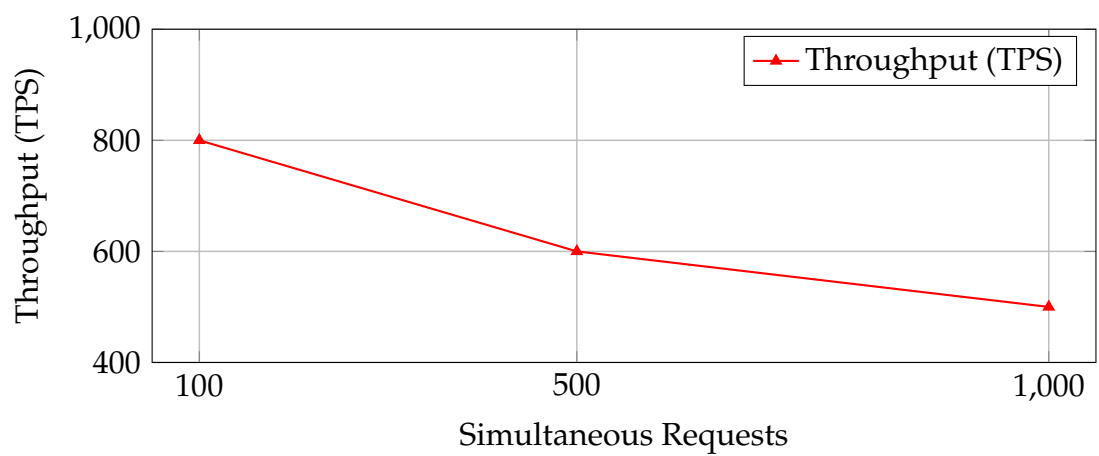
\begin{figure}[H] 
  \centering
  \caption{Blockchain Requests - Throughput under Varying Loads}

    \begin{tikzpicture}
\begin{axis}[
    width=14cm,
    height=6cm,
    xlabel={Simultaneous Requests},
    ylabel={Throughput (TPS)},
    ymin=400, ymax=1000,
    xmin=50, xmax=1050,
    xtick={100, 500, 1000},
    ytick={400, 600, 800, 1000},
    grid=both,
    major grid style={line width=0.2mm,draw=gray!50},
    minor grid style={line width=0.1mm,draw=gray!20},
    legend style={at={(0.5,-0.3)},anchor=north,legend columns=-1}
    every axis plot/.append style={ultra thick}
]

\addplot[mark=triangle*,red,thick] 
coordinates {
    (100, 800)
    (500, 600)
    (1000, 500)
};
\addlegendentry{Throughput (TPS)}

\end{axis}
\end{tikzpicture}

\end{figure}

Despite its strong performance, blockchain latency is inherently influenced by network infrastructure, including node configurations, geographic distribution, and hardware capabilities. For example, in a test environment with geographically distributed nodes, transaction times increased slightly to 2.5–3 seconds, reflecting the additional overhead of cross-region communication. These findings emphasize the importance of deploying the blockchain infrastructure strategically, such as clustering nodes in proximity to major user bases, to minimize latency.

\newpage

\section{IPFS}

One more important aspect to evaluate was the integration with IPFS: the performance evaluation of IPFS focuses on its ability to handle document storage and retrieval within the DocChain system. Tests were conducted to measure key metrics such as file upload and retrieval latency, storage costs, and system scalability under various network conditions.

The performance evaluation of IPFS was conducted under realistic conditions, focusing on file storage and retrieval times, cost efficiency, and scalability. Tests involved documents of varying sizes (1 MB, 10 MB, and 50 MB) across different scenarios.

The following table compares IPFS with blockchain-only and centralized storage solutions across key metrics:

\begin{table}[H]
\caption{IPFS Comparison with Alternative Storage Approaches}
\begin{tabular}{|p{3.2cm}|p{3.2cm}|p{4cm}|p{3cm}|}
\hline
\rowcolor[HTML]{EFEFEF} 
\multicolumn{1}{|p{3.2cm}|}{\cellcolor[HTML]{EFEFEF}Feature} & 
\multicolumn{1}{p{3.2cm}|}{\cellcolor[HTML]{EFEFEF}IPFS} & 
\multicolumn{1}{p{4cm}|}{\cellcolor[HTML]{EFEFEF}Blockchain-Only Storage} & 
\multicolumn{1}{p{3cm}|}{\cellcolor[HTML]{EFEFEF}Centralized Storage} \\ \hline
Censorship Resistance                                          & High                                                       & High                                                                          & Low                                                                       \\ \hline
Decentralized Replication                                      & Yes (on-demand)                                            & Full (all nodes replicate)                                                    & No                                                                        \\ \hline
Latency                                                        & Moderate (500–3,800 ms)                                    & High (5,000–8,000 ms)                                                         & Low (30–60 ms)                                                            \\ \hline
Data Persistence                                               & Requires pinning                                           & Permanent                                                                     & Dependent on provider                                                     \\ \hline
Storage Cost (1 MB file)                                       & \$0.04–\$0.06                                                & \$4.50–\$5.50                                                                   & \$0.02–\$0.10                                                               \\ \hline
Scalability                                                    & High                                                       & Limited (due to storage cost)                                                 & High                                                                      \\ \hline
\end{tabular}
\end{table}

\section{Storage and Processing Costs}
\index{Storage and Processing Costs|(}

Understanding the storage and processing costs of the proposed solution is crucial for evaluating its feasibility and comparing it with existing centralized and blockchain-based document management solutions. The system relies on a hybrid storage approach, combining Pinata for document storage and GoQuorum for blockchain-based notarization. The cost breakdown is as follows:

The system relies on a hybrid storage approach, combining Pinata for document storage and GoQuorum for blockchain-based notarization. The cost breakdown is detailed as follows.

Pinata offers consumer-based pricing plans, including a cost of \$20 per month for 1TB of storage and \$100 per month for 5TB of storage, translating to an approximate cost of \$0.02 per GB per month, or \$0.24 per GB per year. Enterprise pricing is custom, but estimating from these figures, a bulk storage option could likely reduce costs to between \$0.015 and \$0.02 per GB per year for larger-scale usage. If we assume that an average document size is 100 KB, storing one million documents, which requires approximately 100 GB of storage, would cost between \$24 and \$30 per year, while storing ten million documents (1TB) would cost between \$240 and \$300 per year under consumer pricing plans.

For blockchain storage, GoQuorum is chosen as the preferred private blockchain solution to eliminate transaction gas fees associated with public blockchains. Instead of paying per transaction, GoQuorum requires infrastructure maintenance. The estimated annual infrastructure cost for running a single-node GoQuorum setup on AWS EC2 \texttt{t3.medium} (2 vCPUs, 4GB RAM) is approximately \$40 per month, totaling \$480 per year. A more robust multi-node setup using AWS EC2 \texttt{m5.large} (2 vCPUs, 8GB RAM) would cost approximately \$100 per month, or \$1,200 per year. If an enterprise-grade five-node redundant setup is required, the cost would be around \$500 per month, leading to an estimated total of \$6,000 per year. While these costs scale with the redundancy requirements of the system, the per-document transaction cost remains negligible.

\subsection{Processing Costs}
The system's computational needs involve three primary tasks: data extraction, document obfuscation and encryption, and blockchain transactions. Since the solution is intended to operate in the cloud, AWS EC2 is chosen as the primary processing platform, with Azure Virtual Machines (VMs) also considered for comparative analysis.

For data extraction using Optical Character Recognition (OCR) and AI-based processing, an AWS EC2 \texttt{c5.large} instance (2 vCPUs, 4GB RAM) is estimated to cost approximately \$0.10 per document, while an equivalent Azure VM costs around \$0.12 per document. Document encryption and obfuscation require similar computational power, with estimated processing costs of approximately \$0.05 per document on AWS EC2 \texttt{t3.medium} instances and around \$0.06 per document on an Azure equivalent. Blockchain transactions, which include notarization and metadata storage, have an estimated computational cost of approximately \$0.001 per transaction on AWS and approximately \$0.002 per transaction on Azure. Combining these factors, the total estimated processing cost per document using AWS EC2 falls between \$0.15 and \$0.30, while the equivalent cost using Azure Virtual Machines ranges between \$0.18 and \$0.35.

\subsection{Cost Comparison with Similar Solutions}
To better understand the economic feasibility of the proposed system, a comparative analysis is performed between different document storage and notarization solutions. 

\begin{table}[H]
\centering
\caption{Cost Comparison Table}
\begin{tabular}{|p{5cm}|p{2.7cm}|p{3.4cm}|p{3.1cm}|}
\hline
\rowcolor[HTML]{EFEFEF} 
\textbf{Solution}                                & \textbf{Storage Cost (per GB/year)} & \textbf{Transaction Cost (per document)} & \textbf{Processing Cost (per document)}     \\ 
\hline
Proposed System (Pinata + GoQuorum + AWS EC2)    & \$0.015 - \$0.024                   & \$0.001 (negligible)                     & \$0.15 - \$0.30                              \\ 
\hline
Proposed System (Pinata + GoQuorum + Azure VM)   & \$0.015 - \$0.024                   & \$0.002 (negligible)                     & \$0.18 - \$0.35                              \\ 
\hline
Centralized Cloud Storage (Google Drive, AWS S3) and Enterprise Blockchain (Hyperledger, AWS Managed) & \$0.23 - \$0.30                     & \$0.01 - \$0.05                                       & \$0.20 - \$0.50                             \\ 
\hline
Traditional Paper Documents                      & N/A                                 & N/A                                      & Unknown labor costs  \\
\hline
\end{tabular}
\end{table}



The proposed system, which leverages Pinata for IPFS storage and GoQuorum for notarization, achieves storage costs between \$0.015 and \$0.024 per GB per year, with per-document transaction costs as low as \$0.001. The processing cost per document, when using AWS EC2, is estimated between \$0.15 and \$0.30, while Azure incurs slightly higher costs, ranging from \$0.18 to \$0.35. In comparison, centralized cloud storage solutions such as Google Drive and AWS S3 incur storage costs between \$0.23 and \$0.30 per GB per year but do not provide native notarization mechanisms. Enterprise blockchain solutions such as Hyperledger and AWS Managed Blockchain typically incur transaction costs between \$0.01 and \$0.05 per document, with processing costs ranging from \$0.20 to \$0.50 per document. Traditional paper-based document storage does not involve digital storage costs, but it incurs high labor costs associated with manual processing, estimated at over \$1.00 per document.

By leveraging Pinata for IPFS storage and GoQuorum for private blockchain notarization, the proposed system achieves low storage and transaction costs while maintaining decentralization and security. Compared to centralized solutions, block-chain-based notarization provides cost-efficient scalability, while AWS EC2 and Azure processing ensure computational efficiency. While Azure's pricing is slightly higher than AWS, both cloud providers offer scalable, on-demand processing solutions suitable for large-scale document notarization and verification.

\subsection{Sustainability and Cost Model}

Although IPFS and blockchain technologies offer strong guarantees of availability and integrity, they introduce operational costs that must be addressed explicitly. In the current implementation, the system assumes that a centralized operator or consortium maintains IPFS pinning nodes and covers blockchain transaction fees required for fact anchoring. This model simplifies user onboarding, but it does not scale in fully decentralized deployments.

For long-term sustainability, multiple approaches can be considered:

\begin{itemize}
    \item Users or institutions may pre-pay credits to cover their storage and notarization costs.
    \item Public infrastructure grants or university-hosted nodes can subsidize initial usage.  
    \item Integration with decentralized storage incentivization models (e.g., Filecoin, Arweave) may allow for automatic compensation of hosting nodes. 
\end{itemize}

Currently, the system does not operate autonomously, and its deployment requires administrative coordination to ensure data persistence and transaction funding. Future work should explore mechanisms for decentralized governance, cost-sharing, and self-sustaining operation. 

\section{Security}
\index{Security|(}

DocChain incorporates multiple layers of security designed to protect sensitive data, prevent fraud, and ensure document authenticity. These measures include:

\begin{itemize}
  \item 
\textbf{Data Minimization}: No personally identifiable information (PII) is stored. Extracted facts are used transiently and discarded after processing.
  \item 
\textbf{Blockchain Anchoring}: Document hashes are anchored to GoQuorum, making them tamper-evident and verifiable by third parties.
  \item 
\textbf{Obfuscation and Encryption}: Sensitive regions are masked using layered obfuscation and, where applicable, pixel-wise encryption. Keys are stored client-side, making interception ineffective.
  \item 
\textbf{User Verification}: Submission portals can require biometric verification or signed login tokens (e.g., MetaMask) to reduce fraud.
\end{itemize}

These mechanisms support GDPR, CCPA, and zero-trust principles while ensuring independently verifiable credential flow.



    
    
    
    

\section{Comparison with Similar Solutions}
\index{Comp|(}

In this chapter, we will compare our system with the identified similar solutions from the second chapter.

\subsection{DocuSign}

DocChain and DocuSign represent two distinct approaches to document management and verification, each optimized for specific applications. DocuSign is a widely adopted platform for managing and signing digital agreements, offering robust document workflows and legal enforceability. In contrast, DocChain prioritizes privacy and minimal data retention, focusing on extracting and verifying “facts” from documents without storing sensitive information.

The core distinction lies in their handling of data. DocChain is designed to extract specific information (e.g., verifying a user is over 18 without exposing their ID) and store it as immutable proofs on a blockchain. By discarding sensitive data after extraction, DocChain ensures compliance with privacy regulations like GDPR and reduces the risks associated with data breaches. DocuSign, on the other hand, retains entire documents securely, employing encryption and audit trails to protect their integrity, making it suitable for use cases where the full content of a document needs to be accessible.

From a technological perspective, DocChain’s blockchain-based architecture ensures tamper-proof verification and decentralization, aligning with privacy-sensitive industries such as healthcare and compliance auditing. Meanwhile, DocuSign relies on centralized infrastructure optimized for scalability and integration with enterprise workflows, excelling in scenarios like real estate and financial agreements where end-to-end document visibility is critical.

In terms of fraud prevention, DocChain’s approach centers on immutable blockchain records and the absence of sensitive data storage, while DocuSign ensures document authenticity through cryptographic methods and an extensive audit trail. However, DocuSign’s retention of full documents introduces greater exposure to potential breaches compared to DocChain’s privacy-first philosophy.

Ultimately, DocChain and DocuSign serve complementary purposes. DocChain’s innovative fact-based verification model is ideal for scenarios requiring strict data minimization, whereas DocuSign’s comprehensive agreement management capabilities are indispensable for industries needing legally binding digital signatures and full document workflows.

\subsection{BlockCerts}

Blockchain is a cornerstone of secure systems, and DocChain and BlockCerts are examples of its potential in document management. While they share a common foundation, the systems differ considerably in focus and implementation. BlockCerts tackles credential verification by creating immutable records of digital certificates on public blockchains. Meanwhile, DocChain prioritizes privacy, extracting discrete facts from documents while ensuring sensitive information remains protected.

In BlockCerts, complete credentials — such as degrees or professional certifications — are recorded directly on the blockchain. This ensures their authenticity and accessibility, allowing anyone to verify them without relying on centralized authorities. However, this approach often requires users to store full certificates locally, which raises privacy concerns and complicates recovery if the local copy is lost. In contrast, DocChain extracts only the essential information needed for verification. For instance, instead of storing a full identity document, DocChain generates and records abstracted facts, such as confirming that someone is over 18 years old. Once these facts are verified, the original sensitive data is discarded, significantly reducing exposure to privacy risks.

The two systems diverge most sharply in their approach to privacy. While BlockCerts provides transparency and accessibility for credential verification, it does not inherently obscure sensitive details within the certificates. In comparison, DocChain is designed with privacy in mind, ensuring that personal data is neither stored nor shared unnecessarily, making it ideal for contexts such as age or residency verification, where only specific facts are needed, and the underlying document can remain confidential.

From an application perspective, the systems cater to different needs. BlockCerts excels in education and professional environments, where the ability to publicly verify credentials is critical. Its straightforward approach to recording and accessing digital certificates has made it a reliable solution in these domains. DocChain, however, is more versatile, supporting use cases in healthcare, legal compliance, and financial services. By focusing on fact-based verification and privacy preservation, DocChain addresses the needs of industries where data minimization and regulatory compliance are paramount.

In conclusion, although both systems utilize blockchain to ensure trust and data integrity, their strengths lie in different areas. BlockCerts is optimized for creating and verifying immutable credentials, while DocChain provides a flexible, privacy-focused solution for securely sharing facts derived from sensitive documents. These differences highlight how blockchain can be tailored to address diverse challenges in secure data management.

\subsection{SignRequest}

Digital systems for document management and verification continue to evolve, addressing unique challenges in security, usability, and compliance. While both DocChain and SignRequest contribute to this field, their goals and methodologies reflect distinct priorities. DocChain emphasizes privacy-preserving, fact-based verification, leveraging blockchain technology to securely abstract and share key information. In contrast, SignRequest focuses on streamlining electronic signatures and providing secure document workflows, ensuring compliance with legal and regulatory requirements.

SignRequest operates within a centralized framework, offering tools for signing, sending, and managing documents. Its core strength lies in its simplicity and ease of integration, allowing businesses to embed its functionality into their existing workflows using APIs. Documents processed through SignRequest are encrypted and include tamper-evident audit trails, ensuring their integrity and compliance with regulations like GDPR and eIDAS. However, the system retains the full content of documents, which may pose privacy risks in highly sensitive applications.

DocChain, on the other hand, takes a decentralized approach to document verification. Instead of storing entire documents, it extracts specific facts, such as verifying an individual's age or residency, and records these on a blockchain. Once a fact is validated, the original sensitive data is discarded, ensuring that no private information is retained or exposed. This makes DocChain particularly well-suited for industries where data minimization and privacy are paramount, such as healthcare, finance, and legal compliance.

A key distinction between the two systems lies in their handling of data. SignRequest is designed for scenarios where full document access and end-to-end workflows are essential, such as signing contracts or managing approvals. Its centralized architecture facilitates these processes but inherently requires retaining sensitive information. DocChain, in contrast, focuses on generating immutable proofs for specific facts, enabling trustless verification without compromising user privacy. This aligns it with use cases like proving eligibility or compliance without exposing the underlying document.

Both systems also address security, though through different mechanisms. SignRequest relies on encryption and tamper-evident logs to protect documents and provide an audit trail for verification. DocChain, by leveraging blockchain, ensures that stored facts are immutable and cannot be tampered with, offering a decentralized alternative that reduces reliance on centralized storage and processing.

Ultimately, the choice between these systems depends on the intended application. SignRequest is ideal for workflows that require complete document handling and legally binding signatures, particularly in corporate or contractual settings. DocChain, with its innovative focus on privacy-preserving verification, offers a compelling solution for situations where proving specific facts without revealing sensitive information is critical. 

\subsection{Jumio}

Rise of digital identity verification/document management systems such as DocChain and Jumio address critical issues of security, trust and compliance. These solutions, while aiming at guaranteeing the authenticity of the users' information, differ quite substantially amongst each other. DocChain is about decentralized, privacy-preserving verification based on extracting and sharing abstracted facts from documents while Jumio is about centralized identity verification based on biometrics, artificial intelligence and document validation.

Jumio serves as a centralized identity verification platform that authenticates user identities by analyzing government-issued ID documents against facial recognition data. Its system works with worldwide databases to validate documents in real time. A strength of Jumio is its ability to detect fraudulent activity via AI-driven anomaly detection - an ideal solution for banking, e-commerce and healthcare, but its reliance on centralized data processing \cite{Pai1995} and storage creates privacy risks as personal information must be retained for verification purposes.

DocChain does the opposite - it uses decentralized data verification. It does not retain sensitive personal details but extracts facts like age or residency status from documents and stores them on a blockchain. After facts are confirmed, original data is discarded so no private information remains in the system. With this approach privacy \& regulatory compliance are the top priorities, so DocChain is suited for use cases with low data exposure like compliance audits or eligibility checks.

The two systems are also different in their architectural focus. Jumio uses centralized infrastructure for real-time identity verification enabling scalability for big applications but requiring the retention and processing of private user data. DocChain, for its part, uses blockchain to decentralise data storage and ensure immutability of the extracted facts. 

As far as application goes, Jumio is most useful when instantaneous and complex identity verification is required - for example, onboarding customers in financial services or confirming identities for e-commerce platforms. It combines AI and biometrics for high security and usability at the expense of maintaining sensitive data. DocChain, for its part, is more concerned with situations where proving facts without disclosing the document behind them is critical. That extends to privacy sensitive industries similar to healthcare where there is a priority to expose minimal data. The choice between such solutions is ultimately a matter of the requirements of the use case. Jumio's centralized, AI-driven system is ideal for organizations looking for fast, full-feature identity verification workflows. To the contrary, DocChain is a decentralized, privacy-first alternative for secure fact verification without storing personal data. Both systems reflect the moving target of digital identity and document management, and how various approaches can accommodate varying security and compliance requirements.

\subsection{Onfido}

Ensuring the authenticity and privacy of user data is a common goal for both DocChain and Onfido, but their methods reflect very different priorities. While DocChain adopts a decentralized, privacy-preserving verification approach, Onfido operates as a centralized identity verification platform, using artificial intelligence and biometrics to authenticate individuals. These contrasting designs demonstrate different strategies for addressing identity and document-related challenges.

Onfido combines document analysis with biometric authentication for real-time identity verification. The system checks the consistency of an ID document and a selfie submitted by a user, utilizing AI to detect fraud or anomalies. This enables Onfido to support high-volume identity checks for industries such as finance, insurance, and the gig economy. However, its centralized model requires storing sensitive user data, which may pose privacy risks and increase exposure in the event of a security breach.

In contrast, DocChain prioritizes data minimization and user privacy. Instead of authenticating an entire identity document, it extracts and validates specific facts, such as whether a person is over 18 or resides in a specific jurisdiction. These facts are immutably stored on a blockchain, and the original data is discarded after processing. This approach eliminates the need to store sensitive information, aligning with privacy regulations like GDPR and addressing concerns about data breaches.

The two systems also differ significantly in their technological architecture. Onfido operates as a centralized service, offering scalability and integration into enterprise workflows via APIs. This allows for real-time processing but requires trust in the platform’s centralized storage and processing capabilities. Conversely, DocChain uses blockchain to decentralize data storage and validation, reducing reliance on centralized servers while enhancing transparency and security.

In terms of application scope, Onfido is designed for use cases requiring full identity verification, such as customer onboarding, KYC compliance, and fraud prevention. It excels in speed and precision but retains sensitive data as part of its operations. DocChain, on the other hand, is best suited for situations where only specific facts need to be verified without disclosing the underlying document, such as eligibility confirmation or compliance checks in privacy-sensitive industries like healthcare and finance.

In summary, Onfido provides a robust, centralized solution for end-to-end identity verification, emphasizing speed and scalability. DocChain, with its decentralized and privacy-first approach, offers a flexible alternative for securely sharing facts derived from documents without compromising user privacy. These differences highlight how distinct architectural and operational priorities can address varying needs in the evolving landscape of identity and document verification.

\subsection{Trulioo}

Modern identity verification and document management systems are designed to address challenges such as compliance, scalability, and privacy. DocChain and Trulioo represent two specialized solutions tailored to these needs, but their methodologies and focuses differ significantly. DocChain emphasizes decentralized, privacy-preserving fact verification, while Trulioo leverages centralized infrastructure and global data networks to deliver large-scale identity verification services.

Trulioo connects organizations to over 400 data sources across more than 195 countries, enabling real-time user identity verification. This extensive coverage makes it an industry leader in handling cross-border compliance requirements, including Know Your Customer (KYC) and Anti-Money Laundering (AML) regulations. Its GlobalGateway product allows businesses to validate a wide range of data, from government-issued IDs to address records and phone numbers. However, its centralized approach requires the collection and processing of sensitive user data, raising concerns about long-term privacy and data security.

In contrast, DocChain adopts a decentralized and privacy-first model. Instead of storing private user information, DocChain extracts specific facts from documents—such as confirming a user’s age or residency status—and records these facts immutably on a blockchain. After processing, the system discards the original data, ensuring compliance with privacy regulations like GDPR and minimizing the risk of data breaches. This approach allows DocChain to address use cases requiring precise, fact-based verification without exposing sensitive details, making it ideal for privacy-conscious industries such as healthcare and legal compliance.

The architectural differences between the two systems are stark. Trulioo’s centralized infrastructure enables seamless scalability for high-volume, real-time identity checks, particularly in industries with significant compliance needs, such as finance and e-commerce. However, this model relies on the aggregation and storage of large amounts of personal data, which could be vulnerable in the event of a breach. On the other hand, DocChain decentralizes data validation, ensuring that facts remain tamper-proof and independently verifiable without requiring sensitive information to be stored or shared.

Trulioo’s greatest strength lies in its ability to handle diverse identity verification scenarios across international jurisdictions, offering a comprehensive solution for both individuals and businesses. It excels in providing reliable, large-scale identity checks but does so by relying on repositories of sensitive data. DocChain, meanwhile, is better suited for scenarios where minimal data exposure is critical, such as proving eligibility or compliance without disclosing full personal information.

In summary, while both systems aim to ensure trust and compliance in identity verification, their approaches cater to different priorities. Trulioo’s centralized infrastructure provides comprehensive, global identity checks, making it ideal for industries requiring large-scale solutions. DocChain, with its decentralized, privacy-preserving design, offers a targeted and secure alternative for verifying specific facts without compromising user privacy. Together, these solutions demonstrate the diversity of approaches in modern identity and document management technologies.

\subsection{Comparative Feature Table}

The following table provides a comparative analysis of various document management and verification systems, including DocuSign, BlockCerts, SignRequest, Jumio, Onfido, Trulioo, and our proposed solution, DocChain. Its primary aim is to highlight the distinguishing features of each system, emphasizing aspects such as decentralization, privacy preservation, data minimization, and fact-based verification. By systematically comparing these features, this demonstrates how DocChain uniquely addresses privacy and compliance requirements through its decentralized, blockchain-based architecture. In contrast, other solutions often rely on centralized infrastructures that require full document retention, potentially compromising user privacy. This comparison helps to contextualize the need for DocChain, showcasing its strengths in privacy preservation, data minimization, and secure fact verification.

\begin{table}[H]
\centering
\caption{Comparative Feature Analysis of Document Management and Verification Systems}
\begin{tabular}{|p{4cm}|p{1.4cm}|p{1.2cm}|p{1.6cm}|p{1.8cm}|p{1.4cm}|p{1.9cm}|}
\rowcolor[HTML]{EFEFEF} 
\hline
\textbf{Feature} & \textbf{Docu Sign} & \textbf{Block Certs} & \textbf{Sign Request} & \textbf{Jumio/ Onfido} & \textbf{Trulioo} & \textbf{DocChain} \\ \hline
\textbf{Decentralization} & No & Yes & No & No  & No & Yes \\ \hline
\textbf{Privacy Preservation} & No & No & No & No  & No & Yes \\ \hline
\textbf{Data Minimization} & No & No & No & No  & No & Yes \\ \hline
\textbf{Immutable Proofs on Blockchain} & No & Yes & No & No  & No & Yes \\ \hline
\textbf{Full Document Retention} & Yes & Yes & Yes & No & No & No \\ \hline
\textbf{Fact-based Verification} & No & No & No & No  & No & Yes \\ \hline
\textbf{Tamper-Proof Audit Trails} & Yes & Yes & Yes & Yes  & Yes & Yes \\ \hline
\textbf{Centralized Architecture} & Yes & No & Yes & Yes  & Yes & No \\ \hline
\textbf{Compliance with Privacy Laws} & Yes & No & Yes & No  & No & Yes \\ \hline
\textbf{Scalable Enterprise Integration} & Yes & No & Yes & Yes & Yes & Yes \\ \hline
\textbf{Use Case Flexibility} & Medium & Low & Medium & High & High & High \\ \hline
\end{tabular}%
\end{table}

\subsection{General View}

To provide additional context, we created a comparison table with similar systems in order to better view the differences between DocChain and it's main rivals.

\begin{table}[H]
\caption{DocChain Comparison with Alternative Solutions}
\begin{tabular}{|p{3cm}|p{4cm}|p{3.3cm}|p{3.3cm}|}
\hline
\rowcolor[HTML]{EFEFEF} 
\multicolumn{1}{|c|}{\cellcolor[HTML]{EFEFEF}Metric} & \multicolumn{1}{c|}{\cellcolor[HTML]{EFEFEF}DocChain}                         & \multicolumn{1}{c|}{\cellcolor[HTML]{EFEFEF}BlockCerts} & \multicolumn{1}{c|}{\cellcolor[HTML]{EFEFEF}Trulioo} \\ \hline
Extraction Time (1MB)                                         & \begin{tabular}[c]{@{}l@{}}300 ms (template - \\
based)  420 ms \\
(LLM-based)\end{tabular} & Not applicable (focuses on credential issuance)                  & 500 ms (structured data extraction)                           \\ \hline
Obfuscation Speed (1MB)                                       & 500 ms (multi-zone obfuscation)                                                        & Not supported                                                    & Not supported                                                 \\ \hline
Blockchain Latency                                            & 2 seconds (GoQuorum private chain, 1,000 requests)                                     & 4 seconds (public blockchain)                                    & Not applicable                                                \\ \hline
Scalability                                                   & Linear scalability (up to 50MB files)                                                  & Scalable, but limited by blockchain performance                  & Limited to structured/ID verification tasks                   \\ \hline
Security Mechanisms                                           & Blockchain immutability, multi-layer encryption, selective obfuscation                 & Blockchain immutability, public key cryptography                 & End-to-end encryption, identity verification                  \\ \hline
Accuracy                                                      & 98\% (reversible obfuscation techniques)                                               & High (focused on issuing tamper-proof certificates)              & 96\% (identity verification accuracy)                         \\ \hline
Cost Efficiency                                               & Medium (due to blockchain and IPFS use)                                                & Medium-High (focuses on low-cost certificate issuance)           & High (optimized for SaaS scalability)                         \\ \hline
Primary Use Case                                              & Document processing, verification, and secure storage                                  & Credential issuance and verification                             & Identity verification and global compliance                   \\ \hline
\end{tabular}
\end{table}

\section{Conclusions}
\index{Concl|(}

The evaluation of DocChain highlights its strengths as a privacy-focused, blockchain-based document processing system that offers a balance between security, efficiency, and scalability. Through rigorous performance testing, DocChain demonstrated its capability to extract, obfuscate, and verify document data with high accuracy and predictable processing times. The dual approach to data extraction—leveraging both template-based and LLM-based methods—ensures adaptability to various document types, while the obfuscation techniques provide flexible privacy controls for different use cases.

The system's integration with blockchain and IPFS enables decentralized, tamper-resistant data storage and retrieval, making it a viable alternative to traditional document management solutions. Compared to centralized competitors like DocuSign, Jumio, and Trulioo, DocChain prioritizes minimal data retention and user privacy, reducing the risks associated with data breaches while ensuring compliance with regulations like GDPR. However, scalability remains a key challenge, as blockchain transaction latency and IPFS retrieval speeds can impact real-world performance.

Despite its advantages, the widespread adoption of DocChain is contingent upon addressing regulatory challenges, user trust, and integration with existing enterprise systems. The implementation of Layer 2 scaling solutions, hybrid architectures, and user-friendly key management strategies can help mitigate these challenges and facilitate broader adoption.

Overall, DocChain presents a compelling alternative to traditional document verification systems, particularly in privacy-sensitive industries such as healthcare, finance, and legal compliance.  

\section{Legal and Regulatory Considerations}
\index{Legal and Regulatory Considerations|(}

While the technical architecture of the system addresses the security, integrity, and availability of digital documents, legal recognition of blockchain-based notarization remains a critical factor for adoption.

Within the European Union, the eIDAS regulation establishes a legal framework for electronic signatures and trust services. Notably, eIDAS distinguishes between electronic signatures, advanced electronic signatures (AdES), and qualified electronic signatures (QES), only the latter being considered equivalent to handwritten signatures.

In Romania, Law no. 455/2001 on electronic signature aligns with eIDAS but currently lacks explicit provisions recognizing blockchain-based notarization or decentralized identity systems. Therefore, for full legal recognition, the system would need to integrate a qualified trust service provider (QTSP) or collaborate with such entities to issue QES.

Furthermore, all data handling mechanisms must be GDPR-compliant, especially regarding the right to erasure, data minimization, and purpose limitation. While blockchain offers immutability, this may conflict with the GDPR’s dynamic privacy requirements, which presents a regulatory paradox that must be considered in real-world deployments.

Future iterations of the platform should explore off-chain storage of personal data with on-chain references, combined with zero-knowledge proofs, to reconcile GDPR with blockchain's permanent nature. 
\if@openright\cleardoublepage\else\clearpage\fi
\thispagestyle{empty}
\mbox{}

    \chapter{Conclusions}

\section{Achieved Aims and Objectives}

This research proposed a decentralized infrastructure using blockchain, microservices, and IPFS to revolutionize digital notarization, signing, and document sharing. In this thesis, each objective was addressed methodically, culminating in a system that makes significant contributions to document management.

The primary goal of developing a modular microservice-based architecture was achieved. The architecture integrates core functionalities such as data extraction, obfuscation, and notarization while maintaining scalability and flexibility. Each microservice operates independently, enabling seamless updates and parallel processing, which is essential for handling large transaction volumes.

Privacy preservation, another key element of this work, was enabled through robust obfuscation mechanisms. These methods allow selective sharing of sensitive information in compliance with global privacy standards such as GDPR. By retaining only essential facts and discarding sensitive data, the system minimizes exposure to privacy breaches.

The integration of blockchain technology provided transparency and immutability. By recording cryptographic hashes and verified facts on the blockchain, the system guarantees tamper-proof records. This trustless verification mechanism eliminates the need for intermediaries, fostering confidence among both users and verifiers.

Scalability and usability challenges were addressed by designing intuitive user interfaces and employing advanced orchestration techniques. These features ensure that even non-technical users can interact with the system effectively. Additionally, the adoption of hybrid storage solutions, combining blockchain with IPFS, reduced storage costs while maintaining decentralized access.

By achieving these objectives, this research contributes a robust framework for modernizing bureaucratic processes, enhancing efficiency, and safeguarding user privacy.

\section{Critique and Limitations}

While the proposed system offers substantial benefits, several limitations warrant careful examination. One of the most significant challenges lies in the trade-offs associated with the chosen technologies. For instance, while IPFS provides decentralized storage and cost efficiency, its reliance on pinning strategies for data persistence introduces operational complexities. Unpinned files may experience longer retrieval times, which could negatively impact user experience in latency-sensitive scenarios. Similarly, the immutability of the blockchain, while advantageous for transparency, imposes scalability constraints. High transaction volumes can lead to network congestion and increased latency, potentially hindering system performance in high-demand use cases. Although these limitations are partially mitigated by offloading large files to IPFS, the underlying issues remain areas for further optimization.

Another critique pertains to the user dependency on trust during the initial verification of facts. While the system ensures integrity post-verification, errors or biases in the initial stages could propagate through the system, potentially undermining trust. Addressing this issue requires enhanced validation mechanisms, possibly involving AI-driven anomaly detection to flag inconsistencies during the verification process.

Lastly, adoption barriers, including technological literacy and infrastructure requirements, pose significant challenges. Not all users or institutions may possess the technical resources or understanding to fully leverage the system’s capabilities. Overcoming these barriers will require extensive education and outreach, as well as the development of simplified, accessible interfaces.

\section{Future Work}

The potential for enhancing and extending this system is vast. Future work should focus on addressing the identified limitations, with particular emphasis on improving scalability and performance. One promising avenue involves exploring hybrid architectures that combine IPFS with edge caching mechanisms, reducing retrieval times for frequently accessed files.

To enhance blockchain scalability, the adoption of layer-2 solutions such as rollups or sidechains could be investigated. These approaches can significantly reduce transaction costs and improve throughput, making the system more efficient for high-volume use cases.

Another area for future exploration is the integration of advanced machine learning models into the verification pipeline. By employing AI-driven techniques for anomaly detection and document analysis, the system could further automate and enhance the accuracy of fact verification.

Expanding the system's applicability to new domains, such as healthcare records, real estate transactions, or cross-border trade, could unlock additional value. Each of these domains presents unique challenges, requiring tailored solutions to meet specific regulatory and operational requirements.

Additionally, research into privacy-preserving cryptographic techniques, such as zero-knowledge proofs, could enable even greater privacy without compromising usability. This would allow users to prove specific facts without revealing the underlying data, further reducing exposure to privacy risks.

Finally, efforts should be directed toward improving accessibility and adoption. This includes developing lightweight versions of the system for resource-constrained environments, as well as conducting outreach programs to educate stakeholders about the benefits and functionality of the system.

\section{Final Remarks}

The work presented in this thesis represents a significant step forward in the domain of decentralized document management. By combining blockchain technology with microservices and IPFS, this research addresses critical challenges in privacy and efficiency, offering a novel solution that is both robust and scalable.

While limitations and challenges remain, the contributions of this work lay a strong foundation for future innovation. The system’s potential to transform bureaucratic processes, enhance transparency, and empower users to take control of their data underscores its broader societal impact.

As digital transformation accelerates across industries, the principles and technologies explored in this thesis will only grow in relevance. By addressing the identified limitations and pursuing the outlined future directions, this research paves the way for a user-centric digital future.
\if@openright\cleardoublepage\else\clearpage\fi
\thispagestyle{empty}
\mbox{}



{\backmatter
    \if@openright\cleardoublepage\else\clearpage\fi

    \begingroup
    \renewcommand{\bibfont}{\normalsize} 
    \renewcommand{\baselinestretch}{1.5} 
    \large 

    \bibliographystyle{unsrt}
    {
    \footnotesize\bibliography{biblio}
    }
    \endgroup
}

\end{document}